%% file: Competition_Between_Regulation-Providing_and_Fixed-Power_Charging_Stations_for_Electric_Vehicles.tex
\documentclass[conference]{IEEEtran}

\input{Preamble.tex}

\begin{document}

\title{Competition Between Regulation-Providing and Fixed-Power Charging Stations for Electric Vehicles}

\author{Wenjing~Shuai, Patrick~Maill\'e, Alexander~Pelov\\Institut Mines-Telecom/Telecom Bretagne
\thanks{The three authors are with the Networks, Security, Multimedia Department, Institut Mines-Telecom/Telecom Bretagne, 2 rue de la Chataigneraie CS 17607, 35576 Cesson-S\'evign\'e, FRANCE e-mail: \{first\}.\{last\}@telecom-bretagne.eu}
}

\maketitle

\begin{abstract}
\short{This paper models a non-cooperative game between two EV charging stations. One is a fixed-power charging station who purchases electricity from the grid at wholesale price and resell the energy to EV owners at a higher retail price. The other is a regulation-providing who varies the recharging power level of its clients to provide regulation services to the grid, so its profit comes from both the EV owners who get energy and the grid who benefits from the service. 

We find the Nash equilibrium and examine the performance at the equilibrium, in terms of user welfare, station revenue and electricity prices. As expected, competing stations offer users with lower prices than the monopolistic revenue maximizing station do. Moreover, the amount of regulation services increase significantly than that in the monopolistic case. }
This paper models a non-cooperative game between two EV charging stations. One is a fixed-power charging station purchasing electricity from the grid at wholesale price and reselling the energy to EV owners at a higher retail price; the other is regulation-providing and varies the recharging power level of its clients to provide regulation services to the grid, so its profit comes from both EV owners (who buy energy) and the grid (which pays for regulation services). Users are reluctant to charging power variations and prefer shorter overall charging times, hence regulation-providing charging has to be cheaper than fixed-power charging.

We analyze the competition among those charging providers, and examine the performance at the equilibrium in terms of user welfare, station revenue and electricity prices. As expected, competing stations provide users with lower charging prices than when both charging solutions are offered by a monopolistic provider. Moreover, while competition benefits users, it also benefits the grid in that the amount of regulation services increases significantly with respect to the monopolistic case. 

\end{abstract}


%
\IEEEpeerreviewmaketitle

\section{Introduction}

Among the main difficulties of the penetration of Electric Vehicles (EVs) in the smart city is the associated energy equation: how can the power grid accommodate the corresponding demand?~\cite{IEA2013EV_Outlook}. And together with the technical limitations, the question of economic incentives to elicit the most efficient use of resources needs also to be considered (see~\cite{shuai2016charging} and references therein).
But EVs do not \emph{use} the energy in real time, they just \emph{store} it in their batteries until leaving the charging station. This particularity of EV charging demand can be leveraged, in particular for \emph{regulation} purposes, i.e., by adapting the consumption at a very fast pace (within seconds) to the current energy production and consumption state of the grid. In practice, when there is extra (resp. a lack of) energy production with respect to demand, the grid can send a ``down'' (resp., ``up'') regulation signal so that the production side--or here, the consumption side--reacts accordingly.
In this paper, we consider the economic aspects of such an option, from the point of view of EV owners and charging stations.

Previous work focuses on fairness issues among users in terms of final state-of-charge~\cite{Cover2012IT}; on incentivizing EV owners to contribute to regulation~\cite{Wu2012V2G,Wu2012Wind}; or on the resulting user welfare~\cite{Sun2013RegulationAllocation,Sun2014RegulationAllocation}.
The closest works to ours are~\cite{Gao2013ContractRegulation,shuai2015incentivizing}, where the focus is on the pricing strategies of the charging stations: in~\cite{Gao2013ContractRegulation}, Gao \emph{et al.} consider a regulator designing contracts to incentivize EVs to participate so that the station profit is maximized. On the other hand, in~\cite{shuai2015incentivizing} we considered a charging station offering two simple options, namely a fixed-power charging (no regulation) and a varying-power charging (following regulation signals). Another difference is that~\cite{Gao2013ContractRegulation} allows vehicle-to-grid energy exchanges while in~\cite{shuai2015incentivizing} the regulation services are just provided by varying the current charging power. 
But both of those works, as well as~\cite{Sortomme2011ChargingStrategies}, assume a monopolistic revenue-maximizing charging station. 

In this paper, we focus on the effect of \emph{competition}, by considering two competing charging stations, one implementing only regulation-based charging and the other only fixed-power charging. When compared to monopolistic situations, we expect competition to benefit to users, through lower recharging prices. But also, we investigate the viability of each competitor: indeed, regulation is rewarded through financial incentives, and providing regulation during charging may not yield sufficient revenues if those incentives are not large enough. Hence some regions of reward values where EV-charging-based regulation can occur; this was investigated in the monopolistic case in~\cite{shuai2015incentivizing}, here we study the effect of competition on that aspect as well.

Our results indicate that, as expected, competition is beneficial to users, through lower recharging prices. Also, competition appears to be better from the grid perspective, since both the region of rewards for which regulation is viable and the amount of regulation offered are larger in the competition setting. 
The remainder of the paper is organized as follows. Section~\ref{sec:Model} presents our model; Section~\ref{sec:Nash} analyzes the price competition and the resulting Nash equilibrium; In Section~\ref{sec:Compare} we compare the performance of the competition with the monopolistic case and Section~\ref{sec:Conclusion} concludes the paper.


\section{Model description}\label{sec:Model}  

According to a national household travel survey of the United States~\cite{Micro2012Vehicle,Survey2009VehicleUse}, a passenger vehicle spends on average 75 minutes a day on journey, hence is parked most of the time. We assume this to remain true for EVs, i.e., the time during which they are available for charging largely exceeds that needed to fully recharge their batteries. So this is interesting, at least for some EV drivers, to accept longer charging durations for cheaper energy. This opens an opportunity for charging stations to increase revenue through the rewards offered to regulation-contributing entities,  as well as for EV owners to save on their energy bill.

Conventional recharging services are provided by what we will call an \simplecharging{} station, purchasing electricity at the low wholesale unit price $t$\euro/kWh and reselling it to EV owners at a higher price; whereas in a \regcharging{}, charging power is not guaranteed but subject to variations over time, as a response to regulation requests issued by grid operators. We model the interactions among both stations (or sets of stations, each set controlled by a separate entity) as a noncooperative game since they compete over prices to attract EV owners. User preferences between price and charging power variations are assumed heterogeneous, so each station seeks the best tradeoff between market shares and per-client profit in order to maximize its expected revenue. 
 
\subsection{Regulation mechanism}
Frequency regulation, depending on the response time, is mainly divided into: primary, secondary, and tertiary control, with the response time increasing from seconds, to minutes and finally to half an hour respectively~\cite{NERC2011Frequency}. In our proposal, the \regcharging{} station modulates the EV charging power to provide the secondary control: one regulation time slot lasts for $\Delta$ hours, with $\Delta$ typically between $0.1$ (6 minutes) and $0.25$ (15 minutes). Periodically, the grid operator, buyer of the regulation service, sends a regulation request to the \regcharging{} station specifying its demand, which can be regulation-up, -down or -null. 
Receiving the signal, the \regcharging{} station sets the EV recharging power to be 0 kW\footnote{We do not allow here EVs to deliver energy to the grid (the so-called vehicle-to-grid transfer).}, $P_d$ kW, or $P_n$ kW respectively: $P_d$ is the maximum acceptable power level allowed by the EV supply equipment in the station, and $P_n$ is the default recharging power ($0\le P_n \le P_d$) defined by the \regcharging{} station itself, when no regulation is needed, namely regulation null. Note that this mechanism increases (decreases) the EV consumption responding to regulation-down (-up). This counter-intuitive naming stems from conventional regulation services, where providers are \emph{generation units} whereas the task is given to \emph{consumers} here.
For later convenience we will use the notation $x\define\frac{P_n}{P_d}$, so that $x\in[0,1]$.

At the \simplecharging{} station, EVs are always charged at full speed $P_d$ kW. Figure~\ref{fig:Power_Energy.tex} illustrates the charging power profiles for the two stations as well as the energy accumulated in an EV battery being charged for a given scenario of regulation requests.
We denote by $C_B$ the energy demand of an EV, and by $\rho_u$ ($\rho_d$) the probability of occurrence of regulation-up (-down) at each time slot, those signals being assumed independent at each regulation period in this paper. 
\begin{figure}[htbp]
   \centering
   \input{Power_Energy.tex} 
   \label{fig:Power_Energy.tex}
\end{figure}

There may be concerns that varying the charging power for all EVs in \regcharging{} station(s) simultaneously and drastically following this ``ping-pong" policy can lead to an oversupply of regulation, i.e., the aggregated increase or decrease in power is larger than that actually needed by the grid operator. This is hardly possible since in the scale of a grid operator, the disposable regulation capacity scattered in EVs is non-dominant if not negligible given the current penetration levels. 
For example data from RTE (R\'eseau de transport d'electricit\' e), the biggest independent system operator in France, show that the regulation-down demand in 30 minutes\footnote{\url{http://clients.rte-france.com/lang/fr/visiteurs/vie/mecanisme/jour/volume.jsp}} can easily go over 100 MWh, a quantity that could only be absorbed by at least ten thousand EVs doing level 2 recharging (19.2kW~\cite{SAE_J1772}) at the same time. 
Since the whole country has an EV population of 30 thousand, sharing 8600 public charging facilities, the regulation oversupply problem is not of concern so far. 
\short{But it can rapidly become one if EV penetration increases; nevertheless we expect that in this case, the incentives to provide regulation will be adjusted (regulation being rewarded less) so that market mechanisms will reduce supply. In addition, demand for regulation is likely to increase in the next future, with the development of renewable energy production which cannot be controlled like fossil-based electricity plants can: the overall supply-demand balance will be more difficult to maintain, hence a probable larger need for ancillary services such as regulation.}

\subsection{Regulation incentives}

In return for providing regulation, the \regcharging{} station receives monetary incentives, with respect to the default wholesale price $t$.
\begin{itemize}
\item In \emph{regulation-null} periods, it charges each plugged EV with power $P_n$ kW, and pays $\Delta t P_n$ monetary units (no compensation) over such a duration-$\Delta$;
\item in \emph{regulation-up} periods, the grid operator ``re-buys" the energy saved at a unit price $r_u t$, hence the station pays $\Delta t (1-r_u)P_n$ monetary units over such a period (note that we can expect to have $r_u\geq 1$, although it is not always the case in practice);
\item similarly, in a \emph{regulation-down}, the \regcharging{} station pays for the extra energy it consumes at a discount price $t(1-r_d)$ monetary units per kWh, thus a total price paid $\Delta \big(P_n t + (P_d-P_n) t (1-r_d)\big)$ monetary units per EV.
\end{itemize}
Together with the probabilities of regulation-up ($\rho_u$) and down ($\rho_d$), the expected net revenue (possibly negative) over one regulation slot is:
{\tiny
\begin{equation}\label{eq:E_r_slot}
E_\Delta = t\Delta(\rho_u r_u P_n-\rho_d(1- r_d)(P_d-P_n)-P_n)
\end{equation}}

\subsection{User preferences}

We assume that each EV owner needs $C_B$ kWh of energy, say, per day, the owner can choose to charge at the constant power $P_d$ in the \simplecharging{} station, or to charge at a variable power in the \regcharging{} station. They can also choose neither solution (a \nocharging{} choice) if they consider both too expensive. Naturally, users are assumed to:
\begin{itemize}
\item prefer to recharge faster, i.e., at higher power rate;
\item be reluctant to \emph{uncertainty} in the recharging power caused by regulations. Additionally, batteries can be sensitive to power variations in the recharging process, another reason for EV owners to be reluctant to contributing to regulation.
\end{itemize}

Following these criteria, we define the user utility (willingness-to-pay minus price paid) for a recharging option as being of the form
{\tiny
\begin{equation*}\label{eq:Valuation}
V = \theta (\bar P-\gamma\delta (P)) - T C_B
\end{equation*}}
where $\bar P$ is the expected charging power, and $\delta (P)$ its standard deviation. $\theta$ is user-specific: we assume it to be exponentially distributed, with mean $\bar\theta$, over the EV owner population. The parameter $\gamma$ is the reluctance toward power fluctuations, and is assumed the same for all users. Finally, $T$ represents the unit energy price set by the charging station chosen by the user. (We take $T=0$ for users who choose \nocharging). 

Let us define $P_A$ as the value of $\bar P - \gamma\delta(P)$ for the \regcharging{} option, which can easily be expressed from $P_d$, $P_n$, $\rho_d$ and $\rho_u$. The parameter ($\gamma$) we choose always guarantees $P_A\ge0$ which means that this proposal does not target users with a too high sensitivity to power fluctuation.
The probability $\alpha_r$ (resp., $\alpha_s$) that a user chooses the \regcharging{} (resp., \simplecharging{}) station can then be expressed as

{\tiny
\begin{align}
\alpha_r &=\begin{cases}\label{eq:a_r}
1-\exp(-\frac{C_B(T_s-T_r)}{\bar\theta(P_d-P_A)}) & \mbox{ if } T_r < 0 \\
\exp(-\frac{C_BT_r}{\bar\theta P_A})-\exp(-\frac{C_B(T_s-T_r)}{\bar\theta(P_d-P_A)}) & \mbox{ if }0 \le T_r\le \frac{P_A}{P_d}T_s\\
0 & \mbox{ otherwise}\end{cases}
\\
\alpha_s &=\begin{cases}\label{eq:a_s}
\exp(-\frac{C_BT_s}{\bar\theta P_d}) &\mbox{ if } T_s \le \frac{P_d}{P_A}T_r\\
\exp(-\frac{C_B(T_s-T_r)}{\bar\theta(P_d-P_A)}) &\mbox{ otherwise.}\end{cases}
\end{align}}

Note that we allow negative charging prices with the \regcharging{} station: indeed, since that station can make money from the grid thanks to EV owners, the corresponding rewards could be so large that the station would be willing to attract a large number of EVs, even by paying them. This case is for completeness of the model, we think it is not very likely to occur but we cover it in this paper.

Following the classical backward induction method, we first assume $P_n$ (or equivalently $x$) fixed and analyze the pricing game (defined bellow). The outcome is dependent on $x$ so the \regcharging{} station can maximize its profit through altering its value. We examine the first part analytically while the second numerically due to complexity. 

\begin{definition}\label{game:Def}
The pricing game between the \simplecharging{} station and the \regcharging{} station as a collection: $\langle \mathcal{N}, \mathcal{T} , (R_i)\rangle$, where the player set $\mathcal{N}$ consists of the two stations, the price profile $\mathcal{T}$ is a vector $(T_s, T_r)$ on the semi-plane $\mathbb{R}_{\ge 0}\times\mathbb{R}$, and the payoff function $R_i : \mathcal{T} \to \mathbb{R}$ gives each station's expected revenue obtained from one EV.
\end{definition}

Table~\ref{tab:notations} summarizes the notations used in our model.
\begin{table}[htbp]
\begin{center}
\caption{Model notations}
\begin{tabular}{c|p{.35\textwidth}}
$t$ & unit price of energy paid by stations (unit:  \euro/kWh)\\
$r_u$& remuneration ratio for regulation-up (no unit)\\
$r_d$& discount ratio for regulation-down (no unit)\\
$\rho_u$ (resp. $\rho_d$)& probability of an ``up'' (resp. ``down'')  regulation signal\\
$C_B$& average energy recharged per EV per day\\
$\theta$& user sensitivity to recharging power (including variability)\\
$\bar\theta$& average value of $\theta$ among users\\
$\gamma$& user reluctance to power variation\\
$P_n$ (resp. $P_d$)& default (resp. ``regulation-down'') recharging power\\
$x$& $\frac{P_n}{P_d}$\\
$\bar P$& $\rho_dP_d+(1-\rho_u-\rho_d)P_n $\\
$\delta (P)$& $\sqrt{\rho_u\bar P^2+\rho_d(P_d-\bar P)^2+(1-\rho_u-\rho_d)(P_n-\bar P)^2)}$\\
$P_A(x)$, or $P_A$& $\bar P-\gamma\delta (P)$ ( $>0$ )\\
\end{tabular}
\label{tab:notations}
\end{center}
\end{table}

\section{Analysis of the game}\label{sec:Nash}
In this section, we analyze the non-cooperative strategic game defined in~\ref{game:Def}. We derive their respective best-response prices, to characterize the Nash equilibria.

\subsection{Best-response prices}
\subsubsection{\simplecharging{} station revenue and best-response price $T_s^{br}$}
For the \simplecharging{} station owner, its average income $R_s$ depends on the market share $\alpha_s$, and the unit price $T_s$ it offers:
{\tiny
\begin{subequations}\label{eq:R_s}
   \begin{align}[left = {R_s = C_B(T_s-t)\alpha_s =\empheqlbrace}]
   & C_B (T_s-t) \exp(-\frac{C_BT_s}{\bar\theta P_d}) & T_s \le \frac{P_d}{P_A}T_r \label{Rs_a}\\
   & C_B (T_s-t) \exp(-\frac{C_B(T_s-T_r)}{\bar\theta(P_d-P_A)}) & T_s >\frac{P_d}{P_A}T_r.\label{Rs_b}
   \end{align}
\end{subequations}}

Depending on its opponent's strategy $T_r$, the price $T_s$ that maximizes $R_s$ provides the best-response price.

\begin{theorem}\label{theo:T_s}
The \simplecharging{} station has a unique best-response price as follows:
{\tiny
\begin{subequations}\label{eq:prop1}
  \begin{align}[left ={T_s^{br}(T_r)  =\empheqlbrace}] 
 	 & t+(P_d-P_A)\frac{\bar\theta}{C_B}  & \mbox{ if }  T_r < (t+(P_d-P_A)\frac{\bar\theta}{C_B})\frac{P_A}{P_d} \label{eq:Ts_a}\\
	 & t+P_d\frac{\bar\theta}{C_B} & \mbox{ if } T_r > (t+P_d\frac{\bar\theta}{C_B})\frac{P_A}{P_d} \label{eq:Ts_b}\\
	 & T_r\frac{P_d}{P_A}  & \mbox{ otherwise} \label{eq:Ts_c}
   \end{align}
\end{subequations}}  
\end{theorem} 

Proof in Appendix~\ref{appx:proof1}.

Figure~\ref{fig:Rs_surf} illustrates the \simplecharging{} station revenue as a function of $T_r$, and the best-response $T_s^{br}(T_r)$. 
\begin{figure}[htbp] 
   \centering
   \input{Rs_surf_enveloped.tex} 
   \label{fig:Rs_surf}
\end{figure}

\subsubsection{\regcharging{} station revenue and best-response price $T_r^{br}(T_s)$}
Let us now consider the \regcharging{} station owner, having to decide its price $T_r$. 

To estimate how much net renumeration the \regcharging{} station gets from recharging EVs through regulation, we multiply the regulation revenue per slot, i.e. $E_\Delta$ in~\eqref{eq:E_r_slot}, by the average number of slots a regulating EV remains plugged-in before its battery is fully recharged, i.e. $C_B/(\Delta\bar P)$. To facilitate the writing we further divide the product, which has a unit of \euro, by the EV energy demand ($C_B$ kWh), so that its final unit is \euro/kWh and has a form of 
{\tiny
\begin{equation*}
E_r\define t(\rho_u r_u x-\rho_d(1- r_d)(1-x)-x)\frac{P_d}{\bar P}.
\end{equation*}}

The average \regcharging{} station revenue consists of renumeration from providing regulation and income from charging EVs:
{\tiny
\begin{equation}\label{eq:R_r}
R_r=
   \begin{cases}
    C_B(T_r + E_r)[1-\exp(-\frac{C_B(T_s-T_r)}{\bar\theta(P_d-P_A)})] \!\!\!&T_r < 0\\
    C_B(T_r + E_r)[\exp(-\frac{C_BT_r}{\bar\theta P_A})-\exp(-\frac{C_B(T_s-T_r)}{\bar\theta(P_d-P_A)})] \!\!\!& 0 \le T_r\le \frac{P_A}{P_d}T_s \\
    0 \!\!\!&\mbox{ otherwise}\end{cases}
\end{equation}}

The following result summarizes the optimal \regcharging{} station reaction to its competitor.

\begin{theorem}\label{theo:T_r}
The \regcharging{} station has a unique best-response price as follows:

{\tiny
\begin{subequations}\label{eq:prop2}
\begin{align}[left = {T_r^{br}(T_s) =\empheqlbrace}]
&T_s\frac{P_A}{P_d}  &\!\!\!\!\!\!\!\!\!\!\!\!\!\!\!\!\!\!\!\!\!\!\!\!\!\!\!\!\!\!\!\!\!\!\!\!\!\!\!\!\!\!\!\!\!\!\!\!\!\!\!\!\!\!\!\!\!\!\!\!\!\!\!\!\!\!\!\!\!\!\!\!\!\!\!\!\!\!\!\!\!\!\!\!\!\!\!\!\mbox{ if }T_s \le -E_r\frac{P_d}{P_A} \label{eq:Tr_a}\\
&0&\!\!\!\!\!\!\!\!\!\!\!\!\!\!\!\!\!\!\!\!\!\!\!\!\!\!\!\!\!\!\!\!\!\!\!\!\!\!\!\!\!\!\!\!\!\!\!\!\!\!\!\!\!\!\!\!\!\!\!\!\!\!\!\!\!\!\!\!\!\!\!\!\!\!\!\!\!\!\!\!\!\!\!\!\!\!\!\!\mbox{ if } T_s\in \{ T_s : E_{r,1}(T_s) \le E_r \le E_{r,2}(T_s) \} \label{eq:Tr_b}\\
&\{ T_r\in\mathbb{R} : \frac{\partial R_r}{\partial T_r}=0 \} &\nonumber\\
&\subset\big(\min\{0,-E_r\},\max\{0,\min\{\frac{\bar\theta P_A}{C_B}-E_r,\frac{P_A}{P_d}T_s\}\}\big) &\!\!\!\!\!\!\!\!\!\!\!\!\!\!\!\!\!\!\!\!\!\!\!\!\!\!\!\!\!\!\!\!\!\!\!\!\!\!\!\!\!\!\!\!\!\!\!\!\!\!\!\!\!\!\!\!\!\!\!\!\!\!\!\!\!\!\!\!\!\!\!\!\!\!\!\!\!\!\!\!\!\!\!\!\!\!\!\!\mbox{otherwise}\label{eq:Tr_c}
\end{align}
\end{subequations}}

{\tiny\begin{align}
\text{where } &\nonumber\\
&E_{r,1}(T_s) = \frac{\bar\theta(P_d-P_A)\big(1-\exp(-\frac{C_BT_s}{\bar\theta (P_d-P_A)})\big)}{C_B\big(\frac{P_d}{P_A}-1+\exp(\frac{C_BT_s}{\bar\theta(P_d-P_A)})\big)}\nonumber\\
&E_{r,2}(T_s) = E_{r,1}(T_s)\big(1+(\frac{P_d}{P_A}-1)\exp(\frac{C_BT_s}{\bar\theta(P_d-P_A)})\big).\nonumber
\end{align}}
\end{theorem}

Proof in Appendix~\ref{appx:proof2}.

Figure~\ref{fig:Rr_surf} shows the \regcharging{} station revenue as well as the best-response price $T_r^{br}(T_s)$, as a function of $T_s$.
\begin{figure}[htbp] 
   \centering
   \input{Rr_surf_enveloped.tex} 
   \label{fig:Rr_surf}
\end{figure}

\subsection{Nash equilibrium}\label{subsec:Nash}

\begin{theorem}\label{theo:Nash}

The pricing game defined in~\ref{game:Def} has either a unique Nash equilibrium or a unique Pareto-dominant one when there exist infinite number of Nash equilibria. The equilibria prices in different circumstances are:
{\tiny
\begin{subequations}
\begin{align}[left ={N^E: \empheqlbrace}]
& T_r = -E_r;~T_s = -\frac{P_d}{P_A}E_r \nonumber \\
&~~~ \mbox{ if } E_r \le -\frac{P_A}{P_d}[t+(P_d-P_A)\frac{\bar\theta}{C_B}] \label{eq:N1}\\
& T_r \in(0,\min\{\frac{\theta P_A}{C_B}-E_r,\frac{P_A}{P_d}T_s)\};~T_s = t+(P_d-P_A)\frac{\bar\theta}{C_B} \nonumber\\
&~~~ \mbox{ if }-\frac{P_A}{P_d}[t+(P_d-P_A)\frac{\bar\theta}{C_B}] < E_r < E_{r,1}\big(t+(P_d-P_A)\frac{\bar\theta}{C_B}\big)\label{eq:N2}\\
& T_r = 0;~T_s = t+(P_d-P_A)\frac{\bar\theta}{C_B} \nonumber\\
&~~~ \mbox{ if }E_{r,1}\big(t+(P_d-P_A)\frac{\bar\theta}{C_B}\big) \le E_r \le E_{r,2}\big(t+(P_d-P_A)\frac{\bar\theta}{C_B}\big) \label{eq:N3}\\
& T_r \in(-E_r,0);~T_s = t+(P_d-P_A)\frac{\bar\theta}{C_B} \nonumber \\
&~~~ \mbox{ if }E_{r,2}\big(t+(P_d-P_A)\frac{\bar\theta}{C_B}\big) < E_r \label{eq:N4}
\end{align}
\end{subequations}}

\end{theorem}

Proof in Appendix~\ref{appx:proof3}.

Note that $N^E$\eqref{eq:N1} which occurs when $E_r\le-\frac{P_A}{P_d}[t+(P_d-P_A)\frac{\theta}{C_B}]$ is not profitable for the \regcharging{} stations since zero revenue is obtained, and that the condition for a positive \regcharging{} station revenue is $-\frac{P_A}{P_d}[t+(P_d-P_A)\frac{\theta}{C_B}] < E_r$. We will refer to this condition in Section~\ref{subsec:Market}.

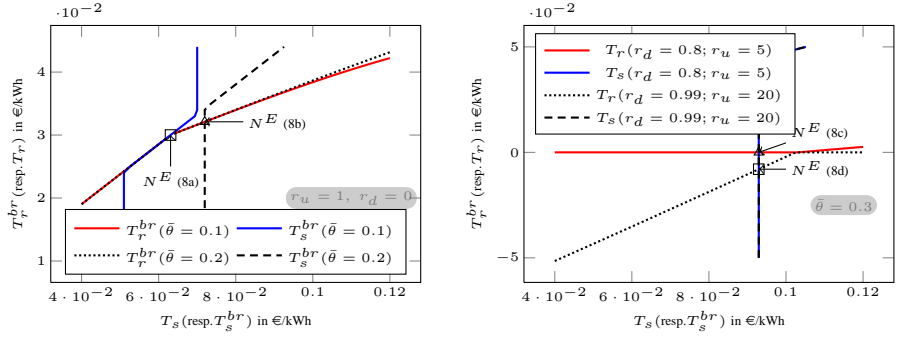
\begin{figure*}[htbp]
	\input{FourNash.tex}
	\caption{Four Nash equilibria in different cases}
	\label{fig:FourNash}
\end{figure*}

Figure~\ref{fig:FourNash} illustrates best-response prices and resulting Nash equilibria in four different circumstances.

\subsubsection{Optimization of $P_n$}
The pricing game defined in~\ref{game:Def} is played given a fixed $P_n$, set by the \regcharging{} station, who can afterwards modify its value to pursue a higher equilibrium revenue. Due to the complexity of the equilibrium price profile we resort to numerical search for the optimal $P_n$.

\section{Comparison between the Nash equilibrium and the Monopolistic case}\label{sec:Compare}
In this section we compare the competition model with the monopolistic case where a single manager sets both the \simplecharging{} price as well as the \regcharging{} price to maximize its overall revenue. We do not repeat the results in~\cite{shuai2015incentivizing} for the monopolistic case due to space limit, but simply compare their performances under the same parameters. 

\subsection{User welfare}
User welfare is the average user utility over the distribution of user preference parameter, i.e. $\theta$. The following formula works for both the monopolistic case and the Nash equilibrium. 
{\tiny
\begin{align*}
U =&\int_{\frac{T_rC_B}{P_A}}^{\frac{(T_s-T_r)C_B}{P_d-P_A}}(\theta P_A-T_rC_B)\frac{1}{\bar\theta}\exp(-\frac{\theta}{\bar\theta}) \text{d}\theta \\\nonumber
&+\int_{\frac{(T_s-T_r)C_B}{P_d-P_A}}^{+\infty}(\theta P_d-T_sC_B)\frac{1}{\bar\theta}\exp(-\frac{\theta}{\bar\theta}) \text{d}\theta \\\nonumber
=& \alpha_r\bar\theta P_A + \alpha_s\bar\theta P_d
\end{align*}}
The first column in Figure~\ref{fig:CompareRevenue} shows a significant increase of user welfare ($U^M$ for monopoly and $U^E$ for equilibrium) after breaking a monopolistic station into two competing ones. Although the total station revenue decrease, the social welfare which is the user utility plus station revenue has a net increase of over 20\%. The second column illustrate an increase of EVs being served, thanks to a decrease of energy prices depicted in the third column.
\begin{figure*}[htbp]
	\input{CompareRevenue.tex}
	\label{fig:CompareRevenue}
\end{figure*}
%

\subsection{Application in a real world market}\label{subsec:Market}
Figure~\ref{fig:Compare_rd_ru_nash_mono} compares the regions for rewards $\{ r_d, r_u \}$ where offering \regcharging{} is profitable. At equilibria (second row), the black zones where \regcharging{} is not preferred is remarkably smaller than those in the monopolistic case. This is because that in a monopoly, feasible region for rewards $\{ r_d, r_u \}$ is composed of those that make the following equation of $x$ solvable in the interval of $[0,1]$~\cite{shuai2015incentivizing}:
{\tiny
\begin{equation*}\label{eq:E_r_mono}
\rho_u r_u x - \rho_d (1-r_d)(1-x) -x - \bar P(x) P_A(x) P_d^{-2} > 0 
\end{equation*}}
Whereas in competition, \eqref{eq:N1} and \eqref{eq:N2} give the condition of:
{\tiny
\begin{align*}
& t(\rho_u r_u x - \rho_d (1-r_d)(1-x) -x - \bar P P_A P_d^{-2}) \\ \nonumber
+ & \frac{1}{t}\bar P P_A P_d^{-2}[P_d-P_A]\frac{\bar\theta}{C_B} > 0 \nonumber
\end{align*}}
Comparing these two we find that the competition enlarges the viable region of $\{ r_d, r_u \}$. The blue and red areas in Figure~\ref{fig:Compare_rd_ru_nash_mono} are referring to the optimal default recharging power $P_n$ in these regions, i.e. the optimal $x$ after exhaustive search. In most combinations of $\{ r_d, r_u \}$, this optimal $x$ is either $0$ or $1$, except for a few $\{ r_d, r_u \}$ observed in the gap between the blue region and the red, in the figures on the third column where average user preference on power is smaller: $\bar\theta = 0.1$ and user sensitivity to variation is greater: $\gamma = 0.5$. We also plot the actual $\{ r_d, r_u \}$ offered by a French operator RTE on these figures. The blue circles correspond to the 48 $\{ r_d, r_u \}$ pairs on the day of 20/07/2015 and the red rectangles are showing the daily averages during the week from 20/07/2015 to 26/07/2015. 
\begin{figure*}[htbp]
	\input{Compare_rd_ru_nash_mono.tex}
	\label{fig:Compare_rd_ru_nash_mono}
\end{figure*}

\short{Figure~\ref{fig:ConcavityRrOnX} shows that for some particular combination of $\{ r_d, r_u \}$, $x$ is neither 0 nor 1, but this happens quite rare, only a narrow region between the red and blue parts in Figure~\ref{fig:Compare_rd_ru_nash_mono}.
\begin{figure}[htbp]
\includegraphics[width=8cm]{Graphics/ConcavityRrOnX}
\label{fig:ConcavityRrOnX}
\end{figure}}

\section{Conclusion and perspective}\label{sec:Conclusion}
This paper considers a competition between two self-interested charging stations. At the Nash equilibrium of this non-cooperative game, both stations tends to offered lower prices to EV owners than a monopolistic controller would do, thus more clients are attracted and greater regulation services is provided to the grid operator. This work can be extended in several ways including: bring in more actors such as charging stations with private renewable energy sources; considering the actor of a ``Grid" who can play with the wholesale electricity price imposed on both \regcharging{} and \simplecharging{} station; or differentiate two charging stations by their locations, which effect users' preferences among them.

\section*{Acknowledgement}

This work has been partially funded by the Fondation Telecom through the Futur\&Ruptures program.

\bibliographystyle{IEEEtran}
\bibliography{IEEEabrv,Biblio/RegulationRecharging}
\input{Appendix_Proof.tex}
\end{document}

%% file: Preamble.tex
\usepackage{comment}

\ifCLASSINFOpdf

\else
\fi
\usepackage{blindtext, graphicx}
\graphicspath{{./Graphics/}}
\DeclareGraphicsExtensions{.pdf,.jpeg,.png,.jpg,.gif}
\usepackage{caption}
\usepackage{multirow}

\usepackage{eurosym}
\usepackage{tikz}
\usepackage{tikz-inet,pgf,pgfplots,rotating}
\pgfplotsset{compat=newest}
\usepgfplotslibrary{groupplots}
\usetikzlibrary{pgfplots.groupplots}
\usetikzlibrary{arrows,automata,snakes}
\gdef\figwidth{8cm}
\gdef\figheight{5cm}
\pgfplotsset{compat=1.3}

\pgfcreateplotcyclelist{\mylist}{
	{red,mark=none,thick}, 
	{blue,mark=none,thick}, 
	{yellow,mark=none,thick},
	{mark=none,thick,densely dotted}, 
	{mark=none,thick,densely dashed},
	{mark=none,thick,dash pattern=on 2pt off 2pt on 6pt off 2pt}}

\pgfcreateplotcyclelist{\nostyle}{
	{mark=none,black},
}

\usepackage[cmex10]{amsmath}
\usepackage{amsfonts,amssymb,amsthm,algorithmic}
\usepackage{cases}
\usepackage[overload]{empheq} 

\newtheorem{theorem}{Proposition}[section]
\newtheorem{definition}[theorem]{Definition}

\usepackage{url}

\newif\ifnotes\notestrue
\newcommand\short[1]{}

\gdef\simplecharging{\emph{S-charging}{}}
\gdef\regcharging{\emph{R-charging}}
\gdef\nocharging{\emph{no\_charging}{}}

\gdef\define{:=}

%% file: Power_Energy.tex
{\footnotesize\begin{tikzpicture}\begin{axis}[width=\figwidth,height=4.6cm,cycle list name=\mylist,xlabel=Time (hours),
ylabel={Power (kW)},
ymax=75,
ymin=-1,
ytick={16,20},yticklabels={$P_n$,$P_d$},
xtick={0,10,20,30,40,50,60},xticklabels={0,1,2,3,4,5,6},
legend columns=2, 
legend style={at={(0.03,1)},anchor=north west},
every axis legend/.append style={nodes={right}}, 
legend entries = {Power (no reg.), Power (regul.), Energy (no reg.), Energy (regul.)},
]
\addplot +[const plot]coordinates{
(1,20)(2,20)(3,20)(4,20)(5,20)(6,20)(7,20)(8,20)(9,20)(10,20)(11,20)(12,20)(13,20)(14,20)(15,20)(16,20)(17,20)(18,20)(19,20)(20,20)(21,20)(22,20)(23,20)(24,20)(25,20)(26,0)(27,0)(28,0)(29,0)(30,0)(31,0)(32,0)(33,0)(34,0)(35,0)(36,0)(37,0)(38,0)(39,0)(40,0)(41,0)(42,0)(43,0)(44,0)(45,0)(46,0)(47,0)(48,0)(49,0)(50,0)(51,0)(52,0)(53,0)(54,0)(55,0)(56,0)(57,0)(58,0)(59,0)(60,0)};
\addplot +[const plot]coordinates{
(1,20)(2,0)(3,0)(4,20)(5,0)(6,16)(7,16)(8,20)(9,16)(10,0)(11,20)(12,20)(13,16)(14,0)(15,0)(16,20)(17,0)(18,0)(19,20)(20,0)(21,0)(22,20)(23,0)(24,20)(25,0)(26,20)(27,20)(28,0)(29,0)(30,0)(31,0)(32,0)(33,0)(34,20)(35,0)(36,20)(37,0)(38,20)(39,0)(40,20)(41,20)(42,0)(43,0)(44,20)(45,20)(46,0)(47,20)(48,0)(49,20)(50,20)(51,0)(52,20)(53,0)(54,0)(55,0)(56,0)(57,0)(58,0)(59,0)(60,0)};
\addplot coordinates{
(1,0)(2,2)(3,4)(4,6)(5,8)(6,10)(7,12)(8,14)(9,16)(10,18)(11,20)(12,22)(13,24)(14,26)(15,28)(16,30)(17,32)(18,34)(19,36)(20,38)(21,40)(22,42)(23,44)(24,46)(25,48)(26,50)(27,50)(28,50)(29,50)(30,50)(31,50)(32,50)(33,50)(34,50)(35,50)(36,50)(37,50)(38,50)(39,50)(40,50)(41,50)(42,50)(43,50)(44,50)(45,50)(46,50)(47,50)(48,50)(49,50)(50,50)(51,50)(52,50)(53,50)(54,50)(55,50)(56,50)(57,50)(58,50)(59,50)(60,50)};
\addplot coordinates{
(1,0)(2,2)(3,2)(4,2)(5,4)(6,4)(7,5.6)(8,7.2)(9,9.2)(10,10.8)(11,10.8)(12,12.8)(13,14.8)(14,16.4)(15,16.4)(16,16.4)(17,18.4)(18,18.4)(19,18.4)(20,20.4)(21,20.4)(22,20.4)(23,22.4)(24,22.4)(25,24.4)(26,24.4)(27,26.4)(28,28.4)(29,28.4)(30,28.4)(31,28.4)(32,28.4)(33,28.4)(34,28.4)(35,30.4)(36,30.4)(37,32.4)(38,32.4)(39,34.4)(40,34.4)(41,36.4)(42,38.4)(43,38.4)(44,38.4)(45,40.4)(46,42.4)(47,42.4)(48,44.4)(49,44.4)(50,46.4)(51,48.4)(52,48.4)(53,50)(54,50)(55,50)(56,50)(57,50)(58,50)(59,50)(60,50)};
\end{axis}
\begin{axis}
[width=\figwidth,height=4.6cm,
axis y line*=right,
axis x line=none,
ymax=75,
ymin=-1,
ytick={50},yticklabels={$C_B$},
ylabel = {Energy (kWh)}]
\addplot[opacity = 0] coordinates{
(1,0)(60,50)};
\end{axis}
\end{tikzpicture}}
\caption{Power and cumulated energy an EV obtained with and without regulation (simulation with 
$C_B = $50kWh, $P_d = $20kW, $P_n = $16kW, $\Delta = $0.1hour, $\rho_u=\rho_d=$0.45~\cite{shuai2015incentivizing})
}

%% file: Rs_surf_enveloped.tex
{\tiny\begin{tikzpicture}\begin{axis}[
view/h={-75},
width=\figwidth,
height=\figheight,
xlabel=$T_r$(\euro/kWh),
ylabel=$T_s$(\euro/kWh),
zlabel=\simplecharging{} station revenue (\euro/EV),
zmax=1.73,zmin=1.0,
legend style={at={(.3,1.003)},anchor=south west},
]



\addplot3 [blue] coordinates{
(0.04,0.09152,1.3313)(0.041,0.09152,1.3532)(0.042,0.09152,1.3753)(0.043,0.09152,1.3979)(0.044,0.09152,1.4208)(0.044601,0.09152,1.4347)(0.045,0.09234,1.4439)(0.046,0.094392,1.4662)(0.047,0.096444,1.4872)(0.048,0.098496,1.5072)(0.049,0.10055,1.526)(0.05,0.1026,1.5438)(0.051,0.10465,1.5605)(0.052,0.1067,1.5762)(0.053,0.10876,1.5909)(0.054,0.11081,1.6047)(0.055,0.11286,1.6176)(0.056,0.11491,1.6295)(0.057,0.11696,1.6406)(0.058,0.11902,1.6508)(0.059,0.12107,1.6603)(0.06,0.12312,1.6689)(0.061,0.12517,1.6767)(0.062,0.12722,1.6839)(0.063,0.12928,1.6902)(0.064,0.13133,1.6959)(0.065,0.13338,1.7009)(0.066,0.13543,1.7053)(0.067,0.13748,1.709)(0.068,0.13954,1.7121)(0.069,0.14159,1.7146)(0.07,0.14364,1.7165)(0.071,0.14569,1.7179)(0.072,0.14774,1.7187)(0.073,0.1498,1.719)(0.073099,0.15,1.719)(0.074,0.15,1.719)(0.075,0.15,1.719)(0.076,0.15,1.719)(0.077,0.15,1.719)(0.078,0.15,1.719)(0.079,0.15,1.719)(0.08,0.15,1.719)};
\addlegendentry{Revenue at best-response price $T_s^{br}(T_r)$};

\addplot3 [dashed,line width=2pt] coordinates{
(0.04,0.09152,1)
(0.041,0.09152,1)
(0.042,0.09152,1)
(0.043,0.09152,1)
(0.044,0.09152,1)
(0.044601,0.09152,1)
(0.045,0.09234,1)
(0.046,0.094392,1)
(0.047,0.096444,1)
(0.048,0.098496,1)
(0.049,0.10055,1)
(0.05,0.1026,1)
(0.051,0.10465,1)
(0.052,0.1067,1)
(0.053,0.10876,1)
(0.054,0.11081,1)
(0.055,0.11286,1)
(0.056,0.11491,1)
(0.057,0.11696,1)
(0.058,0.11902,1)
(0.059,0.12107,1)
(0.06,0.12312,1)
(0.061,0.12517,1)
(0.062,0.12722,1)
(0.063,0.12928,1)
(0.064,0.13133,1)
(0.065,0.13338,1)
(0.066,0.13543,1)
(0.067,0.13748,1)
(0.068,0.13954,1)
(0.069,0.14159,1)
(0.07,0.14364,1)
(0.071,0.14569,1)
(0.072,0.14774,1)
(0.073,0.1498,1)
(0.073099,0.15,1)
(0.074,0.15,1)
(0.075,0.15,1)
(0.076,0.15,1)
(0.077,0.15,1)
(0.078,0.15,1)
(0.079,0.15,1)
(0.08,0.15,1)};
\addlegendentry{Best-response price $T_s^{br}(T_r)$};

\addplot3 [surf,opacity=0.5] coordinates{
(0.04,0.09,1.3309)(0.04,0.0909,1.3313)(0.04,0.0918,1.3313)(0.04,0.0927,1.3311)(0.04,0.0936,1.3306)(0.04,0.0945,1.3298)(0.04,0.0954,1.3288)(0.04,0.0963,1.3275)(0.04,0.0972,1.326)(0.04,0.0981,1.3242)(0.04,0.099,1.3223)(0.04,0.0999,1.3201)(0.04,0.1008,1.3176)(0.04,0.1017,1.315)(0.04,0.1026,1.3122)(0.04,0.1035,1.3091)(0.04,0.1044,1.3059)(0.04,0.1053,1.3025)(0.04,0.1062,1.299)(0.04,0.1071,1.2952)(0.04,0.108,1.2913)(0.04,0.1089,1.2872)(0.04,0.1098,1.283)(0.04,0.1107,1.2786)(0.04,0.1116,1.2741)(0.04,0.1125,1.2695)(0.04,0.1134,1.2647)(0.04,0.1143,1.2598)(0.04,0.1152,1.2547)(0.04,0.1161,1.2496)(0.04,0.117,1.2443)(0.04,0.1179,1.2389)(0.04,0.1188,1.2334)(0.04,0.1197,1.2278)(0.04,0.1206,1.2221)(0.04,0.1215,1.2163)(0.04,0.1224,1.2105)(0.04,0.1233,1.2045)(0.04,0.1242,1.1985)(0.04,0.1251,1.1923)(0.04,0.126,1.1861)(0.04,0.1269,1.1799)(0.04,0.1278,1.1735)(0.04,0.1287,1.1671)(0.04,0.1296,1.1607)(0.04,0.1305,1.1542)(0.04,0.1314,1.1476)(0.04,0.1323,1.141)(0.04,0.1332,1.1343)(0.04,0.1341,1.1276)(0.04,0.135,1.1208)(0.04,0.1359,1.114)(0.04,0.1368,1.1071)(0.04,0.1377,1.1002)(0.04,0.1386,1.0933)(0.04,0.1395,1.0864)(0.04,0.1404,1.0794)(0.04,0.1413,1.0724)(0.04,0.1422,1.0654)(0.04,0.1431,1.0583)(0.04,0.144,1.0512)(0.04,0.1449,1.0442)(0.04,0.1458,1.037)(0.04,0.1467,1.0299)(0.04,0.1476,1.0228)(0.04,0.1485,1.0157)(0.04,0.1494,1.0085)(0.04,0.1503,1.0014)(0.04,0.1512,0.9942)(0.04,0.1521,0.98704)(0.04,0.153,0.97987)(0.04,0.1539,0.97271)(0.04,0.1548,0.96554)(0.04,0.1557,0.95838)(0.04,0.1566,0.95123)(0.04,0.1575,0.94408)(0.04,0.1584,0.93693)(0.04,0.1593,0.9298)(0.04,0.1602,0.92267)(0.04,0.1611,0.91556)(0.04,0.162,0.90845)(0.04,0.1629,0.90137)(0.04,0.1638,0.89429)(0.04,0.1647,0.88723)(0.04,0.1656,0.88019)(0.04,0.1665,0.87316)(0.04,0.1674,0.86616)(0.04,0.1683,0.85917)(0.04,0.1692,0.8522)(0.04,0.1701,0.84525)(0.04,0.171,0.83833)(0.04,0.1719,0.83143)(0.04,0.1728,0.82455)(0.04,0.1737,0.8177)(0.04,0.1746,0.81087)(0.04,0.1755,0.80407)(0.04,0.1764,0.79729)(0.04,0.1773,0.79054)(0.04,0.1782,0.78382)(0.04,0.1791,0.77713)(0.04,0.18,0.77046)

(0.042,0.09,1.3749)(0.042,0.0909,1.3753)(0.042,0.0918,1.3753)(0.042,0.0927,1.3751)(0.042,0.0936,1.3746)(0.042,0.0945,1.3738)(0.042,0.0954,1.3727)(0.042,0.0963,1.3714)(0.042,0.0972,1.3698)(0.042,0.0981,1.368)(0.042,0.099,1.366)(0.042,0.0999,1.3637)(0.042,0.1008,1.3612)(0.042,0.1017,1.3585)(0.042,0.1026,1.3555)(0.042,0.1035,1.3524)(0.042,0.1044,1.3491)(0.042,0.1053,1.3456)(0.042,0.1062,1.3419)(0.042,0.1071,1.338)(0.042,0.108,1.334)(0.042,0.1089,1.3298)(0.042,0.1098,1.3254)(0.042,0.1107,1.3209)(0.042,0.1116,1.3162)(0.042,0.1125,1.3114)(0.042,0.1134,1.3065)(0.042,0.1143,1.3014)(0.042,0.1152,1.2962)(0.042,0.1161,1.2908)(0.042,0.117,1.2854)(0.042,0.1179,1.2798)(0.042,0.1188,1.2742)(0.042,0.1197,1.2684)(0.042,0.1206,1.2625)(0.042,0.1215,1.2565)(0.042,0.1224,1.2505)(0.042,0.1233,1.2443)(0.042,0.1242,1.2381)(0.042,0.1251,1.2317)(0.042,0.126,1.2253)(0.042,0.1269,1.2189)(0.042,0.1278,1.2123)(0.042,0.1287,1.2057)(0.042,0.1296,1.199)(0.042,0.1305,1.1923)(0.042,0.1314,1.1855)(0.042,0.1323,1.1787)(0.042,0.1332,1.1718)(0.042,0.1341,1.1648)(0.042,0.135,1.1578)(0.042,0.1359,1.1508)(0.042,0.1368,1.1437)(0.042,0.1377,1.1366)(0.042,0.1386,1.1294)(0.042,0.1395,1.1223)(0.042,0.1404,1.1151)(0.042,0.1413,1.1078)(0.042,0.1422,1.1006)(0.042,0.1431,1.0933)(0.042,0.144,1.086)(0.042,0.1449,1.0787)(0.042,0.1458,1.0713)(0.042,0.1467,1.064)(0.042,0.1476,1.0566)(0.042,0.1485,1.0492)(0.042,0.1494,1.0418)(0.042,0.1503,1.0344)(0.042,0.1512,1.0271)(0.042,0.1521,1.0197)(0.042,0.153,1.0122)(0.042,0.1539,1.0048)(0.042,0.1548,0.99745)(0.042,0.1557,0.99005)(0.042,0.1566,0.98266)(0.042,0.1575,0.97527)(0.042,0.1584,0.96789)(0.042,0.1593,0.96052)(0.042,0.1602,0.95316)(0.042,0.1611,0.94581)(0.042,0.162,0.93847)(0.042,0.1629,0.93115)(0.042,0.1638,0.92384)(0.042,0.1647,0.91655)(0.042,0.1656,0.90927)(0.042,0.1665,0.90201)(0.042,0.1674,0.89478)(0.042,0.1683,0.88756)(0.042,0.1692,0.88036)(0.042,0.1701,0.87318)(0.042,0.171,0.86603)(0.042,0.1719,0.8589)(0.042,0.1728,0.8518)(0.042,0.1737,0.84472)(0.042,0.1746,0.83766)(0.042,0.1755,0.83063)(0.042,0.1764,0.82363)(0.042,0.1773,0.81666)(0.042,0.1782,0.80972)(0.042,0.1791,0.80281)(0.042,0.18,0.79592)

(0.044,0.09,1.4171)(0.044,0.0909,1.4207)(0.044,0.0918,1.4208)(0.044,0.0927,1.4205)(0.044,0.0936,1.42)(0.044,0.0945,1.4192)(0.044,0.0954,1.4181)(0.044,0.0963,1.4167)(0.044,0.0972,1.4151)(0.044,0.0981,1.4132)(0.044,0.099,1.4111)(0.044,0.0999,1.4087)(0.044,0.1008,1.4062)(0.044,0.1017,1.4033)(0.044,0.1026,1.4003)(0.044,0.1035,1.3971)(0.044,0.1044,1.3937)(0.044,0.1053,1.39)(0.044,0.1062,1.3862)(0.044,0.1071,1.3822)(0.044,0.108,1.3781)(0.044,0.1089,1.3737)(0.044,0.1098,1.3692)(0.044,0.1107,1.3645)(0.044,0.1116,1.3597)(0.044,0.1125,1.3547)(0.044,0.1134,1.3496)(0.044,0.1143,1.3444)(0.044,0.1152,1.339)(0.044,0.1161,1.3335)(0.044,0.117,1.3279)(0.044,0.1179,1.3221)(0.044,0.1188,1.3163)(0.044,0.1197,1.3103)(0.044,0.1206,1.3042)(0.044,0.1215,1.2981)(0.044,0.1224,1.2918)(0.044,0.1233,1.2854)(0.044,0.1242,1.279)(0.044,0.1251,1.2724)(0.044,0.126,1.2658)(0.044,0.1269,1.2591)(0.044,0.1278,1.2524)(0.044,0.1287,1.2455)(0.044,0.1296,1.2387)(0.044,0.1305,1.2317)(0.044,0.1314,1.2247)(0.044,0.1323,1.2176)(0.044,0.1332,1.2105)(0.044,0.1341,1.2033)(0.044,0.135,1.1961)(0.044,0.1359,1.1888)(0.044,0.1368,1.1815)(0.044,0.1377,1.1742)(0.044,0.1386,1.1668)(0.044,0.1395,1.1594)(0.044,0.1404,1.1519)(0.044,0.1413,1.1444)(0.044,0.1422,1.1369)(0.044,0.1431,1.1294)(0.044,0.144,1.1219)(0.044,0.1449,1.1143)(0.044,0.1458,1.1067)(0.044,0.1467,1.0991)(0.044,0.1476,1.0915)(0.044,0.1485,1.0839)(0.044,0.1494,1.0763)(0.044,0.1503,1.0686)(0.044,0.1512,1.061)(0.044,0.1521,1.0533)(0.044,0.153,1.0457)(0.044,0.1539,1.0381)(0.044,0.1548,1.0304)(0.044,0.1557,1.0228)(0.044,0.1566,1.0151)(0.044,0.1575,1.0075)(0.044,0.1584,0.99987)(0.044,0.1593,0.99226)(0.044,0.1602,0.98466)(0.044,0.1611,0.97706)(0.044,0.162,0.96948)(0.044,0.1629,0.96192)(0.044,0.1638,0.95437)(0.044,0.1647,0.94683)(0.044,0.1656,0.93932)(0.044,0.1665,0.93182)(0.044,0.1674,0.92434)(0.044,0.1683,0.91689)(0.044,0.1692,0.90945)(0.044,0.1701,0.90204)(0.044,0.171,0.89465)(0.044,0.1719,0.88728)(0.044,0.1728,0.87994)(0.044,0.1737,0.87263)(0.044,0.1746,0.86534)(0.044,0.1755,0.85808)(0.044,0.1764,0.85085)(0.044,0.1773,0.84365)(0.044,0.1782,0.83648)(0.044,0.1791,0.82933)(0.044,0.18,0.82222)

(0.046,0.09,1.4171)(0.046,0.0909,1.4276)(0.046,0.0918,1.4379)(0.046,0.0927,1.4479)(0.046,0.0936,1.4577)(0.046,0.0945,1.4661)(0.046,0.0954,1.4649)(0.046,0.0963,1.4635)(0.046,0.0972,1.4618)(0.046,0.0981,1.4599)(0.046,0.099,1.4577)(0.046,0.0999,1.4553)(0.046,0.1008,1.4526)(0.046,0.1017,1.4497)(0.046,0.1026,1.4466)(0.046,0.1035,1.4433)(0.046,0.1044,1.4397)(0.046,0.1053,1.436)(0.046,0.1062,1.432)(0.046,0.1071,1.4279)(0.046,0.108,1.4236)(0.046,0.1089,1.4191)(0.046,0.1098,1.4144)(0.046,0.1107,1.4096)(0.046,0.1116,1.4046)(0.046,0.1125,1.3995)(0.046,0.1134,1.3942)(0.046,0.1143,1.3888)(0.046,0.1152,1.3833)(0.046,0.1161,1.3776)(0.046,0.117,1.3718)(0.046,0.1179,1.3658)(0.046,0.1188,1.3598)(0.046,0.1197,1.3536)(0.046,0.1206,1.3473)(0.046,0.1215,1.3409)(0.046,0.1224,1.3345)(0.046,0.1233,1.3279)(0.046,0.1242,1.3212)(0.046,0.1251,1.3145)(0.046,0.126,1.3077)(0.046,0.1269,1.3007)(0.046,0.1278,1.2938)(0.046,0.1287,1.2867)(0.046,0.1296,1.2796)(0.046,0.1305,1.2724)(0.046,0.1314,1.2651)(0.046,0.1323,1.2578)(0.046,0.1332,1.2505)(0.046,0.1341,1.2431)(0.046,0.135,1.2356)(0.046,0.1359,1.2281)(0.046,0.1368,1.2205)(0.046,0.1377,1.213)(0.046,0.1386,1.2053)(0.046,0.1395,1.1977)(0.046,0.1404,1.19)(0.046,0.1413,1.1822)(0.046,0.1422,1.1745)(0.046,0.1431,1.1667)(0.046,0.144,1.1589)(0.046,0.1449,1.1511)(0.046,0.1458,1.1433)(0.046,0.1467,1.1354)(0.046,0.1476,1.1276)(0.046,0.1485,1.1197)(0.046,0.1494,1.1118)(0.046,0.1503,1.1039)(0.046,0.1512,1.096)(0.046,0.1521,1.0882)(0.046,0.153,1.0803)(0.046,0.1539,1.0724)(0.046,0.1548,1.0645)(0.046,0.1557,1.0566)(0.046,0.1566,1.0487)(0.046,0.1575,1.0408)(0.046,0.1584,1.0329)(0.046,0.1593,1.025)(0.046,0.1602,1.0172)(0.046,0.1611,1.0093)(0.046,0.162,1.0015)(0.046,0.1629,0.9937)(0.046,0.1638,0.9859)(0.046,0.1647,0.97812)(0.046,0.1656,0.97036)(0.046,0.1665,0.96261)(0.046,0.1674,0.95489)(0.046,0.1683,0.94718)(0.046,0.1692,0.9395)(0.046,0.1701,0.93184)(0.046,0.171,0.92421)(0.046,0.1719,0.9166)(0.046,0.1728,0.90902)(0.046,0.1737,0.90146)(0.046,0.1746,0.89394)(0.046,0.1755,0.88644)(0.046,0.1764,0.87897)(0.046,0.1773,0.87153)(0.046,0.1782,0.86412)(0.046,0.1791,0.85674)(0.046,0.18,0.84939)

(0.048,0.09,1.4171)(0.048,0.0909,1.4276)(0.048,0.0918,1.4379)(0.048,0.0927,1.4479)(0.048,0.0936,1.4577)(0.048,0.0945,1.4673)(0.048,0.0954,1.4767)(0.048,0.0963,1.4858)(0.048,0.0972,1.4947)(0.048,0.0981,1.5034)(0.048,0.099,1.5059)(0.048,0.0999,1.5034)(0.048,0.1008,1.5006)(0.048,0.1017,1.4976)(0.048,0.1026,1.4944)(0.048,0.1035,1.491)(0.048,0.1044,1.4873)(0.048,0.1053,1.4834)(0.048,0.1062,1.4793)(0.048,0.1071,1.4751)(0.048,0.108,1.4706)(0.048,0.1089,1.466)(0.048,0.1098,1.4612)(0.048,0.1107,1.4562)(0.048,0.1116,1.4511)(0.048,0.1125,1.4458)(0.048,0.1134,1.4403)(0.048,0.1143,1.4347)(0.048,0.1152,1.429)(0.048,0.1161,1.4231)(0.048,0.117,1.4171)(0.048,0.1179,1.4109)(0.048,0.1188,1.4047)(0.048,0.1197,1.3983)(0.048,0.1206,1.3918)(0.048,0.1215,1.3853)(0.048,0.1224,1.3786)(0.048,0.1233,1.3718)(0.048,0.1242,1.3649)(0.048,0.1251,1.3579)(0.048,0.126,1.3509)(0.048,0.1269,1.3437)(0.048,0.1278,1.3365)(0.048,0.1287,1.3292)(0.048,0.1296,1.3219)(0.048,0.1305,1.3144)(0.048,0.1314,1.3069)(0.048,0.1323,1.2994)(0.048,0.1332,1.2918)(0.048,0.1341,1.2841)(0.048,0.135,1.2764)(0.048,0.1359,1.2687)(0.048,0.1368,1.2609)(0.048,0.1377,1.253)(0.048,0.1386,1.2452)(0.048,0.1395,1.2372)(0.048,0.1404,1.2293)(0.048,0.1413,1.2213)(0.048,0.1422,1.2133)(0.048,0.1431,1.2053)(0.048,0.144,1.1972)(0.048,0.1449,1.1892)(0.048,0.1458,1.1811)(0.048,0.1467,1.173)(0.048,0.1476,1.1648)(0.048,0.1485,1.1567)(0.048,0.1494,1.1486)(0.048,0.1503,1.1404)(0.048,0.1512,1.1323)(0.048,0.1521,1.1241)(0.048,0.153,1.1159)(0.048,0.1539,1.1078)(0.048,0.1548,1.0996)(0.048,0.1557,1.0915)(0.048,0.1566,1.0833)(0.048,0.1575,1.0752)(0.048,0.1584,1.067)(0.048,0.1593,1.0589)(0.048,0.1602,1.0508)(0.048,0.1611,1.0427)(0.048,0.162,1.0346)(0.048,0.1629,1.0265)(0.048,0.1638,1.0185)(0.048,0.1647,1.0104)(0.048,0.1656,1.0024)(0.048,0.1665,0.99442)(0.048,0.1674,0.98644)(0.048,0.1683,0.97848)(0.048,0.1692,0.97055)(0.048,0.1701,0.96264)(0.048,0.171,0.95475)(0.048,0.1719,0.94689)(0.048,0.1728,0.93906)(0.048,0.1737,0.93125)(0.048,0.1746,0.92347)(0.048,0.1755,0.91573)(0.048,0.1764,0.90801)(0.048,0.1773,0.90032)(0.048,0.1782,0.89267)(0.048,0.1791,0.88505)(0.048,0.18,0.87746)

(0.05,0.09,1.4171)(0.05,0.0909,1.4276)(0.05,0.0918,1.4379)(0.05,0.0927,1.4479)(0.05,0.0936,1.4577)(0.05,0.0945,1.4673)(0.05,0.0954,1.4767)(0.05,0.0963,1.4858)(0.05,0.0972,1.4947)(0.05,0.0981,1.5034)(0.05,0.099,1.5119)(0.05,0.0999,1.5202)(0.05,0.1008,1.5283)(0.05,0.1017,1.5361)(0.05,0.1026,1.5438)(0.05,0.1035,1.5402)(0.05,0.1044,1.5364)(0.05,0.1053,1.5324)(0.05,0.1062,1.5282)(0.05,0.1071,1.5238)(0.05,0.108,1.5192)(0.05,0.1089,1.5144)(0.05,0.1098,1.5095)(0.05,0.1107,1.5043)(0.05,0.1116,1.499)(0.05,0.1125,1.4935)(0.05,0.1134,1.4879)(0.05,0.1143,1.4821)(0.05,0.1152,1.4762)(0.05,0.1161,1.4701)(0.05,0.117,1.4639)(0.05,0.1179,1.4576)(0.05,0.1188,1.4511)(0.05,0.1197,1.4445)(0.05,0.1206,1.4378)(0.05,0.1215,1.431)(0.05,0.1224,1.4241)(0.05,0.1233,1.4171)(0.05,0.1242,1.41)(0.05,0.1251,1.4028)(0.05,0.126,1.3955)(0.05,0.1269,1.3881)(0.05,0.1278,1.3807)(0.05,0.1287,1.3731)(0.05,0.1296,1.3655)(0.05,0.1305,1.3579)(0.05,0.1314,1.3501)(0.05,0.1323,1.3423)(0.05,0.1332,1.3345)(0.05,0.1341,1.3266)(0.05,0.135,1.3186)(0.05,0.1359,1.3106)(0.05,0.1368,1.3025)(0.05,0.1377,1.2944)(0.05,0.1386,1.2863)(0.05,0.1395,1.2781)(0.05,0.1404,1.2699)(0.05,0.1413,1.2617)(0.05,0.1422,1.2534)(0.05,0.1431,1.2451)(0.05,0.144,1.2368)(0.05,0.1449,1.2285)(0.05,0.1458,1.2201)(0.05,0.1467,1.2117)(0.05,0.1476,1.2033)(0.05,0.1485,1.1949)(0.05,0.1494,1.1865)(0.05,0.1503,1.1781)(0.05,0.1512,1.1697)(0.05,0.1521,1.1613)(0.05,0.153,1.1528)(0.05,0.1539,1.1444)(0.05,0.1548,1.136)(0.05,0.1557,1.1275)(0.05,0.1566,1.1191)(0.05,0.1575,1.1107)(0.05,0.1584,1.1023)(0.05,0.1593,1.0939)(0.05,0.1602,1.0855)(0.05,0.1611,1.0772)(0.05,0.162,1.0688)(0.05,0.1629,1.0605)(0.05,0.1638,1.0521)(0.05,0.1647,1.0438)(0.05,0.1656,1.0355)(0.05,0.1665,1.0273)(0.05,0.1674,1.019)(0.05,0.1683,1.0108)(0.05,0.1692,1.0026)(0.05,0.1701,0.99444)(0.05,0.171,0.9863)(0.05,0.1719,0.97818)(0.05,0.1728,0.97009)(0.05,0.1737,0.96202)(0.05,0.1746,0.95399)(0.05,0.1755,0.94599)(0.05,0.1764,0.93801)(0.05,0.1773,0.93007)(0.05,0.1782,0.92217)(0.05,0.1791,0.91429)(0.05,0.18,0.90645)

(0.052,0.09,1.4171)(0.052,0.0909,1.4276)(0.052,0.0918,1.4379)(0.052,0.0927,1.4479)(0.052,0.0936,1.4577)(0.052,0.0945,1.4673)(0.052,0.0954,1.4767)(0.052,0.0963,1.4858)(0.052,0.0972,1.4947)(0.052,0.0981,1.5034)(0.052,0.099,1.5119)(0.052,0.0999,1.5202)(0.052,0.1008,1.5283)(0.052,0.1017,1.5361)(0.052,0.1026,1.5438)(0.052,0.1035,1.5512)(0.052,0.1044,1.5585)(0.052,0.1053,1.5656)(0.052,0.1062,1.5724)(0.052,0.1071,1.5742)(0.052,0.108,1.5694)(0.052,0.1089,1.5645)(0.052,0.1098,1.5593)(0.052,0.1107,1.554)(0.052,0.1116,1.5485)(0.052,0.1125,1.5429)(0.052,0.1134,1.5371)(0.052,0.1143,1.5311)(0.052,0.1152,1.525)(0.052,0.1161,1.5187)(0.052,0.117,1.5123)(0.052,0.1179,1.5057)(0.052,0.1188,1.4991)(0.052,0.1197,1.4923)(0.052,0.1206,1.4853)(0.052,0.1215,1.4783)(0.052,0.1224,1.4712)(0.052,0.1233,1.4639)(0.052,0.1242,1.4566)(0.052,0.1251,1.4491)(0.052,0.126,1.4416)(0.052,0.1269,1.434)(0.052,0.1278,1.4263)(0.052,0.1287,1.4185)(0.052,0.1296,1.4107)(0.052,0.1305,1.4027)(0.052,0.1314,1.3947)(0.052,0.1323,1.3867)(0.052,0.1332,1.3786)(0.052,0.1341,1.3704)(0.052,0.135,1.3622)(0.052,0.1359,1.3539)(0.052,0.1368,1.3456)(0.052,0.1377,1.3372)(0.052,0.1386,1.3288)(0.052,0.1395,1.3204)(0.052,0.1404,1.3119)(0.052,0.1413,1.3034)(0.052,0.1422,1.2948)(0.052,0.1431,1.2863)(0.052,0.144,1.2777)(0.052,0.1449,1.269)(0.052,0.1458,1.2604)(0.052,0.1467,1.2518)(0.052,0.1476,1.2431)(0.052,0.1485,1.2344)(0.052,0.1494,1.2257)(0.052,0.1503,1.217)(0.052,0.1512,1.2083)(0.052,0.1521,1.1996)(0.052,0.153,1.1909)(0.052,0.1539,1.1822)(0.052,0.1548,1.1735)(0.052,0.1557,1.1648)(0.052,0.1566,1.1561)(0.052,0.1575,1.1474)(0.052,0.1584,1.1387)(0.052,0.1593,1.1301)(0.052,0.1602,1.1214)(0.052,0.1611,1.1128)(0.052,0.162,1.1041)(0.052,0.1629,1.0955)(0.052,0.1638,1.0869)(0.052,0.1647,1.0783)(0.052,0.1656,1.0698)(0.052,0.1665,1.0612)(0.052,0.1674,1.0527)(0.052,0.1683,1.0442)(0.052,0.1692,1.0357)(0.052,0.1701,1.0273)(0.052,0.171,1.0189)(0.052,0.1719,1.0105)(0.052,0.1728,1.0021)(0.052,0.1737,0.99381)(0.052,0.1746,0.98551)(0.052,0.1755,0.97724)(0.052,0.1764,0.96901)(0.052,0.1773,0.96081)(0.052,0.1782,0.95264)(0.052,0.1791,0.94451)(0.052,0.18,0.93641)

(0.054,0.09,1.4171)(0.054,0.0909,1.4276)(0.054,0.0918,1.4379)(0.054,0.0927,1.4479)(0.054,0.0936,1.4577)(0.054,0.0945,1.4673)(0.054,0.0954,1.4767)(0.054,0.0963,1.4858)(0.054,0.0972,1.4947)(0.054,0.0981,1.5034)(0.054,0.099,1.5119)(0.054,0.0999,1.5202)(0.054,0.1008,1.5283)(0.054,0.1017,1.5361)(0.054,0.1026,1.5438)(0.054,0.1035,1.5512)(0.054,0.1044,1.5585)(0.054,0.1053,1.5656)(0.054,0.1062,1.5724)(0.054,0.1071,1.5791)(0.054,0.108,1.5856)(0.054,0.1089,1.5919)(0.054,0.1098,1.5981)(0.054,0.1107,1.604)(0.054,0.1116,1.5997)(0.054,0.1125,1.5939)(0.054,0.1134,1.5879)(0.054,0.1143,1.5817)(0.054,0.1152,1.5754)(0.054,0.1161,1.5689)(0.054,0.117,1.5622)(0.054,0.1179,1.5555)(0.054,0.1188,1.5486)(0.054,0.1197,1.5416)(0.054,0.1206,1.5344)(0.054,0.1215,1.5272)(0.054,0.1224,1.5198)(0.054,0.1233,1.5123)(0.054,0.1242,1.5047)(0.054,0.1251,1.497)(0.054,0.126,1.4893)(0.054,0.1269,1.4814)(0.054,0.1278,1.4734)(0.054,0.1287,1.4654)(0.054,0.1296,1.4573)(0.054,0.1305,1.4491)(0.054,0.1314,1.4408)(0.054,0.1323,1.4325)(0.054,0.1332,1.4241)(0.054,0.1341,1.4157)(0.054,0.135,1.4072)(0.054,0.1359,1.3986)(0.054,0.1368,1.39)(0.054,0.1377,1.3814)(0.054,0.1386,1.3727)(0.054,0.1395,1.364)(0.054,0.1404,1.3552)(0.054,0.1413,1.3464)(0.054,0.1422,1.3376)(0.054,0.1431,1.3288)(0.054,0.144,1.3199)(0.054,0.1449,1.311)(0.054,0.1458,1.3021)(0.054,0.1467,1.2931)(0.054,0.1476,1.2842)(0.054,0.1485,1.2752)(0.054,0.1494,1.2662)(0.054,0.1503,1.2572)(0.054,0.1512,1.2483)(0.054,0.1521,1.2393)(0.054,0.153,1.2303)(0.054,0.1539,1.2213)(0.054,0.1548,1.2123)(0.054,0.1557,1.2033)(0.054,0.1566,1.1943)(0.054,0.1575,1.1853)(0.054,0.1584,1.1764)(0.054,0.1593,1.1674)(0.054,0.1602,1.1585)(0.054,0.1611,1.1495)(0.054,0.162,1.1406)(0.054,0.1629,1.1317)(0.054,0.1638,1.1228)(0.054,0.1647,1.114)(0.054,0.1656,1.1051)(0.054,0.1665,1.0963)(0.054,0.1674,1.0875)(0.054,0.1683,1.0787)(0.054,0.1692,1.07)(0.054,0.1701,1.0613)(0.054,0.171,1.0526)(0.054,0.1719,1.0439)(0.054,0.1728,1.0353)(0.054,0.1737,1.0267)(0.054,0.1746,1.0181)(0.054,0.1755,1.0095)(0.054,0.1764,1.001)(0.054,0.1773,0.99256)(0.054,0.1782,0.98412)(0.054,0.1791,0.97572)(0.054,0.18,0.96735)

(0.056,0.09,1.4171)(0.056,0.0909,1.4276)(0.056,0.0918,1.4379)(0.056,0.0927,1.4479)(0.056,0.0936,1.4577)(0.056,0.0945,1.4673)(0.056,0.0954,1.4767)(0.056,0.0963,1.4858)(0.056,0.0972,1.4947)(0.056,0.0981,1.5034)(0.056,0.099,1.5119)(0.056,0.0999,1.5202)(0.056,0.1008,1.5283)(0.056,0.1017,1.5361)(0.056,0.1026,1.5438)(0.056,0.1035,1.5512)(0.056,0.1044,1.5585)(0.056,0.1053,1.5656)(0.056,0.1062,1.5724)(0.056,0.1071,1.5791)(0.056,0.108,1.5856)(0.056,0.1089,1.5919)(0.056,0.1098,1.5981)(0.056,0.1107,1.604)(0.056,0.1116,1.6098)(0.056,0.1125,1.6154)(0.056,0.1134,1.6208)(0.056,0.1143,1.626)(0.056,0.1152,1.6274)(0.056,0.1161,1.6207)(0.056,0.117,1.6139)(0.056,0.1179,1.6069)(0.056,0.1188,1.5998)(0.056,0.1197,1.5925)(0.056,0.1206,1.5851)(0.056,0.1215,1.5776)(0.056,0.1224,1.57)(0.056,0.1233,1.5623)(0.056,0.1242,1.5544)(0.056,0.1251,1.5465)(0.056,0.126,1.5385)(0.056,0.1269,1.5303)(0.056,0.1278,1.5221)(0.056,0.1287,1.5138)(0.056,0.1296,1.5054)(0.056,0.1305,1.497)(0.056,0.1314,1.4884)(0.056,0.1323,1.4798)(0.056,0.1332,1.4712)(0.056,0.1341,1.4625)(0.056,0.135,1.4537)(0.056,0.1359,1.4449)(0.056,0.1368,1.436)(0.056,0.1377,1.427)(0.056,0.1386,1.4181)(0.056,0.1395,1.4091)(0.056,0.1404,1.4)(0.056,0.1413,1.3909)(0.056,0.1422,1.3818)(0.056,0.1431,1.3727)(0.056,0.144,1.3635)(0.056,0.1449,1.3543)(0.056,0.1458,1.3451)(0.056,0.1467,1.3359)(0.056,0.1476,1.3266)(0.056,0.1485,1.3173)(0.056,0.1494,1.3081)(0.056,0.1503,1.2988)(0.056,0.1512,1.2895)(0.056,0.1521,1.2802)(0.056,0.153,1.2709)(0.056,0.1539,1.2616)(0.056,0.1548,1.2523)(0.056,0.1557,1.243)(0.056,0.1566,1.2338)(0.056,0.1575,1.2245)(0.056,0.1584,1.2152)(0.056,0.1593,1.206)(0.056,0.1602,1.1967)(0.056,0.1611,1.1875)(0.056,0.162,1.1783)(0.056,0.1629,1.1691)(0.056,0.1638,1.1599)(0.056,0.1647,1.1508)(0.056,0.1656,1.1416)(0.056,0.1665,1.1325)(0.056,0.1674,1.1234)(0.056,0.1683,1.1144)(0.056,0.1692,1.1053)(0.056,0.1701,1.0963)(0.056,0.171,1.0873)(0.056,0.1719,1.0784)(0.056,0.1728,1.0695)(0.056,0.1737,1.0606)(0.056,0.1746,1.0517)(0.056,0.1755,1.0429)(0.056,0.1764,1.0341)(0.056,0.1773,1.0254)(0.056,0.1782,1.0166)(0.056,0.1791,1.008)(0.056,0.18,0.99931)

(0.058,0.09,1.4171)(0.058,0.0909,1.4276)(0.058,0.0918,1.4379)(0.058,0.0927,1.4479)(0.058,0.0936,1.4577)(0.058,0.0945,1.4673)(0.058,0.0954,1.4767)(0.058,0.0963,1.4858)(0.058,0.0972,1.4947)(0.058,0.0981,1.5034)(0.058,0.099,1.5119)(0.058,0.0999,1.5202)(0.058,0.1008,1.5283)(0.058,0.1017,1.5361)(0.058,0.1026,1.5438)(0.058,0.1035,1.5512)(0.058,0.1044,1.5585)(0.058,0.1053,1.5656)(0.058,0.1062,1.5724)(0.058,0.1071,1.5791)(0.058,0.108,1.5856)(0.058,0.1089,1.5919)(0.058,0.1098,1.5981)(0.058,0.1107,1.604)(0.058,0.1116,1.6098)(0.058,0.1125,1.6154)(0.058,0.1134,1.6208)(0.058,0.1143,1.626)(0.058,0.1152,1.6311)(0.058,0.1161,1.636)(0.058,0.117,1.6408)(0.058,0.1179,1.6454)(0.058,0.1188,1.6498)(0.058,0.1197,1.6451)(0.058,0.1206,1.6375)(0.058,0.1215,1.6298)(0.058,0.1224,1.6219)(0.058,0.1233,1.6139)(0.058,0.1242,1.6058)(0.058,0.1251,1.5976)(0.058,0.126,1.5893)(0.058,0.1269,1.5809)(0.058,0.1278,1.5724)(0.058,0.1287,1.5638)(0.058,0.1296,1.5552)(0.058,0.1305,1.5464)(0.058,0.1314,1.5376)(0.058,0.1323,1.5287)(0.058,0.1332,1.5198)(0.058,0.1341,1.5108)(0.058,0.135,1.5017)(0.058,0.1359,1.4926)(0.058,0.1368,1.4834)(0.058,0.1377,1.4742)(0.058,0.1386,1.4649)(0.058,0.1395,1.4556)(0.058,0.1404,1.4463)(0.058,0.1413,1.4369)(0.058,0.1422,1.4275)(0.058,0.1431,1.418)(0.058,0.144,1.4085)(0.058,0.1449,1.399)(0.058,0.1458,1.3895)(0.058,0.1467,1.38)(0.058,0.1476,1.3704)(0.058,0.1485,1.3609)(0.058,0.1494,1.3513)(0.058,0.1503,1.3417)(0.058,0.1512,1.3321)(0.058,0.1521,1.3225)(0.058,0.153,1.3129)(0.058,0.1539,1.3033)(0.058,0.1548,1.2937)(0.058,0.1557,1.2841)(0.058,0.1566,1.2745)(0.058,0.1575,1.265)(0.058,0.1584,1.2554)(0.058,0.1593,1.2458)(0.058,0.1602,1.2363)(0.058,0.1611,1.2267)(0.058,0.162,1.2172)(0.058,0.1629,1.2077)(0.058,0.1638,1.1982)(0.058,0.1647,1.1888)(0.058,0.1656,1.1794)(0.058,0.1665,1.1699)(0.058,0.1674,1.1606)(0.058,0.1683,1.1512)(0.058,0.1692,1.1419)(0.058,0.1701,1.1325)(0.058,0.171,1.1233)(0.058,0.1719,1.114)(0.058,0.1728,1.1048)(0.058,0.1737,1.0956)(0.058,0.1746,1.0865)(0.058,0.1755,1.0774)(0.058,0.1764,1.0683)(0.058,0.1773,1.0592)(0.058,0.1782,1.0502)(0.058,0.1791,1.0413)(0.058,0.18,1.0323)

(0.06,0.09,1.4171)(0.06,0.0909,1.4276)(0.06,0.0918,1.4379)(0.06,0.0927,1.4479)(0.06,0.0936,1.4577)(0.06,0.0945,1.4673)(0.06,0.0954,1.4767)(0.06,0.0963,1.4858)(0.06,0.0972,1.4947)(0.06,0.0981,1.5034)(0.06,0.099,1.5119)(0.06,0.0999,1.5202)(0.06,0.1008,1.5283)(0.06,0.1017,1.5361)(0.06,0.1026,1.5438)(0.06,0.1035,1.5512)(0.06,0.1044,1.5585)(0.06,0.1053,1.5656)(0.06,0.1062,1.5724)(0.06,0.1071,1.5791)(0.06,0.108,1.5856)(0.06,0.1089,1.5919)(0.06,0.1098,1.5981)(0.06,0.1107,1.604)(0.06,0.1116,1.6098)(0.06,0.1125,1.6154)(0.06,0.1134,1.6208)(0.06,0.1143,1.626)(0.06,0.1152,1.6311)(0.06,0.1161,1.636)(0.06,0.117,1.6408)(0.06,0.1179,1.6454)(0.06,0.1188,1.6498)(0.06,0.1197,1.6541)(0.06,0.1206,1.6582)(0.06,0.1215,1.6621)(0.06,0.1224,1.6659)(0.06,0.1233,1.6672)(0.06,0.1242,1.6589)(0.06,0.1251,1.6504)(0.06,0.126,1.6418)(0.06,0.1269,1.6331)(0.06,0.1278,1.6244)(0.06,0.1287,1.6155)(0.06,0.1296,1.6066)(0.06,0.1305,1.5975)(0.06,0.1314,1.5884)(0.06,0.1323,1.5793)(0.06,0.1332,1.57)(0.06,0.1341,1.5607)(0.06,0.135,1.5513)(0.06,0.1359,1.5419)(0.06,0.1368,1.5324)(0.06,0.1377,1.5229)(0.06,0.1386,1.5133)(0.06,0.1395,1.5037)(0.06,0.1404,1.4941)(0.06,0.1413,1.4844)(0.06,0.1422,1.4746)(0.06,0.1431,1.4649)(0.06,0.144,1.4551)(0.06,0.1449,1.4453)(0.06,0.1458,1.4354)(0.06,0.1467,1.4256)(0.06,0.1476,1.4157)(0.06,0.1485,1.4058)(0.06,0.1494,1.3959)(0.06,0.1503,1.386)(0.06,0.1512,1.3761)(0.06,0.1521,1.3662)(0.06,0.153,1.3563)(0.06,0.1539,1.3464)(0.06,0.1548,1.3365)(0.06,0.1557,1.3266)(0.06,0.1566,1.3167)(0.06,0.1575,1.3068)(0.06,0.1584,1.2969)(0.06,0.1593,1.287)(0.06,0.1602,1.2771)(0.06,0.1611,1.2673)(0.06,0.162,1.2574)(0.06,0.1629,1.2476)(0.06,0.1638,1.2378)(0.06,0.1647,1.2281)(0.06,0.1656,1.2183)(0.06,0.1665,1.2086)(0.06,0.1674,1.1989)(0.06,0.1683,1.1892)(0.06,0.1692,1.1796)(0.06,0.1701,1.17)(0.06,0.171,1.1604)(0.06,0.1719,1.1508)(0.06,0.1728,1.1413)(0.06,0.1737,1.1318)(0.06,0.1746,1.1224)(0.06,0.1755,1.113)(0.06,0.1764,1.1036)(0.06,0.1773,1.0942)(0.06,0.1782,1.0849)(0.06,0.1791,1.0757)(0.06,0.18,1.0664)

(0.062,0.09,1.4171)(0.062,0.0909,1.4276)(0.062,0.0918,1.4379)(0.062,0.0927,1.4479)(0.062,0.0936,1.4577)(0.062,0.0945,1.4673)(0.062,0.0954,1.4767)(0.062,0.0963,1.4858)(0.062,0.0972,1.4947)(0.062,0.0981,1.5034)(0.062,0.099,1.5119)(0.062,0.0999,1.5202)(0.062,0.1008,1.5283)(0.062,0.1017,1.5361)(0.062,0.1026,1.5438)(0.062,0.1035,1.5512)(0.062,0.1044,1.5585)(0.062,0.1053,1.5656)(0.062,0.1062,1.5724)(0.062,0.1071,1.5791)(0.062,0.108,1.5856)(0.062,0.1089,1.5919)(0.062,0.1098,1.5981)(0.062,0.1107,1.604)(0.062,0.1116,1.6098)(0.062,0.1125,1.6154)(0.062,0.1134,1.6208)(0.062,0.1143,1.626)(0.062,0.1152,1.6311)(0.062,0.1161,1.636)(0.062,0.117,1.6408)(0.062,0.1179,1.6454)(0.062,0.1188,1.6498)(0.062,0.1197,1.6541)(0.062,0.1206,1.6582)(0.062,0.1215,1.6621)(0.062,0.1224,1.6659)(0.062,0.1233,1.6696)(0.062,0.1242,1.6731)(0.062,0.1251,1.6765)(0.062,0.126,1.6797)(0.062,0.1269,1.6828)(0.062,0.1278,1.678)(0.062,0.1287,1.6689)(0.062,0.1296,1.6597)(0.062,0.1305,1.6503)(0.062,0.1314,1.6409)(0.062,0.1323,1.6314)(0.062,0.1332,1.6219)(0.062,0.1341,1.6123)(0.062,0.135,1.6026)(0.062,0.1359,1.5929)(0.062,0.1368,1.5831)(0.062,0.1377,1.5732)(0.062,0.1386,1.5633)(0.062,0.1395,1.5534)(0.062,0.1404,1.5434)(0.062,0.1413,1.5334)(0.062,0.1422,1.5234)(0.062,0.1431,1.5133)(0.062,0.144,1.5032)(0.062,0.1449,1.493)(0.062,0.1458,1.4829)(0.062,0.1467,1.4727)(0.062,0.1476,1.4625)(0.062,0.1485,1.4523)(0.062,0.1494,1.4421)(0.062,0.1503,1.4318)(0.062,0.1512,1.4216)(0.062,0.1521,1.4114)(0.062,0.153,1.4011)(0.062,0.1539,1.3909)(0.062,0.1548,1.3806)(0.062,0.1557,1.3704)(0.062,0.1566,1.3602)(0.062,0.1575,1.3499)(0.062,0.1584,1.3397)(0.062,0.1593,1.3295)(0.062,0.1602,1.3193)(0.062,0.1611,1.3092)(0.062,0.162,1.299)(0.062,0.1629,1.2889)(0.062,0.1638,1.2787)(0.062,0.1647,1.2687)(0.062,0.1656,1.2586)(0.062,0.1665,1.2485)(0.062,0.1674,1.2385)(0.062,0.1683,1.2285)(0.062,0.1692,1.2186)(0.062,0.1701,1.2086)(0.062,0.171,1.1987)(0.062,0.1719,1.1889)(0.062,0.1728,1.179)(0.062,0.1737,1.1692)(0.062,0.1746,1.1595)(0.062,0.1755,1.1497)(0.062,0.1764,1.14)(0.062,0.1773,1.1304)(0.062,0.1782,1.1208)(0.062,0.1791,1.1112)(0.062,0.18,1.1017)

(0.064,0.09,1.4171)(0.064,0.0909,1.4276)(0.064,0.0918,1.4379)(0.064,0.0927,1.4479)(0.064,0.0936,1.4577)(0.064,0.0945,1.4673)(0.064,0.0954,1.4767)(0.064,0.0963,1.4858)(0.064,0.0972,1.4947)(0.064,0.0981,1.5034)(0.064,0.099,1.5119)(0.064,0.0999,1.5202)(0.064,0.1008,1.5283)(0.064,0.1017,1.5361)(0.064,0.1026,1.5438)(0.064,0.1035,1.5512)(0.064,0.1044,1.5585)(0.064,0.1053,1.5656)(0.064,0.1062,1.5724)(0.064,0.1071,1.5791)(0.064,0.108,1.5856)(0.064,0.1089,1.5919)(0.064,0.1098,1.5981)(0.064,0.1107,1.604)(0.064,0.1116,1.6098)(0.064,0.1125,1.6154)(0.064,0.1134,1.6208)(0.064,0.1143,1.626)(0.064,0.1152,1.6311)(0.064,0.1161,1.636)(0.064,0.117,1.6408)(0.064,0.1179,1.6454)(0.064,0.1188,1.6498)(0.064,0.1197,1.6541)(0.064,0.1206,1.6582)(0.064,0.1215,1.6621)(0.064,0.1224,1.6659)(0.064,0.1233,1.6696)(0.064,0.1242,1.6731)(0.064,0.1251,1.6765)(0.064,0.126,1.6797)(0.064,0.1269,1.6828)(0.064,0.1278,1.6857)(0.064,0.1287,1.6885)(0.064,0.1296,1.6912)(0.064,0.1305,1.6937)(0.064,0.1314,1.6951)(0.064,0.1323,1.6854)(0.064,0.1332,1.6755)(0.064,0.1341,1.6656)(0.064,0.135,1.6556)(0.064,0.1359,1.6455)(0.064,0.1368,1.6354)(0.064,0.1377,1.6252)(0.064,0.1386,1.615)(0.064,0.1395,1.6047)(0.064,0.1404,1.5944)(0.064,0.1413,1.5841)(0.064,0.1422,1.5737)(0.064,0.1431,1.5633)(0.064,0.144,1.5528)(0.064,0.1449,1.5424)(0.064,0.1458,1.5319)(0.064,0.1467,1.5214)(0.064,0.1476,1.5108)(0.064,0.1485,1.5003)(0.064,0.1494,1.4897)(0.064,0.1503,1.4792)(0.064,0.1512,1.4686)(0.064,0.1521,1.458)(0.064,0.153,1.4474)(0.064,0.1539,1.4368)(0.064,0.1548,1.4263)(0.064,0.1557,1.4157)(0.064,0.1566,1.4051)(0.064,0.1575,1.3945)(0.064,0.1584,1.384)(0.064,0.1593,1.3734)(0.064,0.1602,1.3629)(0.064,0.1611,1.3524)(0.064,0.162,1.3419)(0.064,0.1629,1.3315)(0.064,0.1638,1.321)(0.064,0.1647,1.3106)(0.064,0.1656,1.3002)(0.064,0.1665,1.2898)(0.064,0.1674,1.2794)(0.064,0.1683,1.2691)(0.064,0.1692,1.2588)(0.064,0.1701,1.2486)(0.064,0.171,1.2383)(0.064,0.1719,1.2281)(0.064,0.1728,1.218)(0.064,0.1737,1.2079)(0.064,0.1746,1.1978)(0.064,0.1755,1.1877)(0.064,0.1764,1.1777)(0.064,0.1773,1.1677)(0.064,0.1782,1.1578)(0.064,0.1791,1.1479)(0.064,0.18,1.1381)

(0.066,0.09,1.4171)(0.066,0.0909,1.4276)(0.066,0.0918,1.4379)(0.066,0.0927,1.4479)(0.066,0.0936,1.4577)(0.066,0.0945,1.4673)(0.066,0.0954,1.4767)(0.066,0.0963,1.4858)(0.066,0.0972,1.4947)(0.066,0.0981,1.5034)(0.066,0.099,1.5119)(0.066,0.0999,1.5202)(0.066,0.1008,1.5283)(0.066,0.1017,1.5361)(0.066,0.1026,1.5438)(0.066,0.1035,1.5512)(0.066,0.1044,1.5585)(0.066,0.1053,1.5656)(0.066,0.1062,1.5724)(0.066,0.1071,1.5791)(0.066,0.108,1.5856)(0.066,0.1089,1.5919)(0.066,0.1098,1.5981)(0.066,0.1107,1.604)(0.066,0.1116,1.6098)(0.066,0.1125,1.6154)(0.066,0.1134,1.6208)(0.066,0.1143,1.626)(0.066,0.1152,1.6311)(0.066,0.1161,1.636)(0.066,0.117,1.6408)(0.066,0.1179,1.6454)(0.066,0.1188,1.6498)(0.066,0.1197,1.6541)(0.066,0.1206,1.6582)(0.066,0.1215,1.6621)(0.066,0.1224,1.6659)(0.066,0.1233,1.6696)(0.066,0.1242,1.6731)(0.066,0.1251,1.6765)(0.066,0.126,1.6797)(0.066,0.1269,1.6828)(0.066,0.1278,1.6857)(0.066,0.1287,1.6885)(0.066,0.1296,1.6912)(0.066,0.1305,1.6937)(0.066,0.1314,1.6961)(0.066,0.1323,1.6984)(0.066,0.1332,1.7005)(0.066,0.1341,1.7025)(0.066,0.135,1.7044)(0.066,0.1359,1.6999)(0.066,0.1368,1.6894)(0.066,0.1377,1.6789)(0.066,0.1386,1.6684)(0.066,0.1395,1.6578)(0.066,0.1404,1.6471)(0.066,0.1413,1.6364)(0.066,0.1422,1.6257)(0.066,0.1431,1.6149)(0.066,0.144,1.6042)(0.066,0.1449,1.5933)(0.066,0.1458,1.5825)(0.066,0.1467,1.5716)(0.066,0.1476,1.5608)(0.066,0.1485,1.5499)(0.066,0.1494,1.539)(0.066,0.1503,1.528)(0.066,0.1512,1.5171)(0.066,0.1521,1.5062)(0.066,0.153,1.4952)(0.066,0.1539,1.4843)(0.066,0.1548,1.4734)(0.066,0.1557,1.4625)(0.066,0.1566,1.4515)(0.066,0.1575,1.4406)(0.066,0.1584,1.4297)(0.066,0.1593,1.4188)(0.066,0.1602,1.408)(0.066,0.1611,1.3971)(0.066,0.162,1.3863)(0.066,0.1629,1.3754)(0.066,0.1638,1.3647)(0.066,0.1647,1.3539)(0.066,0.1656,1.3431)(0.066,0.1665,1.3324)(0.066,0.1674,1.3217)(0.066,0.1683,1.3111)(0.066,0.1692,1.3004)(0.066,0.1701,1.2898)(0.066,0.171,1.2793)(0.066,0.1719,1.2687)(0.066,0.1728,1.2582)(0.066,0.1737,1.2478)(0.066,0.1746,1.2374)(0.066,0.1755,1.227)(0.066,0.1764,1.2166)(0.066,0.1773,1.2063)(0.066,0.1782,1.1961)(0.066,0.1791,1.1859)(0.066,0.18,1.1757)

(0.068,0.09,1.4171)(0.068,0.0909,1.4276)(0.068,0.0918,1.4379)(0.068,0.0927,1.4479)(0.068,0.0936,1.4577)(0.068,0.0945,1.4673)(0.068,0.0954,1.4767)(0.068,0.0963,1.4858)(0.068,0.0972,1.4947)(0.068,0.0981,1.5034)(0.068,0.099,1.5119)(0.068,0.0999,1.5202)(0.068,0.1008,1.5283)(0.068,0.1017,1.5361)(0.068,0.1026,1.5438)(0.068,0.1035,1.5512)(0.068,0.1044,1.5585)(0.068,0.1053,1.5656)(0.068,0.1062,1.5724)(0.068,0.1071,1.5791)(0.068,0.108,1.5856)(0.068,0.1089,1.5919)(0.068,0.1098,1.5981)(0.068,0.1107,1.604)(0.068,0.1116,1.6098)(0.068,0.1125,1.6154)(0.068,0.1134,1.6208)(0.068,0.1143,1.626)(0.068,0.1152,1.6311)(0.068,0.1161,1.636)(0.068,0.117,1.6408)(0.068,0.1179,1.6454)(0.068,0.1188,1.6498)(0.068,0.1197,1.6541)(0.068,0.1206,1.6582)(0.068,0.1215,1.6621)(0.068,0.1224,1.6659)(0.068,0.1233,1.6696)(0.068,0.1242,1.6731)(0.068,0.1251,1.6765)(0.068,0.126,1.6797)(0.068,0.1269,1.6828)(0.068,0.1278,1.6857)(0.068,0.1287,1.6885)(0.068,0.1296,1.6912)(0.068,0.1305,1.6937)(0.068,0.1314,1.6961)(0.068,0.1323,1.6984)(0.068,0.1332,1.7005)(0.068,0.1341,1.7025)(0.068,0.135,1.7044)(0.068,0.1359,1.7062)(0.068,0.1368,1.7078)(0.068,0.1377,1.7094)(0.068,0.1386,1.7108)(0.068,0.1395,1.7121)(0.068,0.1404,1.7015)(0.068,0.1413,1.6905)(0.068,0.1422,1.6794)(0.068,0.1431,1.6683)(0.068,0.144,1.6572)(0.068,0.1449,1.646)(0.068,0.1458,1.6348)(0.068,0.1467,1.6236)(0.068,0.1476,1.6123)(0.068,0.1485,1.6011)(0.068,0.1494,1.5898)(0.068,0.1503,1.5785)(0.068,0.1512,1.5672)(0.068,0.1521,1.5559)(0.068,0.153,1.5447)(0.068,0.1539,1.5334)(0.068,0.1548,1.5221)(0.068,0.1557,1.5108)(0.068,0.1566,1.4995)(0.068,0.1575,1.4882)(0.068,0.1584,1.477)(0.068,0.1593,1.4657)(0.068,0.1602,1.4545)(0.068,0.1611,1.4433)(0.068,0.162,1.4321)(0.068,0.1629,1.4209)(0.068,0.1638,1.4097)(0.068,0.1647,1.3986)(0.068,0.1656,1.3875)(0.068,0.1665,1.3764)(0.068,0.1674,1.3654)(0.068,0.1683,1.3544)(0.068,0.1692,1.3434)(0.068,0.1701,1.3324)(0.068,0.171,1.3215)(0.068,0.1719,1.3106)(0.068,0.1728,1.2998)(0.068,0.1737,1.289)(0.068,0.1746,1.2782)(0.068,0.1755,1.2675)(0.068,0.1764,1.2568)(0.068,0.1773,1.2462)(0.068,0.1782,1.2356)(0.068,0.1791,1.2251)(0.068,0.18,1.2145)

(0.07,0.09,1.4171)(0.07,0.0909,1.4276)(0.07,0.0918,1.4379)(0.07,0.0927,1.4479)(0.07,0.0936,1.4577)(0.07,0.0945,1.4673)(0.07,0.0954,1.4767)(0.07,0.0963,1.4858)(0.07,0.0972,1.4947)(0.07,0.0981,1.5034)(0.07,0.099,1.5119)(0.07,0.0999,1.5202)(0.07,0.1008,1.5283)(0.07,0.1017,1.5361)(0.07,0.1026,1.5438)(0.07,0.1035,1.5512)(0.07,0.1044,1.5585)(0.07,0.1053,1.5656)(0.07,0.1062,1.5724)(0.07,0.1071,1.5791)(0.07,0.108,1.5856)(0.07,0.1089,1.5919)(0.07,0.1098,1.5981)(0.07,0.1107,1.604)(0.07,0.1116,1.6098)(0.07,0.1125,1.6154)(0.07,0.1134,1.6208)(0.07,0.1143,1.626)(0.07,0.1152,1.6311)(0.07,0.1161,1.636)(0.07,0.117,1.6408)(0.07,0.1179,1.6454)(0.07,0.1188,1.6498)(0.07,0.1197,1.6541)(0.07,0.1206,1.6582)(0.07,0.1215,1.6621)(0.07,0.1224,1.6659)(0.07,0.1233,1.6696)(0.07,0.1242,1.6731)(0.07,0.1251,1.6765)(0.07,0.126,1.6797)(0.07,0.1269,1.6828)(0.07,0.1278,1.6857)(0.07,0.1287,1.6885)(0.07,0.1296,1.6912)(0.07,0.1305,1.6937)(0.07,0.1314,1.6961)(0.07,0.1323,1.6984)(0.07,0.1332,1.7005)(0.07,0.1341,1.7025)(0.07,0.135,1.7044)(0.07,0.1359,1.7062)(0.07,0.1368,1.7078)(0.07,0.1377,1.7094)(0.07,0.1386,1.7108)(0.07,0.1395,1.7121)(0.07,0.1404,1.7132)(0.07,0.1413,1.7143)(0.07,0.1422,1.7152)(0.07,0.1431,1.7161)(0.07,0.144,1.7119)(0.07,0.1449,1.7004)(0.07,0.1458,1.6888)(0.07,0.1467,1.6772)(0.07,0.1476,1.6656)(0.07,0.1485,1.654)(0.07,0.1494,1.6423)(0.07,0.1503,1.6307)(0.07,0.1512,1.619)(0.07,0.1521,1.6074)(0.07,0.153,1.5957)(0.07,0.1539,1.584)(0.07,0.1548,1.5724)(0.07,0.1557,1.5607)(0.07,0.1566,1.549)(0.07,0.1575,1.5374)(0.07,0.1584,1.5258)(0.07,0.1593,1.5141)(0.07,0.1602,1.5025)(0.07,0.1611,1.491)(0.07,0.162,1.4794)(0.07,0.1629,1.4678)(0.07,0.1638,1.4563)(0.07,0.1647,1.4448)(0.07,0.1656,1.4334)(0.07,0.1665,1.4219)(0.07,0.1674,1.4105)(0.07,0.1683,1.3991)(0.07,0.1692,1.3878)(0.07,0.1701,1.3765)(0.07,0.171,1.3652)(0.07,0.1719,1.354)(0.07,0.1728,1.3428)(0.07,0.1737,1.3316)(0.07,0.1746,1.3205)(0.07,0.1755,1.3094)(0.07,0.1764,1.2984)(0.07,0.1773,1.2874)(0.07,0.1782,1.2764)(0.07,0.1791,1.2655)(0.07,0.18,1.2547)

(0.072,0.09,1.4171)(0.072,0.0909,1.4276)(0.072,0.0918,1.4379)(0.072,0.0927,1.4479)(0.072,0.0936,1.4577)(0.072,0.0945,1.4673)(0.072,0.0954,1.4767)(0.072,0.0963,1.4858)(0.072,0.0972,1.4947)(0.072,0.0981,1.5034)(0.072,0.099,1.5119)(0.072,0.0999,1.5202)(0.072,0.1008,1.5283)(0.072,0.1017,1.5361)(0.072,0.1026,1.5438)(0.072,0.1035,1.5512)(0.072,0.1044,1.5585)(0.072,0.1053,1.5656)(0.072,0.1062,1.5724)(0.072,0.1071,1.5791)(0.072,0.108,1.5856)(0.072,0.1089,1.5919)(0.072,0.1098,1.5981)(0.072,0.1107,1.604)(0.072,0.1116,1.6098)(0.072,0.1125,1.6154)(0.072,0.1134,1.6208)(0.072,0.1143,1.626)(0.072,0.1152,1.6311)(0.072,0.1161,1.636)(0.072,0.117,1.6408)(0.072,0.1179,1.6454)(0.072,0.1188,1.6498)(0.072,0.1197,1.6541)(0.072,0.1206,1.6582)(0.072,0.1215,1.6621)(0.072,0.1224,1.6659)(0.072,0.1233,1.6696)(0.072,0.1242,1.6731)(0.072,0.1251,1.6765)(0.072,0.126,1.6797)(0.072,0.1269,1.6828)(0.072,0.1278,1.6857)(0.072,0.1287,1.6885)(0.072,0.1296,1.6912)(0.072,0.1305,1.6937)(0.072,0.1314,1.6961)(0.072,0.1323,1.6984)(0.072,0.1332,1.7005)(0.072,0.1341,1.7025)(0.072,0.135,1.7044)(0.072,0.1359,1.7062)(0.072,0.1368,1.7078)(0.072,0.1377,1.7094)(0.072,0.1386,1.7108)(0.072,0.1395,1.7121)(0.072,0.1404,1.7132)(0.072,0.1413,1.7143)(0.072,0.1422,1.7152)(0.072,0.1431,1.7161)(0.072,0.144,1.7168)(0.072,0.1449,1.7174)(0.072,0.1458,1.718)(0.072,0.1467,1.7184)(0.072,0.1476,1.7187)(0.072,0.1485,1.7086)(0.072,0.1494,1.6966)(0.072,0.1503,1.6846)(0.072,0.1512,1.6725)(0.072,0.1521,1.6605)(0.072,0.153,1.6484)(0.072,0.1539,1.6364)(0.072,0.1548,1.6243)(0.072,0.1557,1.6123)(0.072,0.1566,1.6002)(0.072,0.1575,1.5882)(0.072,0.1584,1.5762)(0.072,0.1593,1.5642)(0.072,0.1602,1.5522)(0.072,0.1611,1.5402)(0.072,0.162,1.5283)(0.072,0.1629,1.5164)(0.072,0.1638,1.5045)(0.072,0.1647,1.4926)(0.072,0.1656,1.4807)(0.072,0.1665,1.4689)(0.072,0.1674,1.4571)(0.072,0.1683,1.4454)(0.072,0.1692,1.4336)(0.072,0.1701,1.422)(0.072,0.171,1.4103)(0.072,0.1719,1.3987)(0.072,0.1728,1.3871)(0.072,0.1737,1.3756)(0.072,0.1746,1.3641)(0.072,0.1755,1.3527)(0.072,0.1764,1.3413)(0.072,0.1773,1.3299)(0.072,0.1782,1.3186)(0.072,0.1791,1.3073)(0.072,0.18,1.2961)

(0.074,0.09,1.4171)(0.074,0.0909,1.4276)(0.074,0.0918,1.4379)(0.074,0.0927,1.4479)(0.074,0.0936,1.4577)(0.074,0.0945,1.4673)(0.074,0.0954,1.4767)(0.074,0.0963,1.4858)(0.074,0.0972,1.4947)(0.074,0.0981,1.5034)(0.074,0.099,1.5119)(0.074,0.0999,1.5202)(0.074,0.1008,1.5283)(0.074,0.1017,1.5361)(0.074,0.1026,1.5438)(0.074,0.1035,1.5512)(0.074,0.1044,1.5585)(0.074,0.1053,1.5656)(0.074,0.1062,1.5724)(0.074,0.1071,1.5791)(0.074,0.108,1.5856)(0.074,0.1089,1.5919)(0.074,0.1098,1.5981)(0.074,0.1107,1.604)(0.074,0.1116,1.6098)(0.074,0.1125,1.6154)(0.074,0.1134,1.6208)(0.074,0.1143,1.626)(0.074,0.1152,1.6311)(0.074,0.1161,1.636)(0.074,0.117,1.6408)(0.074,0.1179,1.6454)(0.074,0.1188,1.6498)(0.074,0.1197,1.6541)(0.074,0.1206,1.6582)(0.074,0.1215,1.6621)(0.074,0.1224,1.6659)(0.074,0.1233,1.6696)(0.074,0.1242,1.6731)(0.074,0.1251,1.6765)(0.074,0.126,1.6797)(0.074,0.1269,1.6828)(0.074,0.1278,1.6857)(0.074,0.1287,1.6885)(0.074,0.1296,1.6912)(0.074,0.1305,1.6937)(0.074,0.1314,1.6961)(0.074,0.1323,1.6984)(0.074,0.1332,1.7005)(0.074,0.1341,1.7025)(0.074,0.135,1.7044)(0.074,0.1359,1.7062)(0.074,0.1368,1.7078)(0.074,0.1377,1.7094)(0.074,0.1386,1.7108)(0.074,0.1395,1.7121)(0.074,0.1404,1.7132)(0.074,0.1413,1.7143)(0.074,0.1422,1.7152)(0.074,0.1431,1.7161)(0.074,0.144,1.7168)(0.074,0.1449,1.7174)(0.074,0.1458,1.718)(0.074,0.1467,1.7184)(0.074,0.1476,1.7187)(0.074,0.1485,1.7189)(0.074,0.1494,1.719)(0.074,0.1503,1.719)(0.074,0.1512,1.7189)(0.074,0.1521,1.7153)(0.074,0.153,1.7029)(0.074,0.1539,1.6904)(0.074,0.1548,1.678)(0.074,0.1557,1.6655)(0.074,0.1566,1.6531)(0.074,0.1575,1.6407)(0.074,0.1584,1.6283)(0.074,0.1593,1.6159)(0.074,0.1602,1.6035)(0.074,0.1611,1.5911)(0.074,0.162,1.5788)(0.074,0.1629,1.5665)(0.074,0.1638,1.5542)(0.074,0.1647,1.5419)(0.074,0.1656,1.5297)(0.074,0.1665,1.5174)(0.074,0.1674,1.5053)(0.074,0.1683,1.4931)(0.074,0.1692,1.481)(0.074,0.1701,1.4689)(0.074,0.171,1.4569)(0.074,0.1719,1.4449)(0.074,0.1728,1.433)(0.074,0.1737,1.4211)(0.074,0.1746,1.4092)(0.074,0.1755,1.3974)(0.074,0.1764,1.3856)(0.074,0.1773,1.3739)(0.074,0.1782,1.3622)(0.074,0.1791,1.3505)(0.074,0.18,1.339)

(0.076,0.09,1.4171)(0.076,0.0909,1.4276)(0.076,0.0918,1.4379)(0.076,0.0927,1.4479)(0.076,0.0936,1.4577)(0.076,0.0945,1.4673)(0.076,0.0954,1.4767)(0.076,0.0963,1.4858)(0.076,0.0972,1.4947)(0.076,0.0981,1.5034)(0.076,0.099,1.5119)(0.076,0.0999,1.5202)(0.076,0.1008,1.5283)(0.076,0.1017,1.5361)(0.076,0.1026,1.5438)(0.076,0.1035,1.5512)(0.076,0.1044,1.5585)(0.076,0.1053,1.5656)(0.076,0.1062,1.5724)(0.076,0.1071,1.5791)(0.076,0.108,1.5856)(0.076,0.1089,1.5919)(0.076,0.1098,1.5981)(0.076,0.1107,1.604)(0.076,0.1116,1.6098)(0.076,0.1125,1.6154)(0.076,0.1134,1.6208)(0.076,0.1143,1.626)(0.076,0.1152,1.6311)(0.076,0.1161,1.636)(0.076,0.117,1.6408)(0.076,0.1179,1.6454)(0.076,0.1188,1.6498)(0.076,0.1197,1.6541)(0.076,0.1206,1.6582)(0.076,0.1215,1.6621)(0.076,0.1224,1.6659)(0.076,0.1233,1.6696)(0.076,0.1242,1.6731)(0.076,0.1251,1.6765)(0.076,0.126,1.6797)(0.076,0.1269,1.6828)(0.076,0.1278,1.6857)(0.076,0.1287,1.6885)(0.076,0.1296,1.6912)(0.076,0.1305,1.6937)(0.076,0.1314,1.6961)(0.076,0.1323,1.6984)(0.076,0.1332,1.7005)(0.076,0.1341,1.7025)(0.076,0.135,1.7044)(0.076,0.1359,1.7062)(0.076,0.1368,1.7078)(0.076,0.1377,1.7094)(0.076,0.1386,1.7108)(0.076,0.1395,1.7121)(0.076,0.1404,1.7132)(0.076,0.1413,1.7143)(0.076,0.1422,1.7152)(0.076,0.1431,1.7161)(0.076,0.144,1.7168)(0.076,0.1449,1.7174)(0.076,0.1458,1.718)(0.076,0.1467,1.7184)(0.076,0.1476,1.7187)(0.076,0.1485,1.7189)(0.076,0.1494,1.719)(0.076,0.1503,1.719)(0.076,0.1512,1.7189)(0.076,0.1521,1.7188)(0.076,0.153,1.7185)(0.076,0.1539,1.7181)(0.076,0.1548,1.7177)(0.076,0.1557,1.7171)(0.076,0.1566,1.7077)(0.076,0.1575,1.6949)(0.076,0.1584,1.6821)(0.076,0.1593,1.6693)(0.076,0.1602,1.6565)(0.076,0.1611,1.6437)(0.076,0.162,1.6309)(0.076,0.1629,1.6182)(0.076,0.1638,1.6055)(0.076,0.1647,1.5928)(0.076,0.1656,1.5802)(0.076,0.1665,1.5676)(0.076,0.1674,1.555)(0.076,0.1683,1.5425)(0.076,0.1692,1.53)(0.076,0.1701,1.5175)(0.076,0.171,1.5051)(0.076,0.1719,1.4927)(0.076,0.1728,1.4803)(0.076,0.1737,1.468)(0.076,0.1746,1.4557)(0.076,0.1755,1.4435)(0.076,0.1764,1.4314)(0.076,0.1773,1.4193)(0.076,0.1782,1.4072)(0.076,0.1791,1.3952)(0.076,0.18,1.3832)

(0.078,0.09,1.4171)(0.078,0.0909,1.4276)(0.078,0.0918,1.4379)(0.078,0.0927,1.4479)(0.078,0.0936,1.4577)(0.078,0.0945,1.4673)(0.078,0.0954,1.4767)(0.078,0.0963,1.4858)(0.078,0.0972,1.4947)(0.078,0.0981,1.5034)(0.078,0.099,1.5119)(0.078,0.0999,1.5202)(0.078,0.1008,1.5283)(0.078,0.1017,1.5361)(0.078,0.1026,1.5438)(0.078,0.1035,1.5512)(0.078,0.1044,1.5585)(0.078,0.1053,1.5656)(0.078,0.1062,1.5724)(0.078,0.1071,1.5791)(0.078,0.108,1.5856)(0.078,0.1089,1.5919)(0.078,0.1098,1.5981)(0.078,0.1107,1.604)(0.078,0.1116,1.6098)(0.078,0.1125,1.6154)(0.078,0.1134,1.6208)(0.078,0.1143,1.626)(0.078,0.1152,1.6311)(0.078,0.1161,1.636)(0.078,0.117,1.6408)(0.078,0.1179,1.6454)(0.078,0.1188,1.6498)(0.078,0.1197,1.6541)(0.078,0.1206,1.6582)(0.078,0.1215,1.6621)(0.078,0.1224,1.6659)(0.078,0.1233,1.6696)(0.078,0.1242,1.6731)(0.078,0.1251,1.6765)(0.078,0.126,1.6797)(0.078,0.1269,1.6828)(0.078,0.1278,1.6857)(0.078,0.1287,1.6885)(0.078,0.1296,1.6912)(0.078,0.1305,1.6937)(0.078,0.1314,1.6961)(0.078,0.1323,1.6984)(0.078,0.1332,1.7005)(0.078,0.1341,1.7025)(0.078,0.135,1.7044)(0.078,0.1359,1.7062)(0.078,0.1368,1.7078)(0.078,0.1377,1.7094)(0.078,0.1386,1.7108)(0.078,0.1395,1.7121)(0.078,0.1404,1.7132)(0.078,0.1413,1.7143)(0.078,0.1422,1.7152)(0.078,0.1431,1.7161)(0.078,0.144,1.7168)(0.078,0.1449,1.7174)(0.078,0.1458,1.718)(0.078,0.1467,1.7184)(0.078,0.1476,1.7187)(0.078,0.1485,1.7189)(0.078,0.1494,1.719)(0.078,0.1503,1.719)(0.078,0.1512,1.7189)(0.078,0.1521,1.7188)(0.078,0.153,1.7185)(0.078,0.1539,1.7181)(0.078,0.1548,1.7177)(0.078,0.1557,1.7171)(0.078,0.1566,1.7165)(0.078,0.1575,1.7158)(0.078,0.1584,1.715)(0.078,0.1593,1.7141)(0.078,0.1602,1.7112)(0.078,0.1611,1.698)(0.078,0.162,1.6848)(0.078,0.1629,1.6717)(0.078,0.1638,1.6586)(0.078,0.1647,1.6455)(0.078,0.1656,1.6324)(0.078,0.1665,1.6194)(0.078,0.1674,1.6064)(0.078,0.1683,1.5934)(0.078,0.1692,1.5805)(0.078,0.1701,1.5676)(0.078,0.171,1.5548)(0.078,0.1719,1.542)(0.078,0.1728,1.5292)(0.078,0.1737,1.5165)(0.078,0.1746,1.5039)(0.078,0.1755,1.4912)(0.078,0.1764,1.4787)(0.078,0.1773,1.4662)(0.078,0.1782,1.4537)(0.078,0.1791,1.4413)(0.078,0.18,1.4289)

(0.08,0.09,1.4171)(0.08,0.0909,1.4276)(0.08,0.0918,1.4379)(0.08,0.0927,1.4479)(0.08,0.0936,1.4577)(0.08,0.0945,1.4673)(0.08,0.0954,1.4767)(0.08,0.0963,1.4858)(0.08,0.0972,1.4947)(0.08,0.0981,1.5034)(0.08,0.099,1.5119)(0.08,0.0999,1.5202)(0.08,0.1008,1.5283)(0.08,0.1017,1.5361)(0.08,0.1026,1.5438)(0.08,0.1035,1.5512)(0.08,0.1044,1.5585)(0.08,0.1053,1.5656)(0.08,0.1062,1.5724)(0.08,0.1071,1.5791)(0.08,0.108,1.5856)(0.08,0.1089,1.5919)(0.08,0.1098,1.5981)(0.08,0.1107,1.604)(0.08,0.1116,1.6098)(0.08,0.1125,1.6154)(0.08,0.1134,1.6208)(0.08,0.1143,1.626)(0.08,0.1152,1.6311)(0.08,0.1161,1.636)(0.08,0.117,1.6408)(0.08,0.1179,1.6454)(0.08,0.1188,1.6498)(0.08,0.1197,1.6541)(0.08,0.1206,1.6582)(0.08,0.1215,1.6621)(0.08,0.1224,1.6659)(0.08,0.1233,1.6696)(0.08,0.1242,1.6731)(0.08,0.1251,1.6765)(0.08,0.126,1.6797)(0.08,0.1269,1.6828)(0.08,0.1278,1.6857)(0.08,0.1287,1.6885)(0.08,0.1296,1.6912)(0.08,0.1305,1.6937)(0.08,0.1314,1.6961)(0.08,0.1323,1.6984)(0.08,0.1332,1.7005)(0.08,0.1341,1.7025)(0.08,0.135,1.7044)(0.08,0.1359,1.7062)(0.08,0.1368,1.7078)(0.08,0.1377,1.7094)(0.08,0.1386,1.7108)(0.08,0.1395,1.7121)(0.08,0.1404,1.7132)(0.08,0.1413,1.7143)(0.08,0.1422,1.7152)(0.08,0.1431,1.7161)(0.08,0.144,1.7168)(0.08,0.1449,1.7174)(0.08,0.1458,1.718)(0.08,0.1467,1.7184)(0.08,0.1476,1.7187)(0.08,0.1485,1.7189)(0.08,0.1494,1.719)(0.08,0.1503,1.719)(0.08,0.1512,1.7189)(0.08,0.1521,1.7188)(0.08,0.153,1.7185)(0.08,0.1539,1.7181)(0.08,0.1548,1.7177)(0.08,0.1557,1.7171)(0.08,0.1566,1.7165)(0.08,0.1575,1.7158)(0.08,0.1584,1.715)(0.08,0.1593,1.7141)(0.08,0.1602,1.7132)(0.08,0.1611,1.7121)(0.08,0.162,1.711)(0.08,0.1629,1.7098)(0.08,0.1638,1.7085)(0.08,0.1647,1.6999)(0.08,0.1656,1.6864)(0.08,0.1665,1.6729)(0.08,0.1674,1.6595)(0.08,0.1683,1.6461)(0.08,0.1692,1.6327)(0.08,0.1701,1.6194)(0.08,0.171,1.6062)(0.08,0.1719,1.5929)(0.08,0.1728,1.5798)(0.08,0.1737,1.5666)(0.08,0.1746,1.5535)(0.08,0.1755,1.5405)(0.08,0.1764,1.5275)(0.08,0.1773,1.5146)(0.08,0.1782,1.5017)(0.08,0.1791,1.4889)(0.08,0.18,1.4761)
};

\addplot3 [dotted] coordinates{
(0.044601,0.09,1)(0.044601,0.09152,1)(0.073099,0.15,1)(0.073099,0.09,1)
};

\addplot3 [fill=red,fill opacity=0.4,draw opacity=0,line width=0pt] coordinates{(0.04,0.09,1)(0.044601,0.09,1)(0.044601,0.09152,1)(0.04,0.09152,1)};
\addplot3 [fill=yellow,fill opacity=0.4,draw opacity=0,line width=0pt] coordinates{(0.044601,0.09,1)(0.073099,0.09,1)(0.073099,0.15,1)(0.044601,0.09152,1)};
\addplot3 [fill=blue,fill opacity=0.4,draw opacity=0,line width=0pt] coordinates{(0.073099,0.09,1)(0.08,0.09,1)(0.08,0.15,1)(0.073099,0.15,1)};

\end{axis}\end{tikzpicture}
}
\caption{\simplecharging{} station revenue as a function of $T_r$ and $T_s$ ($t=0.03$, $\bar\theta=0.3$, $C_B=50kWh$, $x=0.8$). The red, yellow and blue areas are separated by $T_r=(t+(P_d-P_A)\frac{\bar\theta}{C_B})\frac{P_A}{P_d}$ and $T_r=(t+P_d\frac{\bar\theta}{C_B})\frac{P_A}{P_d}$, referring to~\eqref{eq:prop1}.}

%% file: Rr_surf_enveloped.tex
{\tiny\begin{tikzpicture}\begin{axis}[
view/h={-75},view/v={20},
width=\figwidth,height=\figheight,
xlabel=$T_s$(\euro/kWh),ylabel=$T_r$(\euro/kWh),zlabel=\simplecharging{} station revenue (\euro/EV),
legend style={at={(.3,1.003)},anchor=south west},
zmin=-0.1
]


\addplot3 [dashed,line width=2pt] coordinates{
(0,NaN,-0.1)(0.01,0.003092,-0.1)(0.02,0.0055223,-0.1)(0.03,0.0079431,-0.1)(0.04,0.01035,-0.1)(0.05,0.01274,-0.1)(0.06,0.015108,-0.1)(0.07,0.017449,-0.1)(0.08,0.019761,-0.1)(0.09,0.022037,-0.1)(0.1,0.024274,-0.1)(0.11,0.026468,-0.1)(0.12,0.028613,-0.1)(0.13,0.030706,-0.1)(0.14,0.032741,-0.1)(0.15,0.034714,-0.1)};
\addlegendentry{Best-response price $T_r^{br}(T_s)$};

\addplot3 [blue] coordinates{
(0,NaN,0)(0.01,0.003092,0.0048681)(0.02,0.0055223,0.025171)(0.03,0.0079431,0.057783)(0.04,0.01035,0.09951)(0.05,0.01274,0.14771)(0.06,0.015108,0.20023)(0.07,0.017449,0.25528)(0.08,0.019761,0.31144)(0.09,0.022037,0.36758)(0.1,0.024274,0.42279)(0.11,0.026468,0.47637)(0.12,0.028613,0.52782)(0.13,0.030706,0.57673)(0.14,0.032741,0.62287)(0.15,0.034714,0.66605)};
\addlegendentry{Revenue at best-response price $T_r^{br}(T_s)$};

\addplot3 [surf,opacity=0.5] coordinates{
(0,0,0)(0,0.001,0)(0,0.002,0)(0,0.003,0)(0,0.004,0)(0,0.005,0)(0,0.006,0)(0,0.007,0)(0,0.008,0)(0,0.009,0)(0,0.01,0)(0,0.011,0)(0,0.012,0)(0,0.013,0)(0,0.014,0)(0,0.015,0)(0,0.016,0)(0,0.017,0)(0,0.018,0)(0,0.019,0)(0,0.02,0)(0,0.021,0)(0,0.022,0)(0,0.023,0)(0,0.024,0)(0,0.025,0)(0,0.026,0)(0,0.027,0)(0,0.028,0)(0,0.029,0)(0,0.03,0)(0,0.031,0)(0,0.032,0)(0,0.033,0)(0,0.034,0)(0,0.035,0)(0,0.036,0)(0,0.037,0)(0,0.038,0)(0,0.039,0)(0,0.04,0)(0,0.041,0)(0,0.042,0)(0,0.043,0)(0,0.044,0)(0,0.045,0)(0,0.046,0)(0,0.047,0)(0,0.048,0)(0,0.049,0)(0,0.05,0)

(0.01,0,-0.0098453)(0.01,0.001,-0.0018616)(0.01,0.002,0.0030359)(0.01,0.003,0.0048551)(0.01,0.004,0.0036026)(0.01,0.005,0)(0.01,0.006,0)(0.01,0.007,0)(0.01,0.008,0)(0.01,0.009,0)(0.01,0.01,0)(0.01,0.011,0)(0.01,0.012,0)(0.01,0.013,0)(0.01,0.014,0)(0.01,0.015,0)(0.01,0.016,0)(0.01,0.017,0)(0.01,0.018,0)(0.01,0.019,0)(0.01,0.02,0)(0.01,0.021,0)(0.01,0.022,0)(0.01,0.023,0)(0.01,0.024,0)(0.01,0.025,0)(0.01,0.026,0)(0.01,0.027,0)(0.01,0.028,0)(0.01,0.029,0)(0.01,0.03,0)(0.01,0.031,0)(0.01,0.032,0)(0.01,0.033,0)(0.01,0.034,0)(0.01,0.035,0)(0.01,0.036,0)(0.01,0.037,0)(0.01,0.038,0)(0.01,0.039,0)(0.01,0.04,0)(0.01,0.041,0)(0.01,0.042,0)(0.01,0.043,0)(0.01,0.044,0)(0.01,0.045,0)(0.01,0.046,0)(0.01,0.047,0)(0.01,0.048,0)(0.01,0.049,0)(0.01,0.05,0)

(0.02,0,-0.018214)(0.02,0.001,-0.0038867)(0.02,0.002,0.0075641)(0.02,0.003,0.016152)(0.02,0.004,0.021889)(0.02,0.005,0.024785)(0.02,0.006,0.024848)(0.02,0.007,0.022086)(0.02,0.008,0.016504)(0.02,0.009,0.0081051)(0.02,0.01,0)(0.02,0.011,0)(0.02,0.012,0)(0.02,0.013,0)(0.02,0.014,0)(0.02,0.015,0)(0.02,0.016,0)(0.02,0.017,0)(0.02,0.018,0)(0.02,0.019,0)(0.02,0.02,0)(0.02,0.021,0)(0.02,0.022,0)(0.02,0.023,0)(0.02,0.024,0)(0.02,0.025,0)(0.02,0.026,0)(0.02,0.027,0)(0.02,0.028,0)(0.02,0.029,0)(0.02,0.03,0)(0.02,0.031,0)(0.02,0.032,0)(0.02,0.033,0)(0.02,0.034,0)(0.02,0.035,0)(0.02,0.036,0)(0.02,0.037,0)(0.02,0.038,0)(0.02,0.039,0)(0.02,0.04,0)(0.02,0.041,0)(0.02,0.042,0)(0.02,0.043,0)(0.02,0.044,0)(0.02,0.045,0)(0.02,0.046,0)(0.02,0.047,0)(0.02,0.048,0)(0.02,0.049,0)(0.02,0.05,0)

(0.03,0,-0.025327)(0.03,0.001,-0.005608)(0.03,0.002,0.011413)(0.03,0.003,0.025754)(0.03,0.004,0.037432)(0.03,0.005,0.04646)(0.03,0.006,0.052853)(0.03,0.007,0.056623)(0.03,0.008,0.057778)(0.03,0.009,0.056329)(0.03,0.01,0.052283)(0.03,0.011,0.045644)(0.03,0.012,0.036419)(0.03,0.013,0.02461)(0.03,0.014,0.010218)(0.03,0.015,0)(0.03,0.016,0)(0.03,0.017,0)(0.03,0.018,0)(0.03,0.019,0)(0.03,0.02,0)(0.03,0.021,0)(0.03,0.022,0)(0.03,0.023,0)(0.03,0.024,0)(0.03,0.025,0)(0.03,0.026,0)(0.03,0.027,0)(0.03,0.028,0)(0.03,0.029,0)(0.03,0.03,0)(0.03,0.031,0)(0.03,0.032,0)(0.03,0.033,0)(0.03,0.034,0)(0.03,0.035,0)(0.03,0.036,0)(0.03,0.037,0)(0.03,0.038,0)(0.03,0.039,0)(0.03,0.04,0)(0.03,0.041,0)(0.03,0.042,0)(0.03,0.043,0)(0.03,0.044,0)(0.03,0.045,0)(0.03,0.046,0)(0.03,0.047,0)(0.03,0.048,0)(0.03,0.049,0)(0.03,0.05,0)

(0.04,0,-0.031372)(0.04,0.001,-0.007071)(0.04,0.002,0.014684)(0.04,0.003,0.033916)(0.04,0.004,0.050643)(0.04,0.005,0.064884)(0.04,0.006,0.076657)(0.04,0.007,0.085978)(0.04,0.008,0.092861)(0.04,0.009,0.097318)(0.04,0.01,0.099363)(0.04,0.011,0.099004)(0.04,0.012,0.096252)(0.04,0.013,0.091113)(0.04,0.014,0.083595)(0.04,0.015,0.073701)(0.04,0.016,0.061435)(0.04,0.017,0.046801)(0.04,0.018,0.029798)(0.04,0.019,0.010426)(0.04,0.02,0)(0.04,0.021,0)(0.04,0.022,0)(0.04,0.023,0)(0.04,0.024,0)(0.04,0.025,0)(0.04,0.026,0)(0.04,0.027,0)(0.04,0.028,0)(0.04,0.029,0)(0.04,0.03,0)(0.04,0.031,0)(0.04,0.032,0)(0.04,0.033,0)(0.04,0.034,0)(0.04,0.035,0)(0.04,0.036,0)(0.04,0.037,0)(0.04,0.038,0)(0.04,0.039,0)(0.04,0.04,0)(0.04,0.041,0)(0.04,0.042,0)(0.04,0.043,0)(0.04,0.044,0)(0.04,0.045,0)(0.04,0.046,0)(0.04,0.047,0)(0.04,0.048,0)(0.04,0.049,0)(0.04,0.05,0)

(0.05,0,-0.036511)(0.05,0.001,-0.0083146)(0.05,0.002,0.017465)(0.05,0.003,0.040853)(0.05,0.004,0.061872)(0.05,0.005,0.080544)(0.05,0.006,0.09689)(0.05,0.007,0.11093)(0.05,0.008,0.12268)(0.05,0.009,0.13216)(0.05,0.01,0.13938)(0.05,0.011,0.14436)(0.05,0.012,0.14711)(0.05,0.013,0.14764)(0.05,0.014,0.14596)(0.05,0.015,0.14209)(0.05,0.016,0.13602)(0.05,0.017,0.12777)(0.05,0.018,0.11734)(0.05,0.019,0.10473)(0.05,0.02,0.089958)(0.05,0.021,0.07301)(0.05,0.022,0.053892)(0.05,0.023,0.032604)(0.05,0.024,0.0091425)(0.05,0.025,0)(0.05,0.026,0)(0.05,0.027,0)(0.05,0.028,0)(0.05,0.029,0)(0.05,0.03,0)(0.05,0.031,0)(0.05,0.032,0)(0.05,0.033,0)(0.05,0.034,0)(0.05,0.035,0)(0.05,0.036,0)(0.05,0.037,0)(0.05,0.038,0)(0.05,0.039,0)(0.05,0.04,0)(0.05,0.041,0)(0.05,0.042,0)(0.05,0.043,0)(0.05,0.044,0)(0.05,0.045,0)(0.05,0.046,0)(0.05,0.047,0)(0.05,0.048,0)(0.05,0.049,0)(0.05,0.05,0)

(0.06,0,-0.040879)(0.06,0.001,-0.0093716)(0.06,0.002,0.019829)(0.06,0.003,0.046749)(0.06,0.004,0.071416)(0.06,0.005,0.093854)(0.06,0.006,0.11409)(0.06,0.007,0.13214)(0.06,0.008,0.14803)(0.06,0.009,0.16177)(0.06,0.01,0.17339)(0.06,0.011,0.18291)(0.06,0.012,0.19034)(0.06,0.013,0.19569)(0.06,0.014,0.19897)(0.06,0.015,0.20021)(0.06,0.016,0.19942)(0.06,0.017,0.19659)(0.06,0.018,0.19175)(0.06,0.019,0.18489)(0.06,0.02,0.17604)(0.06,0.021,0.16518)(0.06,0.022,0.15233)(0.06,0.023,0.13749)(0.06,0.024,0.12067)(0.06,0.025,0.10185)(0.06,0.026,0.081054)(0.06,0.027,0.058266)(0.06,0.028,0.033488)(0.06,0.029,0.0067157)(0.06,0.03,0)(0.06,0.031,0)(0.06,0.032,0)(0.06,0.033,0)(0.06,0.034,0)(0.06,0.035,0)(0.06,0.036,0)(0.06,0.037,0)(0.06,0.038,0)(0.06,0.039,0)(0.06,0.04,0)(0.06,0.041,0)(0.06,0.042,0)(0.06,0.043,0)(0.06,0.044,0)(0.06,0.045,0)(0.06,0.046,0)(0.06,0.047,0)(0.06,0.048,0)(0.06,0.049,0)(0.06,0.05,0)

(0.07,0,-0.044591)(0.07,0.001,-0.01027)(0.07,0.002,0.021838)(0.07,0.003,0.051761)(0.07,0.004,0.079529)(0.07,0.005,0.10517)(0.07,0.006,0.1287)(0.07,0.007,0.15016)(0.07,0.008,0.16957)(0.07,0.009,0.18694)(0.07,0.01,0.2023)(0.07,0.011,0.21568)(0.07,0.012,0.22708)(0.07,0.013,0.23652)(0.07,0.014,0.24403)(0.07,0.015,0.24962)(0.07,0.016,0.2533)(0.07,0.017,0.25509)(0.07,0.018,0.25499)(0.07,0.019,0.25303)(0.07,0.02,0.2492)(0.07,0.021,0.24353)(0.07,0.022,0.23601)(0.07,0.023,0.22665)(0.07,0.024,0.21546)(0.07,0.025,0.20245)(0.07,0.026,0.18761)(0.07,0.027,0.17096)(0.07,0.028,0.15249)(0.07,0.029,0.1322)(0.07,0.03,0.11009)(0.07,0.031,0.086162)(0.07,0.032,0.060414)(0.07,0.033,0.032842)(0.07,0.034,0.0034394)(0.07,0.035,0)(0.07,0.036,0)(0.07,0.037,0)(0.07,0.038,0)(0.07,0.039,0)(0.07,0.04,0)(0.07,0.041,0)(0.07,0.042,0)(0.07,0.043,0)(0.07,0.044,0)(0.07,0.045,0)(0.07,0.046,0)(0.07,0.047,0)(0.07,0.048,0)(0.07,0.049,0)(0.07,0.05,0)

(0.08,0,-0.047747)(0.08,0.001,-0.011034)(0.08,0.002,0.023545)(0.08,0.003,0.056021)(0.08,0.004,0.086424)(0.08,0.005,0.11478)(0.08,0.006,0.14113)(0.08,0.007,0.16549)(0.08,0.008,0.18788)(0.08,0.009,0.20834)(0.08,0.01,0.22688)(0.08,0.011,0.24353)(0.08,0.012,0.25831)(0.08,0.013,0.27123)(0.08,0.014,0.28233)(0.08,0.015,0.29162)(0.08,0.016,0.2991)(0.08,0.017,0.30481)(0.08,0.018,0.30875)(0.08,0.019,0.31094)(0.08,0.02,0.31139)(0.08,0.021,0.31012)(0.08,0.022,0.30713)(0.08,0.023,0.30243)(0.08,0.024,0.29604)(0.08,0.025,0.28795)(0.08,0.026,0.27819)(0.08,0.027,0.26674)(0.08,0.028,0.25363)(0.08,0.029,0.23885)(0.08,0.03,0.22241)(0.08,0.031,0.2043)(0.08,0.032,0.18453)(0.08,0.033,0.16311)(0.08,0.034,0.14002)(0.08,0.035,0.11526)(0.08,0.036,0.088846)(0.08,0.037,0.060759)(0.08,0.038,0.030999)(0.08,0.039,0)(0.08,0.04,0)(0.08,0.041,0)(0.08,0.042,0)(0.08,0.043,0)(0.08,0.044,0)(0.08,0.045,0)(0.08,0.046,0)(0.08,0.047,0)(0.08,0.048,0)(0.08,0.049,0)(0.08,0.05,0)

(0.09,0,-0.050429)(0.09,0.001,-0.011683)(0.09,0.002,0.024996)(0.09,0.003,0.059642)(0.09,0.004,0.092285)(0.09,0.005,0.12296)(0.09,0.006,0.15169)(0.09,0.007,0.17851)(0.09,0.008,0.20344)(0.09,0.009,0.22652)(0.09,0.01,0.24776)(0.09,0.011,0.2672)(0.09,0.012,0.28485)(0.09,0.013,0.30074)(0.09,0.014,0.31488)(0.09,0.015,0.32731)(0.09,0.016,0.33803)(0.09,0.017,0.34707)(0.09,0.018,0.35444)(0.09,0.019,0.36017)(0.09,0.02,0.36425)(0.09,0.021,0.36672)(0.09,0.022,0.36758)(0.09,0.023,0.36684)(0.09,0.024,0.36452)(0.09,0.025,0.36063)(0.09,0.026,0.35517)(0.09,0.027,0.34816)(0.09,0.028,0.3396)(0.09,0.029,0.32951)(0.09,0.03,0.31787)(0.09,0.031,0.30472)(0.09,0.032,0.29003)(0.09,0.033,0.27383)(0.09,0.034,0.2561)(0.09,0.035,0.23686)(0.09,0.036,0.21611)(0.09,0.037,0.19383)(0.09,0.038,0.17004)(0.09,0.039,0.14474)(0.09,0.04,0.11791)(0.09,0.041,0.089549)(0.09,0.042,0.059663)(0.09,0.043,0.028243)(0.09,0.044,0)(0.09,0.045,0)(0.09,0.046,0)(0.09,0.047,0)(0.09,0.048,0)(0.09,0.049,0)(0.09,0.05,0)

(0.1,0,-0.052709)(0.1,0.001,-0.012234)(0.1,0.002,0.02623)(0.1,0.003,0.062719)(0.1,0.004,0.097267)(0.1,0.005,0.12991)(0.1,0.006,0.16067)(0.1,0.007,0.18958)(0.1,0.008,0.21667)(0.1,0.009,0.24198)(0.1,0.01,0.26552)(0.1,0.011,0.28732)(0.1,0.012,0.30741)(0.1,0.013,0.32582)(0.1,0.014,0.34255)(0.1,0.015,0.35765)(0.1,0.016,0.37112)(0.1,0.017,0.38299)(0.1,0.018,0.39328)(0.1,0.019,0.402)(0.1,0.02,0.40918)(0.1,0.021,0.41483)(0.1,0.022,0.41896)(0.1,0.023,0.42159)(0.1,0.024,0.42273)(0.1,0.025,0.4224)(0.1,0.026,0.42061)(0.1,0.027,0.41736)(0.1,0.028,0.41268)(0.1,0.029,0.40656)(0.1,0.03,0.39902)(0.1,0.031,0.39007)(0.1,0.032,0.3797)(0.1,0.033,0.36794)(0.1,0.034,0.35477)(0.1,0.035,0.34022)(0.1,0.036,0.32427)(0.1,0.037,0.30694)(0.1,0.038,0.28823)(0.1,0.039,0.26813)(0.1,0.04,0.24665)(0.1,0.041,0.22379)(0.1,0.042,0.19954)(0.1,0.043,0.17391)(0.1,0.044,0.14688)(0.1,0.045,0.11846)(0.1,0.046,0.088651)(0.1,0.047,0.057435)(0.1,0.048,0.02481)(0.1,0.049,0)(0.1,0.05,0)

(0.11,0,-0.054647)(0.11,0.001,-0.012703)(0.11,0.002,0.027279)(0.11,0.003,0.065335)(0.11,0.004,0.1015)(0.11,0.005,0.13581)(0.11,0.006,0.1683)(0.11,0.007,0.19899)(0.11,0.008,0.22792)(0.11,0.009,0.25511)(0.11,0.01,0.28061)(0.11,0.011,0.30442)(0.11,0.012,0.32659)(0.11,0.013,0.34713)(0.11,0.014,0.36607)(0.11,0.015,0.38344)(0.11,0.016,0.39925)(0.11,0.017,0.41353)(0.11,0.018,0.42629)(0.11,0.019,0.43757)(0.11,0.02,0.44737)(0.11,0.021,0.45572)(0.11,0.022,0.46263)(0.11,0.023,0.46812)(0.11,0.024,0.47221)(0.11,0.025,0.4749)(0.11,0.026,0.47622)(0.11,0.027,0.47618)(0.11,0.028,0.47479)(0.11,0.029,0.47205)(0.11,0.03,0.46799)(0.11,0.031,0.46261)(0.11,0.032,0.45592)(0.11,0.033,0.44793)(0.11,0.034,0.43864)(0.11,0.035,0.42807)(0.11,0.036,0.41622)(0.11,0.037,0.40309)(0.11,0.038,0.38868)(0.11,0.039,0.37302)(0.11,0.04,0.35608)(0.11,0.041,0.33789)(0.11,0.042,0.31843)(0.11,0.043,0.29771)(0.11,0.044,0.27574)(0.11,0.045,0.2525)(0.11,0.046,0.228)(0.11,0.047,0.20224)(0.11,0.048,0.17521)(0.11,0.049,0.14691)(0.11,0.05,0.11733)

(0.12,0,-0.056294)(0.12,0.001,-0.013102)(0.12,0.002,0.02817)(0.12,0.003,0.067559)(0.12,0.004,0.1051)(0.12,0.005,0.14083)(0.12,0.006,0.17478)(0.12,0.007,0.20698)(0.12,0.008,0.23747)(0.12,0.009,0.26628)(0.12,0.01,0.29343)(0.12,0.011,0.31896)(0.12,0.012,0.34289)(0.12,0.013,0.36525)(0.12,0.014,0.38606)(0.12,0.015,0.40536)(0.12,0.016,0.42315)(0.12,0.017,0.43948)(0.12,0.018,0.45435)(0.12,0.019,0.46779)(0.12,0.02,0.47983)(0.12,0.021,0.49048)(0.12,0.022,0.49975)(0.12,0.023,0.50767)(0.12,0.024,0.51426)(0.12,0.025,0.51953)(0.12,0.026,0.5235)(0.12,0.027,0.52618)(0.12,0.028,0.52758)(0.12,0.029,0.52772)(0.12,0.03,0.52662)(0.12,0.031,0.52427)(0.12,0.032,0.5207)(0.12,0.033,0.51592)(0.12,0.034,0.50993)(0.12,0.035,0.50274)(0.12,0.036,0.49436)(0.12,0.037,0.4848)(0.12,0.038,0.47407)(0.12,0.039,0.46216)(0.12,0.04,0.4491)(0.12,0.041,0.43487)(0.12,0.042,0.41949)(0.12,0.043,0.40295)(0.12,0.044,0.38526)(0.12,0.045,0.36643)(0.12,0.046,0.34645)(0.12,0.047,0.32532)(0.12,0.048,0.30305)(0.12,0.049,0.27962)(0.12,0.05,0.25505)

(0.13,0,-0.057694)(0.13,0.001,-0.013441)(0.13,0.002,0.028927)(0.13,0.003,0.069448)(0.13,0.004,0.10816)(0.13,0.005,0.1451)(0.13,0.006,0.18029)(0.13,0.007,0.21378)(0.13,0.008,0.2456)(0.13,0.009,0.27577)(0.13,0.01,0.30434)(0.13,0.011,0.33132)(0.13,0.012,0.35674)(0.13,0.013,0.38065)(0.13,0.014,0.40305)(0.13,0.015,0.42399)(0.13,0.016,0.44347)(0.13,0.017,0.46154)(0.13,0.018,0.4782)(0.13,0.019,0.49349)(0.13,0.02,0.50742)(0.13,0.021,0.52002)(0.13,0.022,0.5313)(0.13,0.023,0.54129)(0.13,0.024,0.55001)(0.13,0.025,0.55746)(0.13,0.026,0.56368)(0.13,0.027,0.56867)(0.13,0.028,0.57245)(0.13,0.029,0.57504)(0.13,0.03,0.57644)(0.13,0.031,0.57668)(0.13,0.032,0.57577)(0.13,0.033,0.57371)(0.13,0.034,0.57052)(0.13,0.035,0.56621)(0.13,0.036,0.56079)(0.13,0.037,0.55426)(0.13,0.038,0.54664)(0.13,0.039,0.53794)(0.13,0.04,0.52816)(0.13,0.041,0.5173)(0.13,0.042,0.50538)(0.13,0.043,0.4924)(0.13,0.044,0.47836)(0.13,0.045,0.46327)(0.13,0.046,0.44712)(0.13,0.047,0.42994)(0.13,0.048,0.4117)(0.13,0.049,0.39243)(0.13,0.05,0.37211)

(0.14,0,-0.058883)(0.14,0.001,-0.013729)(0.14,0.002,0.029571)(0.14,0.003,0.071055)(0.14,0.004,0.11076)(0.14,0.005,0.14872)(0.14,0.006,0.18498)(0.14,0.007,0.21956)(0.14,0.008,0.2525)(0.14,0.009,0.28384)(0.14,0.01,0.3136)(0.14,0.011,0.34182)(0.14,0.012,0.36852)(0.14,0.013,0.39374)(0.14,0.014,0.41749)(0.14,0.015,0.43982)(0.14,0.016,0.46074)(0.14,0.017,0.48028)(0.14,0.018,0.49847)(0.14,0.019,0.51532)(0.14,0.02,0.53087)(0.14,0.021,0.54513)(0.14,0.022,0.55812)(0.14,0.023,0.56987)(0.14,0.024,0.58039)(0.14,0.025,0.58971)(0.14,0.026,0.59783)(0.14,0.027,0.60479)(0.14,0.028,0.61059)(0.14,0.029,0.61526)(0.14,0.03,0.6188)(0.14,0.031,0.62123)(0.14,0.032,0.62257)(0.14,0.033,0.62283)(0.14,0.034,0.62202)(0.14,0.035,0.62015)(0.14,0.036,0.61724)(0.14,0.037,0.6133)(0.14,0.038,0.60833)(0.14,0.039,0.60235)(0.14,0.04,0.59535)(0.14,0.041,0.58737)(0.14,0.042,0.57839)(0.14,0.043,0.56843)(0.14,0.044,0.55749)(0.14,0.045,0.54557)(0.14,0.046,0.5327)(0.14,0.047,0.51886)(0.14,0.048,0.50406)(0.14,0.049,0.48831)(0.14,0.05,0.4716)

(0.15,0,-0.059895)(0.15,0.001,-0.013973)(0.15,0.002,0.030119)(0.15,0.003,0.07242)(0.15,0.004,0.11297)(0.15,0.005,0.1518)(0.15,0.006,0.18896)(0.15,0.007,0.22447)(0.15,0.008,0.25837)(0.15,0.009,0.2907)(0.15,0.01,0.32148)(0.15,0.011,0.35075)(0.15,0.012,0.37853)(0.15,0.013,0.40486)(0.15,0.014,0.42977)(0.15,0.015,0.45328)(0.15,0.016,0.47542)(0.15,0.017,0.49622)(0.15,0.018,0.5157)(0.15,0.019,0.53389)(0.15,0.02,0.5508)(0.15,0.021,0.56647)(0.15,0.022,0.58092)(0.15,0.023,0.59416)(0.15,0.024,0.60622)(0.15,0.025,0.61711)(0.15,0.026,0.62686)(0.15,0.027,0.63549)(0.15,0.028,0.64301)(0.15,0.029,0.64944)(0.15,0.03,0.6548)(0.15,0.031,0.6591)(0.15,0.032,0.66235)(0.15,0.033,0.66458)(0.15,0.034,0.66579)(0.15,0.035,0.66601)(0.15,0.036,0.66523)(0.15,0.037,0.66348)(0.15,0.038,0.66076)(0.15,0.039,0.65709)(0.15,0.04,0.65247)(0.15,0.041,0.64692)(0.15,0.042,0.64044)(0.15,0.043,0.63305)(0.15,0.044,0.62474)(0.15,0.045,0.61553)(0.15,0.046,0.60543)(0.15,0.047,0.59444)(0.15,0.048,0.58256)(0.15,0.049,0.5698)(0.15,0.05,0.55617)
};

\addplot3 [fill=red,fill opacity=0.4,draw opacity=0,line width=0pt] coordinates{
(0,0,-0.1)(0.15,0,-0.1)(0.15,0.05,-0.1)(0.1026,0.05,-0.1)
};

\end{axis}\end{tikzpicture}}
\caption{\regcharging{} Station revenue as a function of $T_r$ and $T_s$ ($t=0.03$, $r_d=0.7$, $r_u=2.1$, $C_B=50kWh$, $\rho_d=0.48$, $\rho_u=0.48$, $\gamma=0.05$, $\bar\theta=0.3$, $x=0.8$). Red region corresponds to non-negative revenue, i.e. $\frac{T_r}{P_A}<\frac{T_s}{P_d}$}

%% file: FourNash.tex
\centering
\gdef\figheight{5cm}
\gdef\figwidth{6.5cm}
\begin{tabular}{cc}
	\input{FourCases_case1and2.tex} &  \input{FourCases_4.tex}
\end{tabular}

%% file: FourCases_case1and2.tex
{\tiny\begin{tikzpicture}
\begin{axis}[width=\figwidth,height=\figheight,xlabel=$T_s(\mbox{resp.}T_s^{br})$ in \euro/kWh,ylabel=$T_r^{br}(\mbox{resp.}T_r)$ in \euro/kWh,
legend style={at={(0.5,0.03)},anchor=south},
legend columns = 2,
cycle list name=\mylist]

\addplot coordinates{
(0.04,0.01902)(0.041,0.019496)(0.042,0.019971)(0.043,0.020447)(0.044,0.020922)(0.045,0.021398)(0.046,0.021873)(0.047,0.022349)(0.048,0.022824)(0.049,0.0233)(0.05,0.023775)(0.051,0.024251)(0.052,0.024726)(0.053,0.025202)(0.054,0.025677)(0.055,0.026153)(0.056,0.026628)(0.057,0.027104)(0.058,0.027579)(0.059,0.028055)(0.06,0.02853)(0.061,0.029006)(0.062,0.029481)};

\addlegendentry{$T_r^{br}(\bar\theta=0.1)$};
\addplot coordinates{
(0.05098,0.01)(0.05098,0.011)(0.05098,0.012)(0.05098,0.013)(0.05098,0.014)(0.05098,0.015)(0.05098,0.016)(0.05098,0.017)(0.05098,0.018)(0.05098,0.019)(0.05098,0.02)(0.05098,0.021)(0.05098,0.022)(0.05098,0.023)(0.05098,0.024)(0.052576,0.025)(0.054679,0.026)(0.056782,0.027)(0.058885,0.028)(0.060988,0.029)(0.063091,0.03)(0.065194,0.031)(0.067297,0.032)(0.0694,0.033)(0.07,0.034)(0.07,0.035)(0.07,0.036)(0.07,0.037)(0.07,0.038)(0.07,0.039)(0.07,0.04)(0.07,0.041)(0.07,0.042)(0.07,0.043)(0.07,0.044)};
\addlegendentry{$T_s^{br}(\bar\theta=0.1)$};
\addplot coordinates{};

\addplot coordinates{
(0.04,0.01902)(0.041,0.019496)(0.042,0.019971)(0.043,0.020447)(0.044,0.020922)(0.045,0.021398)(0.046,0.021873)(0.047,0.022349)(0.048,0.022824)(0.049,0.0233)(0.05,0.023775)(0.051,0.024251)(0.052,0.024726)(0.053,0.025202)(0.054,0.025677)(0.055,0.026153)(0.056,0.026628)(0.057,0.027104)(0.058,0.027579)(0.059,0.028055)(0.06,0.02853)(0.061,0.029006)(0.062,0.029481)};

\addlegendentry{$T_r^{br}(\bar\theta=0.2)$};
\addplot coordinates{
(0.07196,0.01)(0.07196,0.011)(0.07196,0.012)(0.07196,0.013)(0.07196,0.014)(0.07196,0.015)(0.07196,0.016)(0.07196,0.017)(0.07196,0.018)(0.07196,0.019)(0.07196,0.02)(0.07196,0.021)(0.07196,0.022)(0.07196,0.023)(0.07196,0.024)(0.07196,0.025)(0.07196,0.026)(0.07196,0.027)(0.07196,0.028)(0.07196,0.029)(0.07196,0.03)(0.07196,0.031)(0.07196,0.032)(0.07196,0.033)(0.07196,0.034)(0.073606,0.035)(0.075709,0.036)(0.077812,0.037)(0.079915,0.038)(0.082018,0.039)(0.084121,0.04)(0.086224,0.041)(0.088327,0.042)(0.09043,0.043)(0.092533,0.044)};
\addlegendentry{$T_s^{br}(\bar\theta=0.2)$};
\addplot coordinates{};

\addplot coordinates{
(0.064,0.030216)(0.066,0.030691)(0.068,0.031165)(0.07,0.031638)(0.072,0.032109)(0.074,0.032578)(0.076,0.033045)(0.078,0.03351)(0.08,0.033973)(0.082,0.034432)(0.084,0.034889)(0.086,0.035342)(0.088,0.035791)(0.09,0.036236)(0.092,0.036677)(0.094,0.037113)(0.096,0.037544)(0.098,0.03797)(0.1,0.03839)(0.102,0.038805)(0.104,0.039213)(0.106,0.039615)(0.108,0.04001)(0.11,0.040398)(0.112,0.040778)(0.114,0.041151)(0.116,0.041516)(0.118,0.041872)(0.12,0.04222)};

\draw (axis cs:0.063090805624839,0.025) node (N1) [anchor=north]{$N^E~\eqref{eq:N1}$};
\draw[->](N1)--(axis cs:0.063090805624839,0.03);
\addplot [only marks, mark=square,black] coordinates{(0.063090805624839,0.03)};
\draw (axis cs:0.081959591794227,0.032104864530290) node (N2) [anchor=west]{$N^E~\eqref{eq:N2}$};
\draw[->](N2)--(axis cs:0.071959591794227,0.032104864530290);
\addplot [only marks, mark=triangle,black]coordinates {(0.071959591794227,0.032104864530290)};
\draw (axis cs: .11,.02) node[rectangle,rounded corners=4pt,fill=gray,fill opacity=.4,text opacity=1]{$r_u=1,~r_d=0$};

\addplot coordinates{(0.064,0.030216)(0.066,0.030691)(0.068,0.031166)(0.07,0.031641)(0.072,0.032114)(0.074,0.032588)(0.076,0.03306)(0.078,0.033532)(0.08,0.034003)(0.082,0.034474)(0.084,0.034943)(0.086,0.035412)(0.088,0.035879)(0.09,0.036345)(0.092,0.03681)(0.094,0.037274)(0.096,0.037736)(0.098,0.038197)(0.1,0.038657)(0.102,0.039114)(0.104,0.039571)(0.106,0.040025)(0.108,0.040478)(0.11,0.040929)(0.112,0.041378)(0.114,0.041825)(0.116,0.04227)(0.118,0.042713)(0.12,0.043154)};

\end{axis}\end{tikzpicture}}

%% file: FourCases_4.tex
{\tiny\begin{tikzpicture}
\begin{axis}[width=\figwidth,height=\figheight,xlabel=$T_s(\mbox{resp.}T_s^{br})$ in \euro/kWh,ylabel=$T_r^{br}(\mbox{resp.}T_r)$ in \euro/kWh,
legend style={at={(0.03,0.97)},anchor=north west},
cycle list name=\mylist]

\addplot coordinates{
(0.04,0)(0.041,0)(0.042,0)(0.043,0)(0.044,0)(0.045,0)(0.046,0)(0.047,0)(0.048,0)(0.049,0)(0.05,0)(0.051,0)(0.052,0)(0.053,0)(0.054,0)(0.055,0)(0.056,0)(0.057,0)(0.058,0)(0.059,0)(0.06,0)(0.061,0)(0.062,0)(0.063,0)(0.064,0)(0.065,0)(0.066,0)(0.067,0)(0.068,0)(0.069,0)(0.07,0)(0.071,0)(0.072,0)(0.073,0)(0.074,0)(0.075,0)(0.076,0)(0.077,0)(0.078,0)(0.079,0)(0.08,0)(0.081,0)(0.082,0)(0.083,0)(0.084,0)(0.085,0)(0.086,0)(0.087,0)(0.088,0)(0.089,0)(0.09,0)(0.091,0)(0.092,0)(0.093,0)(0.094,0)(0.095,0)(0.096,0)(0.097,0)(0.098,0)(0.099,0)(0.1,0)(0.101,0)(0.102,0)(0.103,0)(0.104,0.00012058)(0.105,0.00028016)(0.106,0.00043896)(0.107,0.00059695)(0.108,0.00075416)(0.109,0.00091056)(0.11,0.0010662)(0.111,0.001221)(0.112,0.001375)(0.113,0.0015281)(0.114,0.0016805)(0.115,0.0018321)(0.116,0.0019828)(0.117,0.0021327)(0.118,0.0022818)(0.119,0.0024301)(0.12,0.0025775)};
\addlegendentry{$T_r( r_d = 0.8; r_u = 5)$};
\addplot coordinates{
(0.092939,-0.05)(0.092939,-0.049)(0.092939,-0.048)(0.092939,-0.047)(0.092939,-0.046)(0.092939,-0.045)(0.092939,-0.044)(0.092939,-0.043)(0.092939,-0.042)(0.092939,-0.041)(0.092939,-0.04)(0.092939,-0.039)(0.092939,-0.038)(0.092939,-0.037)(0.092939,-0.036)(0.092939,-0.035)(0.092939,-0.034)(0.092939,-0.033)(0.092939,-0.032)(0.092939,-0.031)(0.092939,-0.03)(0.092939,-0.029)(0.092939,-0.028)(0.092939,-0.027)(0.092939,-0.026)(0.092939,-0.025)(0.092939,-0.024)(0.092939,-0.023)(0.092939,-0.022)(0.092939,-0.021)(0.092939,-0.02)(0.092939,-0.019)(0.092939,-0.018)(0.092939,-0.017)(0.092939,-0.016)(0.092939,-0.015)(0.092939,-0.014)(0.092939,-0.013)(0.092939,-0.012)(0.092939,-0.011)(0.092939,-0.01)(0.092939,-0.009)(0.092939,-0.008)(0.092939,-0.007)(0.092939,-0.006)(0.092939,-0.005)(0.092939,-0.004)(0.092939,-0.003)(0.092939,-0.002)(0.092939,-0.001)(0.092939,0)(0.092939,0.001)(0.092939,0.002)(0.092939,0.003)(0.092939,0.004)(0.092939,0.005)(0.092939,0.006)(0.092939,0.007)(0.092939,0.008)(0.092939,0.009)(0.092939,0.01)(0.092939,0.011)(0.092939,0.012)(0.092939,0.013)(0.092939,0.014)(0.092939,0.015)(0.092939,0.016)(0.092939,0.017)(0.092939,0.018)(0.092939,0.019)(0.092939,0.02)(0.092939,0.021)(0.092939,0.022)(0.092939,0.023)(0.092939,0.024)(0.092939,0.025)(0.092939,0.026)(0.092939,0.027)(0.092939,0.028)(0.092939,0.029)(0.092939,0.03)(0.092939,0.031)(0.092939,0.032)(0.092939,0.033)(0.092939,0.034)(0.092939,0.035)(0.092939,0.036)(0.092939,0.037)(0.092939,0.038)(0.092939,0.039)(0.092939,0.04)(0.092939,0.041)(0.092939,0.042)(0.092939,0.043)(0.092939,0.044)(0.094636,0.045)(0.096739,0.046)(0.098842,0.047)(0.10095,0.048)(0.10305,0.049)(0.10515,0.05)};
\addlegendentry{$T_s( r_d = 0.8; r_u = 5)$};
\addplot coordinates{};

\addplot coordinates{
(0.04,-0.051493)(0.041,-0.050682)(0.042,-0.049871)(0.043,-0.049059)(0.044,-0.048247)(0.045,-0.047434)(0.046,-0.046621)(0.047,-0.045808)(0.048,-0.044994)(0.049,-0.044179)(0.05,-0.043364)(0.051,-0.042549)(0.052,-0.041733)(0.053,-0.040917)(0.054,-0.0401)(0.055,-0.039283)(0.056,-0.038466)(0.057,-0.037648)(0.058,-0.03683)(0.059,-0.036011)(0.06,-0.035192)(0.061,-0.034372)(0.062,-0.033552)(0.063,-0.032731)(0.064,-0.031911)(0.065,-0.031089)(0.066,-0.030268)(0.067,-0.029445)(0.068,-0.028623)(0.069,-0.0278)(0.07,-0.026976)(0.071,-0.026153)(0.072,-0.025329)(0.073,-0.024504)(0.074,-0.023679)(0.075,-0.022854)(0.076,-0.022028)(0.077,-0.021202)(0.078,-0.020375)(0.079,-0.019548)(0.08,-0.018721)(0.081,-0.017893)(0.082,-0.017065)(0.083,-0.016236)(0.084,-0.015407)(0.085,-0.014578)(0.086,-0.013748)(0.087,-0.012918)(0.088,-0.012088)(0.089,-0.011257)(0.09,-0.010426)(0.091,-0.009594)(0.092,-0.008762)(0.093,-0.0079297)(0.094,-0.007097)(0.095,-0.0062639)(0.096,-0.0054305)(0.097,-0.0045967)(0.098,-0.0037625)(0.099,-0.002928)(0.1,-0.0020931)(0.101,-0.0012578)(0.102,-0.00042217)(0.103,0)(0.104,0)(0.105,0)(0.106,0)(0.107,0)(0.108,0)(0.109,0)(0.11,0)(0.111,0)(0.112,0)(0.113,0)(0.114,0)(0.115,0)(0.116,0)(0.117,0)(0.118,0)(0.119,0)(0.12,0)};
\addlegendentry{$T_r( r_d = 0.99; r_u = 20)$};
\addplot coordinates{
(0.092939,-0.05)(0.092939,-0.049)(0.092939,-0.048)(0.092939,-0.047)(0.092939,-0.046)(0.092939,-0.045)(0.092939,-0.044)(0.092939,-0.043)(0.092939,-0.042)(0.092939,-0.041)(0.092939,-0.04)(0.092939,-0.039)(0.092939,-0.038)(0.092939,-0.037)(0.092939,-0.036)(0.092939,-0.035)(0.092939,-0.034)(0.092939,-0.033)(0.092939,-0.032)(0.092939,-0.031)(0.092939,-0.03)(0.092939,-0.029)(0.092939,-0.028)(0.092939,-0.027)(0.092939,-0.026)(0.092939,-0.025)(0.092939,-0.024)(0.092939,-0.023)(0.092939,-0.022)(0.092939,-0.021)(0.092939,-0.02)(0.092939,-0.019)(0.092939,-0.018)(0.092939,-0.017)(0.092939,-0.016)(0.092939,-0.015)(0.092939,-0.014)(0.092939,-0.013)(0.092939,-0.012)(0.092939,-0.011)(0.092939,-0.01)(0.092939,-0.009)(0.092939,-0.008)(0.092939,-0.007)(0.092939,-0.006)(0.092939,-0.005)(0.092939,-0.004)(0.092939,-0.003)(0.092939,-0.002)(0.092939,-0.001)(0.092939,0)(0.092939,0.001)(0.092939,0.002)(0.092939,0.003)(0.092939,0.004)(0.092939,0.005)(0.092939,0.006)(0.092939,0.007)(0.092939,0.008)(0.092939,0.009)(0.092939,0.01)(0.092939,0.011)(0.092939,0.012)(0.092939,0.013)(0.092939,0.014)(0.092939,0.015)(0.092939,0.016)(0.092939,0.017)(0.092939,0.018)(0.092939,0.019)(0.092939,0.02)(0.092939,0.021)(0.092939,0.022)(0.092939,0.023)(0.092939,0.024)(0.092939,0.025)(0.092939,0.026)(0.092939,0.027)(0.092939,0.028)(0.092939,0.029)(0.092939,0.03)(0.092939,0.031)(0.092939,0.032)(0.092939,0.033)(0.092939,0.034)(0.092939,0.035)(0.092939,0.036)(0.092939,0.037)(0.092939,0.038)(0.092939,0.039)(0.092939,0.04)(0.092939,0.041)(0.092939,0.042)(0.092939,0.043)(0.092939,0.044)(0.094636,0.045)(0.096739,0.046)(0.098842,0.047)(0.10095,0.048)(0.10305,0.049)(0.10515,0.05)};
\addlegendentry{$T_s( r_d = 0.99; r_u = 20)$};

\draw (axis cs:0.1,0.01) node (N3) [anchor= west]{$N^E~\eqref{eq:N3}$};
\draw[->](N3)--(axis cs:0.092939387691340,0);
\addplot [only marks, mark=triangle,black]coordinates {(0.092939387691340,0)};

\draw (axis cs:0.1,-0.007980174342918) node (N4) [anchor=west]{$N^E~\eqref{eq:N4}$};
\draw[->](N4)--(axis cs:0.092939387691340,-0.007980174342918);
\addplot [only marks, mark=square,black] coordinates{(0.092939387691340,-0.007980174342918)};
\draw (axis cs: .115,-0.025) node[rectangle,rounded corners=4pt,fill=gray,fill opacity=.4,text opacity=1]{$\bar\theta=0.3$};

\end{axis}\end{tikzpicture}}

%% file: CompareRevenue.tex
\centering
\gdef\figheight{4cm}
\gdef\figwidth{6cm}
\begin{tabular}{ccc}

	
	{\tiny\begin{tikzpicture}
\begin{axis}[width=\figwidth,height=\figheight,xlabel=$\bar\theta$,ylabel=Revenue/Utility (\euro/EV),
legend style={at={(0.03,1)},anchor=north west},
legend entries = {$R_s^M$,$R_r^M$,$U^M$},
cycle list name=\mylist]

\addplot coordinates{
(0.1,0.28181)(0.11,0.33825)(0.12,0.39683)(0.13,0.45717)(0.14,0.519)(0.15,0.58207)(0.16,0.6462)(0.17,0.71125)(0.18,0.77708)(0.19,0.84359)(0.2,0.9107)(0.21,0.97834)(0.22,1.0464)(0.23,1.1149)(0.24,1.1838)(0.25,1.253)(0.26,1.3225)(0.27,1.3923)(0.28,1.4623)(0.29,1.5326)(0.3,1.603)(0.31,1.6736)(0.32,1.7444)(0.33,1.8153)(0.34,1.8864)(0.35,1.9576)(0.36,2.0289)(0.37,2.1004)(0.38,2.1719)(0.39,2.2435)(0.4,2.3153)(0.41,2.3871)(0.42,2.459)(0.43,2.5309)(0.44,2.6029)(0.45,2.675)(0.46,2.7472)(0.47,2.8194)(0.48,2.8917)(0.49,2.964)(0.5,3.0363)};
\addplot coordinates{
(0.1,0.35538)(0.11,0.4169)(0.12,0.47997)(0.13,0.54434)(0.14,0.60976)(0.15,0.67608)(0.16,0.74315)(0.17,0.81086)(0.18,0.87913)(0.19,0.94788)(0.2,1.017)(0.21,1.0866)(0.22,1.1564)(0.23,1.2265)(0.24,1.2969)(0.25,1.3676)(0.26,1.4384)(0.27,1.5094)(0.28,1.5805)(0.29,1.6519)(0.3,1.7233)(0.31,1.7949)(0.32,1.8666)(0.33,1.9384)(0.34,2.0103)(0.35,2.0823)(0.36,2.1543)(0.37,2.2265)(0.38,2.2987)(0.39,2.3709)(0.4,2.4433)(0.41,2.5157)(0.42,2.5881)(0.43,2.6606)(0.44,2.7332)(0.45,2.8057)(0.46,2.8784)(0.47,2.951)(0.48,3.0237)(0.49,3.0964)(0.5,3.1692)};
\addplot coordinates{
(0.1,0.71076)(0.11,0.83379)(0.12,0.95995)(0.13,1.0887)(0.14,1.2195)(0.15,1.3522)(0.16,1.4863)(0.17,1.6217)(0.18,1.7583)(0.19,1.8958)(0.2,2.0341)(0.21,2.1731)(0.22,2.3128)(0.23,2.4531)(0.24,2.5939)(0.25,2.7351)(0.26,2.8767)(0.27,3.0187)(0.28,3.1611)(0.29,3.3037)(0.3,3.4466)(0.31,3.5898)(0.32,3.7332)(0.33,3.8768)(0.34,4.0206)(0.35,4.1645)(0.36,4.3087)(0.37,4.4529)(0.38,4.5973)(0.39,4.7419)(0.4,4.8866)(0.41,5.0313)(0.42,5.1762)(0.43,5.3212)(0.44,5.4663)(0.45,5.6115)(0.46,5.7567)(0.47,5.902)(0.48,6.0474)(0.49,6.1929)(0.5,6.3384)};

\addplot[fill=red,fill opacity=0.8,draw opacity=0,line width=0pt] coordinates{(0.1,0)
(0.1,0.28181)(0.11,0.33825)(0.12,0.39683)(0.13,0.45717)(0.14,0.519)(0.15,0.58207)(0.16,0.6462)(0.17,0.71125)(0.18,0.77708)(0.19,0.84359)(0.2,0.9107)(0.21,0.97834)(0.22,1.0464)(0.23,1.1149)(0.24,1.1838)(0.25,1.253)(0.26,1.3225)(0.27,1.3923)(0.28,1.4623)(0.29,1.5326)(0.3,1.603)(0.31,1.6736)(0.32,1.7444)(0.33,1.8153)(0.34,1.8864)(0.35,1.9576)(0.36,2.0289)(0.37,2.1004)(0.38,2.1719)(0.39,2.2435)(0.4,2.3153)(0.41,2.3871)(0.42,2.459)(0.43,2.5309)(0.44,2.6029)(0.45,2.675)(0.46,2.7472)(0.47,2.8194)(0.48,2.8917)(0.49,2.964)(0.5,3.0363)
(0.5,0)};

\addplot[fill=blue,fill opacity=0.8,draw opacity=0,line width=0pt] coordinates{
(0.1,0.35538)(0.11,0.4169)(0.12,0.47997)(0.13,0.54434)(0.14,0.60976)(0.15,0.67608)(0.16,0.74315)(0.17,0.81086)(0.18,0.87913)(0.19,0.94788)(0.2,1.017)(0.21,1.0866)(0.22,1.1564)(0.23,1.2265)(0.24,1.2969)(0.25,1.3676)(0.26,1.4384)(0.27,1.5094)(0.28,1.5805)(0.29,1.6519)(0.3,1.7233)(0.31,1.7949)(0.32,1.8666)(0.33,1.9384)(0.34,2.0103)(0.35,2.0823)(0.36,2.1543)(0.37,2.2265)(0.38,2.2987)(0.39,2.3709)(0.4,2.4433)(0.41,2.5157)(0.42,2.5881)(0.43,2.6606)(0.44,2.7332)(0.45,2.8057)(0.46,2.8784)(0.47,2.951)(0.48,3.0237)(0.49,3.0964)(0.5,3.1692)
(0.5,3.0363)(0.49,2.964)(0.48,2.8917)(0.47,2.8194)(0.46,2.7472)(0.45,2.675)(0.44,2.6029)(0.43,2.5309)(0.42,2.459)(0.41,2.3871)(0.4,2.3153)(0.39,2.2435)(0.38,2.1719)(0.37,2.1004)(0.36,2.0289)(0.35,1.9576)(0.34,1.8864)(0.33,1.8153)(0.32,1.7444)(0.31,1.6736)(0.3,1.603)(0.29,1.5326)(0.28,1.4623)(0.27,1.3923)(0.26,1.3225)(0.25,1.253)(0.24,1.1838)(0.23,1.1149)(0.22,1.0464)(0.21,0.97834)(0.2,0.9107)(0.19,0.84359)(0.18,0.77708)(0.17,0.71125)(0.16,0.6462)(0.15,0.58207)(0.14,0.519)(0.13,0.45717)(0.12,0.39683)(0.11,0.33825)(0.1,0.28181)
};

\addplot[fill=yellow,fill opacity=0.8,draw opacity=0,line width=0pt] coordinates{
(0.1,0.71076)(0.11,0.83379)(0.12,0.95995)(0.13,1.0887)(0.14,1.2195)(0.15,1.3522)(0.16,1.4863)(0.17,1.6217)(0.18,1.7583)(0.19,1.8958)(0.2,2.0341)(0.21,2.1731)(0.22,2.3128)(0.23,2.4531)(0.24,2.5939)(0.25,2.7351)(0.26,2.8767)(0.27,3.0187)(0.28,3.1611)(0.29,3.3037)(0.3,3.4466)(0.31,3.5898)(0.32,3.7332)(0.33,3.8768)(0.34,4.0206)(0.35,4.1645)(0.36,4.3087)(0.37,4.4529)(0.38,4.5973)(0.39,4.7419)(0.4,4.8866)(0.41,5.0313)(0.42,5.1762)(0.43,5.3212)(0.44,5.4663)(0.45,5.6115)(0.46,5.7567)(0.47,5.902)(0.48,6.0474)(0.49,6.1929)(0.5,6.3384)
(0.5,3.1692)(0.49,3.0964)(0.48,3.0237)(0.47,2.951)(0.46,2.8784)(0.45,2.8057)(0.44,2.7332)(0.43,2.6606)(0.42,2.5881)(0.41,2.5157)(0.4,2.4433)(0.39,2.3709)(0.38,2.2987)(0.37,2.2265)(0.36,2.1543)(0.35,2.0823)(0.34,2.0103)(0.33,1.9384)(0.32,1.8666)(0.31,1.7949)(0.3,1.7233)(0.29,1.6519)(0.28,1.5805)(0.27,1.5094)(0.26,1.4384)(0.25,1.3676)(0.24,1.2969)(0.23,1.2265)(0.22,1.1564)(0.21,1.0866)(0.2,1.017)(0.19,0.94788)(0.18,0.87913)(0.17,0.81086)(0.16,0.74315)(0.15,0.67608)(0.14,0.60976)(0.13,0.54434)(0.12,0.47997)(0.11,0.4169)(0.1,0.35538)};
\end{axis}
\end{tikzpicture}}

	&
	{\tiny\begin{tikzpicture}\begin{axis}[width=\figwidth,height=\figheight,xlabel=$\bar\theta$,ylabel=Participation probability,
legend style={at={(0.5,1)},anchor=north west},
legend entries = {$\alpha_s^M$,$\alpha_r^M$},
cycle list name=\mylist]

\addplot coordinates{
(0.1,0.1409)(0.11,0.15375)(0.12,0.16534)(0.13,0.17583)(0.14,0.18536)(0.15,0.19402)(0.16,0.20194)(0.17,0.20919)(0.18,0.21585)(0.19,0.222)(0.2,0.22767)(0.21,0.23294)(0.22,0.23783)(0.23,0.24238)(0.24,0.24663)(0.25,0.25061)(0.26,0.25433)(0.27,0.25784)(0.28,0.26113)(0.29,0.26423)(0.3,0.26716)(0.31,0.26994)(0.32,0.27256)(0.33,0.27505)(0.34,0.27741)(0.35,0.27966)(0.36,0.28179)(0.37,0.28383)(0.38,0.28578)(0.39,0.28763)(0.4,0.28941)(0.41,0.29111)(0.42,0.29273)(0.43,0.29429)(0.44,0.29579)(0.45,0.29723)(0.46,0.29861)(0.47,0.29994)(0.48,0.30121)(0.49,0.30245)(0.5,0.30363)};
\addplot coordinates{
(0.1,0.21522)(0.11,0.22597)(0.12,0.23533)(0.13,0.24356)(0.14,0.25084)(0.15,0.25733)(0.16,0.26314)(0.17,0.26838)(0.18,0.27312)(0.19,0.27744)(0.2,0.28138)(0.21,0.28499)(0.22,0.28832)(0.23,0.29139)(0.24,0.29423)(0.25,0.29688)(0.26,0.29933)(0.27,0.30163)(0.28,0.30378)(0.29,0.30579)(0.3,0.30768)(0.31,0.30946)(0.32,0.31113)(0.33,0.31272)(0.34,0.31421)(0.35,0.31563)(0.36,0.31698)(0.37,0.31826)(0.38,0.31947)(0.39,0.32063)(0.4,0.32173)(0.41,0.32279)(0.42,0.32379)(0.43,0.32476)(0.44,0.32568)(0.45,0.32656)(0.46,0.32741)(0.47,0.32822)(0.48,0.329)(0.49,0.32975)(0.5,0.33048)};

\addplot[fill=red,fill opacity=0.8,draw opacity=0,line width=0pt] coordinates{(0.1,0)
(0.1,0.1409)(0.11,0.15375)(0.12,0.16534)(0.13,0.17583)(0.14,0.18536)(0.15,0.19402)(0.16,0.20194)(0.17,0.20919)(0.18,0.21585)(0.19,0.222)(0.2,0.22767)(0.21,0.23294)(0.22,0.23783)(0.23,0.24238)(0.24,0.24663)(0.25,0.25061)(0.26,0.25433)(0.27,0.25784)(0.28,0.26113)(0.29,0.26423)(0.3,0.26716)(0.31,0.26994)(0.32,0.27256)(0.33,0.27505)(0.34,0.27741)(0.35,0.27966)(0.36,0.28179)(0.37,0.28383)(0.38,0.28578)(0.39,0.28763)(0.4,0.28941)(0.41,0.29111)(0.42,0.29273)(0.43,0.29429)(0.44,0.29579)(0.45,0.29723)(0.46,0.29861)(0.47,0.29994)(0.48,0.30121)(0.49,0.30245)(0.5,0.30363)
(0.5,0)};

\addplot[fill=blue,fill opacity=0.8,draw opacity=0,line width=0pt] coordinates{
(0.1,0.21522)(0.11,0.22597)(0.12,0.23533)(0.13,0.24356)(0.14,0.25084)(0.15,0.25733)(0.16,0.26314)(0.17,0.26838)(0.18,0.27312)(0.19,0.27744)(0.2,0.28138)(0.21,0.28499)(0.22,0.28832)(0.23,0.29139)(0.24,0.29423)(0.25,0.29688)(0.26,0.29933)(0.27,0.30163)(0.28,0.30378)(0.29,0.30579)(0.3,0.30768)(0.31,0.30946)(0.32,0.31113)(0.33,0.31272)(0.34,0.31421)(0.35,0.31563)(0.36,0.31698)(0.37,0.31826)(0.38,0.31947)(0.39,0.32063)(0.4,0.32173)(0.41,0.32279)(0.42,0.32379)(0.43,0.32476)(0.44,0.32568)(0.45,0.32656)(0.46,0.32741)(0.47,0.32822)(0.48,0.329)(0.49,0.32975)(0.5,0.33048)
(0.5,0.30363)(0.49,0.30245)(0.48,0.30121)(0.47,0.29994)(0.46,0.29861)(0.45,0.29723)(0.44,0.29579)(0.43,0.29429)(0.42,0.29273)(0.41,0.29111)(0.4,0.28941)(0.39,0.28763)(0.38,0.28578)(0.37,0.28383)(0.36,0.28179)(0.35,0.27966)(0.34,0.27741)(0.33,0.27505)(0.32,0.27256)(0.31,0.26994)(0.3,0.26716)(0.29,0.26423)(0.28,0.26113)(0.27,0.25784)(0.26,0.25433)(0.25,0.25061)(0.24,0.24663)(0.23,0.24238)(0.22,0.23783)(0.21,0.23294)(0.2,0.22767)(0.19,0.222)(0.18,0.21585)(0.17,0.20919)(0.16,0.20194)(0.15,0.19402)(0.14,0.18536)(0.13,0.17583)(0.12,0.16534)(0.11,0.15375)(0.1,0.1409)};

\end{axis}\end{tikzpicture}}

	&
	{\tiny\begin{tikzpicture}\begin{axis}[width=\figwidth,height=\figheight,xlabel=$\bar\theta$,ylabel=Unit price (\euro/kWh),
legend style={at={(0.03,1)},anchor=north west},
legend entries = {$T_s^M$,$T_r^M$},
cycle list name=\mylist]

\addplot coordinates{
(0.1,0.07)(0.11,0.074)(0.12,0.078)(0.13,0.082)(0.14,0.086)(0.15,0.09)(0.16,0.094)(0.17,0.098)(0.18,0.102)(0.19,0.106)(0.2,0.11)(0.21,0.114)(0.22,0.118)(0.23,0.122)(0.24,0.126)(0.25,0.13)(0.26,0.134)(0.27,0.138)(0.28,0.142)(0.29,0.146)(0.3,0.15)(0.31,0.154)(0.32,0.158)(0.33,0.162)(0.34,0.166)(0.35,0.17)(0.36,0.174)(0.37,0.178)(0.38,0.182)(0.39,0.186)(0.4,0.19)(0.41,0.194)(0.42,0.198)(0.43,0.202)(0.44,0.206)(0.45,0.21)(0.46,0.214)(0.47,0.218)(0.48,0.222)(0.49,0.226)(0.5,0.23)};

\addplot coordinates{
(0.1,0.030416)(0.11,0.032396)(0.12,0.034376)(0.13,0.036356)(0.14,0.038337)(0.15,0.040317)(0.16,0.042297)(0.17,0.044277)(0.18,0.046257)(0.19,0.048237)(0.2,0.050217)(0.21,0.052197)(0.22,0.054177)(0.23,0.056157)(0.24,0.058137)(0.25,0.060117)(0.26,0.062097)(0.27,0.064078)(0.28,0.066058)(0.29,0.068038)(0.3,0.070018)(0.31,0.071998)(0.32,0.073978)(0.33,0.075958)(0.34,0.077938)(0.35,0.079918)(0.36,0.081898)(0.37,0.083878)(0.38,0.085858)(0.39,0.087839)(0.4,0.089819)(0.41,0.091799)(0.42,0.093779)(0.43,0.095759)(0.44,0.097739)(0.45,0.099719)(0.46,0.1017)(0.47,0.10368)(0.48,0.10566)(0.49,0.10764)(0.5,0.10962)};
\end{axis}\end{tikzpicture}}

	\\
	{\tiny\begin{tikzpicture}\begin{axis}[width=\figwidth,height=\figheight,xlabel=$\bar\theta$,ylabel=Revenue/Utility (\euro/EV),
legend style={at={(0.03,1)},anchor=north west},
legend entries = {$R_s^E$,$R_r^E$,$U^E$},
cycle list name=\mylist]

\addplot coordinates{
(0.1,0.20077)(0.11,0.23885)(0.12,0.27813)(0.13,0.31841)(0.14,0.35952)(0.15,0.40131)(0.16,0.4437)(0.17,0.48658)(0.18,0.52989)(0.19,0.57358)(0.2,0.61759)(0.21,0.66188)(0.22,0.70643)(0.23,0.75119)(0.24,0.79616)(0.25,0.8413)(0.26,0.8866)(0.27,0.93204)(0.28,0.97762)(0.29,1.0233)(0.3,1.0691)(0.31,1.115)(0.32,1.161)(0.33,1.207)(0.34,1.2532)(0.35,1.2994)(0.36,1.3457)(0.37,1.392)(0.38,1.4384)(0.39,1.4848)(0.4,1.5313)(0.41,1.5778)(0.42,1.6244)(0.43,1.671)(0.44,1.7176)(0.45,1.7643)(0.46,1.811)(0.47,1.8578)(0.48,1.9045)(0.49,1.9513)(0.5,1.9981)};
\addplot coordinates{
(0.1,0.27482)(0.11,0.32129)(0.12,0.36887)(0.13,0.41737)(0.14,0.46663)(0.15,0.51652)(0.16,0.56696)(0.17,0.61785)(0.18,0.66914)(0.19,0.72076)(0.2,0.77269)(0.21,0.82488)(0.22,0.8773)(0.23,0.92993)(0.24,0.98273)(0.25,1.0357)(0.26,1.0888)(0.27,1.1421)(0.28,1.1954)(0.29,1.2489)(0.3,1.3025)(0.31,1.3561)(0.32,1.4099)(0.33,1.4637)(0.34,1.5176)(0.35,1.5715)(0.36,1.6255)(0.37,1.6796)(0.38,1.7337)(0.39,1.7878)(0.4,1.842)(0.41,1.8963)(0.42,1.9506)(0.43,2.0049)(0.44,2.0592)(0.45,2.1136)(0.46,2.168)(0.47,2.2224)(0.48,2.2769)(0.49,2.3314)(0.5,2.3859)};
\addplot coordinates{
(0.1,0.88329)(0.11,1.0353)(0.12,1.191)(0.13,1.3496)(0.14,1.5107)(0.15,1.6739)(0.16,1.8388)(0.17,2.0052)(0.18,2.1729)(0.19,2.3417)(0.2,2.5114)(0.21,2.6819)(0.22,2.8532)(0.23,3.0251)(0.24,3.1977)(0.25,3.3707)(0.26,3.5442)(0.27,3.7181)(0.28,3.8924)(0.29,4.067)(0.3,4.2419)(0.31,4.4171)(0.32,4.5926)(0.33,4.7683)(0.34,4.9442)(0.35,5.1203)(0.36,5.2966)(0.37,5.4731)(0.38,5.6497)(0.39,5.8265)(0.4,6.0034)(0.41,6.1804)(0.42,6.3576)(0.43,6.5348)(0.44,6.7122)(0.45,6.8897)(0.46,7.0672)(0.47,7.2449)(0.48,7.4226)(0.49,7.6004)(0.5,7.7783)};

\addplot[fill=red,fill opacity=0.8,draw opacity=0,line width=0pt] coordinates{(0.1,0)
(0.1,0.20077)(0.11,0.23885)(0.12,0.27813)(0.13,0.31841)(0.14,0.35952)(0.15,0.40131)(0.16,0.4437)(0.17,0.48658)(0.18,0.52989)(0.19,0.57358)(0.2,0.61759)(0.21,0.66188)(0.22,0.70643)(0.23,0.75119)(0.24,0.79616)(0.25,0.8413)(0.26,0.8866)(0.27,0.93204)(0.28,0.97762)(0.29,1.0233)(0.3,1.0691)(0.31,1.115)(0.32,1.161)(0.33,1.207)(0.34,1.2532)(0.35,1.2994)(0.36,1.3457)(0.37,1.392)(0.38,1.4384)(0.39,1.4848)(0.4,1.5313)(0.41,1.5778)(0.42,1.6244)(0.43,1.671)(0.44,1.7176)(0.45,1.7643)(0.46,1.811)(0.47,1.8578)(0.48,1.9045)(0.49,1.9513)(0.5,1.9981)
(0.5,0)};

\addplot[fill=blue,fill opacity=0.8,draw opacity=0,line width=0pt] coordinates{
(0.1,0.27482)(0.11,0.32129)(0.12,0.36887)(0.13,0.41737)(0.14,0.46663)(0.15,0.51652)(0.16,0.56696)(0.17,0.61785)(0.18,0.66914)(0.19,0.72076)(0.2,0.77269)(0.21,0.82488)(0.22,0.8773)(0.23,0.92993)(0.24,0.98273)(0.25,1.0357)(0.26,1.0888)(0.27,1.1421)(0.28,1.1954)(0.29,1.2489)(0.3,1.3025)(0.31,1.3561)(0.32,1.4099)(0.33,1.4637)(0.34,1.5176)(0.35,1.5715)(0.36,1.6255)(0.37,1.6796)(0.38,1.7337)(0.39,1.7878)(0.4,1.842)(0.41,1.8963)(0.42,1.9506)(0.43,2.0049)(0.44,2.0592)(0.45,2.1136)(0.46,2.168)(0.47,2.2224)(0.48,2.2769)(0.49,2.3314)(0.5,2.3859)
(0.5,1.9981)(0.49,1.9513)(0.48,1.9045)(0.47,1.8578)(0.46,1.811)(0.45,1.7643)(0.44,1.7176)(0.43,1.671)(0.42,1.6244)(0.41,1.5778)(0.4,1.5313)(0.39,1.4848)(0.38,1.4384)(0.37,1.392)(0.36,1.3457)(0.35,1.2994)(0.34,1.2532)(0.33,1.207)(0.32,1.161)(0.31,1.115)(0.3,1.0691)(0.29,1.0233)(0.28,0.97762)(0.27,0.93204)(0.26,0.8866)(0.25,0.8413)(0.24,0.79616)(0.23,0.75119)(0.22,0.70643)(0.21,0.66188)(0.2,0.61759)(0.19,0.57358)(0.18,0.52989)(0.17,0.48658)(0.16,0.4437)(0.15,0.40131)(0.14,0.35952)(0.13,0.31841)(0.12,0.27813)(0.11,0.23885)(0.1,0.20077)
};

\addplot[fill=yellow,fill opacity=0.8,draw opacity=0,line width=0pt] coordinates{
(0.1,0.88329)(0.11,1.0353)(0.12,1.191)(0.13,1.3496)(0.14,1.5107)(0.15,1.6739)(0.16,1.8388)(0.17,2.0052)(0.18,2.1729)(0.19,2.3417)(0.2,2.5114)(0.21,2.6819)(0.22,2.8532)(0.23,3.0251)(0.24,3.1977)(0.25,3.3707)(0.26,3.5442)(0.27,3.7181)(0.28,3.8924)(0.29,4.067)(0.3,4.2419)(0.31,4.4171)(0.32,4.5926)(0.33,4.7683)(0.34,4.9442)(0.35,5.1203)(0.36,5.2966)(0.37,5.4731)(0.38,5.6497)(0.39,5.8265)(0.4,6.0034)(0.41,6.1804)(0.42,6.3576)(0.43,6.5348)(0.44,6.7122)(0.45,6.8897)(0.46,7.0672)(0.47,7.2449)(0.48,7.4226)(0.49,7.6004)(0.5,7.7783)
(0.5,2.3859)(0.49,2.3314)(0.48,2.2769)(0.47,2.2224)(0.46,2.168)(0.45,2.1136)(0.44,2.0592)(0.43,2.0049)(0.42,1.9506)(0.41,1.8963)(0.4,1.842)(0.39,1.7878)(0.38,1.7337)(0.37,1.6796)(0.36,1.6255)(0.35,1.5715)(0.34,1.5176)(0.33,1.4637)(0.32,1.4099)(0.31,1.3561)(0.3,1.3025)(0.29,1.2489)(0.28,1.1954)(0.27,1.1421)(0.26,1.0888)(0.25,1.0357)(0.24,0.98273)(0.23,0.92993)(0.22,0.8773)(0.21,0.82488)(0.2,0.77269)(0.19,0.72076)(0.18,0.66914)(0.17,0.61785)(0.16,0.56696)(0.15,0.51652)(0.14,0.46663)(0.13,0.41737)(0.12,0.36887)(0.11,0.32129)(0.1,0.27482)};
\end{axis}\end{tikzpicture}}

	&
	{\tiny\begin{tikzpicture}\begin{axis}[width=\figwidth,height=\figheight,xlabel=$\bar\theta$,ylabel=Participation probability,
legend style={at={(0.5,1)},anchor=north west},
legend entries = {$\alpha_s^E$,$\alpha_r^E$},
cycle list name=\mylist]

\addplot coordinates{
(0.1,0.19879)(0.11,0.21499)(0.12,0.22949)(0.13,0.24252)(0.14,0.25427)(0.15,0.2649)(0.16,0.27458)(0.17,0.2834)(0.18,0.29148)(0.19,0.29891)(0.2,0.30575)(0.21,0.31207)(0.22,0.31794)(0.23,0.32338)(0.24,0.32846)(0.25,0.3332)(0.26,0.33764)(0.27,0.3418)(0.28,0.34571)(0.29,0.34938)(0.3,0.35285)(0.31,0.35613)(0.32,0.35923)(0.33,0.36216)(0.34,0.36495)(0.35,0.36759)(0.36,0.37011)(0.37,0.3725)(0.38,0.37479)(0.39,0.37697)(0.4,0.37905)(0.41,0.38104)(0.42,0.38294)(0.43,0.38477)(0.44,0.38652)(0.45,0.3882)(0.46,0.38982)(0.47,0.39137)(0.48,0.39286)(0.49,0.3943)(0.5,0.39568)};
\addplot coordinates{
(0.1,0.4118)(0.11,0.43631)(0.12,0.45786)(0.13,0.47693)(0.14,0.49391)(0.15,0.50912)(0.16,0.52282)(0.17,0.53522)(0.18,0.54648)(0.19,0.55677)(0.2,0.56619)(0.21,0.57485)(0.22,0.58284)(0.23,0.59024)(0.24,0.5971)(0.25,0.60348)(0.26,0.60944)(0.27,0.615)(0.28,0.62022)(0.29,0.62511)(0.3,0.62972)(0.31,0.63405)(0.32,0.63815)(0.33,0.64202)(0.34,0.64568)(0.35,0.64915)(0.36,0.65245)(0.37,0.65559)(0.38,0.65857)(0.39,0.66141)(0.4,0.66413)(0.41,0.66672)(0.42,0.6692)(0.43,0.67157)(0.44,0.67384)(0.45,0.67601)(0.46,0.6781)(0.47,0.68011)(0.48,0.68204)(0.49,0.68389)(0.5,0.68568)};

\addplot[fill=red,fill opacity=0.8,draw opacity=0,line width=0pt] coordinates{(0.1,0)
(0.1,0.19879)(0.11,0.21499)(0.12,0.22949)(0.13,0.24252)(0.14,0.25427)(0.15,0.2649)(0.16,0.27458)(0.17,0.2834)(0.18,0.29148)(0.19,0.29891)(0.2,0.30575)(0.21,0.31207)(0.22,0.31794)(0.23,0.32338)(0.24,0.32846)(0.25,0.3332)(0.26,0.33764)(0.27,0.3418)(0.28,0.34571)(0.29,0.34938)(0.3,0.35285)(0.31,0.35613)(0.32,0.35923)(0.33,0.36216)(0.34,0.36495)(0.35,0.36759)(0.36,0.37011)(0.37,0.3725)(0.38,0.37479)(0.39,0.37697)(0.4,0.37905)(0.41,0.38104)(0.42,0.38294)(0.43,0.38477)(0.44,0.38652)(0.45,0.3882)(0.46,0.38982)(0.47,0.39137)(0.48,0.39286)(0.49,0.3943)(0.5,0.39568)
(0.5,0)
};
\addplot[fill=blue,fill opacity=0.8,draw opacity=0,line width=0pt] coordinates{
(0.1,0.4118)(0.11,0.43631)(0.12,0.45786)(0.13,0.47693)(0.14,0.49391)(0.15,0.50912)(0.16,0.52282)(0.17,0.53522)(0.18,0.54648)(0.19,0.55677)(0.2,0.56619)(0.21,0.57485)(0.22,0.58284)(0.23,0.59024)(0.24,0.5971)(0.25,0.60348)(0.26,0.60944)(0.27,0.615)(0.28,0.62022)(0.29,0.62511)(0.3,0.62972)(0.31,0.63405)(0.32,0.63815)(0.33,0.64202)(0.34,0.64568)(0.35,0.64915)(0.36,0.65245)(0.37,0.65559)(0.38,0.65857)(0.39,0.66141)(0.4,0.66413)(0.41,0.66672)(0.42,0.6692)(0.43,0.67157)(0.44,0.67384)(0.45,0.67601)(0.46,0.6781)(0.47,0.68011)(0.48,0.68204)(0.49,0.68389)(0.5,0.68568)
(0.5,0.39568)(0.49,0.3943)(0.48,0.39286)(0.47,0.39137)(0.46,0.38982)(0.45,0.3882)(0.44,0.38652)(0.43,0.38477)(0.42,0.38294)(0.41,0.38104)(0.4,0.37905)(0.39,0.37697)(0.38,0.37479)(0.37,0.3725)(0.36,0.37011)(0.35,0.36759)(0.34,0.36495)(0.33,0.36216)(0.32,0.35923)(0.31,0.35613)(0.3,0.35285)(0.29,0.34938)(0.28,0.34571)(0.27,0.3418)(0.26,0.33764)(0.25,0.3332)(0.24,0.32846)(0.23,0.32338)(0.22,0.31794)(0.21,0.31207)(0.2,0.30575)(0.19,0.29891)(0.18,0.29148)(0.17,0.2834)(0.16,0.27458)(0.15,0.2649)(0.14,0.25427)(0.13,0.24252)(0.12,0.22949)(0.11,0.21499)(0.1,0.19879)};

\end{axis}\end{tikzpicture}}

	&
	{\tiny\begin{tikzpicture}\begin{axis}[width=\figwidth,height=\figheight,xlabel=$\bar\theta$,ylabel=Unit price (\euro/kWh),
legend style={at={(0.03,1)},anchor=north west},
legend entries = {$T_s^E$,$T_r^E$},
cycle list name=\mylist]

\addplot coordinates{
(0.1,0.050199)(0.11,0.052219)(0.12,0.054239)(0.13,0.056259)(0.14,0.058279)(0.15,0.060299)(0.16,0.062319)(0.17,0.064339)(0.18,0.066359)(0.19,0.068378)(0.2,0.070398)(0.21,0.072418)(0.22,0.074438)(0.23,0.076458)(0.24,0.078478)(0.25,0.080498)(0.26,0.082518)(0.27,0.084538)(0.28,0.086558)(0.29,0.088578)(0.3,0.090598)(0.31,0.092618)(0.32,0.094637)(0.33,0.096657)(0.34,0.098677)(0.35,0.1007)(0.36,0.10272)(0.37,0.10474)(0.38,0.10676)(0.39,0.10878)(0.4,0.1108)(0.41,0.11282)(0.42,0.11484)(0.43,0.11686)(0.44,0.11888)(0.45,0.1209)(0.46,0.12292)(0.47,0.12494)(0.48,0.12696)(0.49,0.12898)(0.5,0.131)};

\addplot coordinates{
(0.1,0.017568)(0.11,0.018065)(0.12,0.018562)(0.13,0.019058)(0.14,0.019555)(0.15,0.02005)(0.16,0.020546)(0.17,0.021041)(0.18,0.021536)(0.19,0.022032)(0.2,0.022526)(0.21,0.023021)(0.22,0.023516)(0.23,0.024011)(0.24,0.024506)(0.25,0.025)(0.26,0.025495)(0.27,0.025989)(0.28,0.026484)(0.29,0.026978)(0.3,0.027473)(0.31,0.027967)(0.32,0.028462)(0.33,0.028956)(0.34,0.02945)(0.35,0.029945)(0.36,0.030439)(0.37,0.030933)(0.38,0.031428)(0.39,0.031922)(0.4,0.032416)(0.41,0.032911)(0.42,0.033405)(0.43,0.033899)(0.44,0.034393)(0.45,0.034888)(0.46,0.035382)(0.47,0.035876)(0.48,0.03637)(0.49,0.036865)(0.5,0.037359)};

\end{axis}\end{tikzpicture}}

\end{tabular}
\caption{Comparison between Monopoly (first row) and Nash equilibrium (second row), with $t = 0.03$, $\bar\theta=0.3$, $C_B = 50$, $\rho_d = \rho_u = 0.48$, $\gamma = 0.05$, $r_d = 0.4$, $r_u = 1.6$ ($r_d$ and $r_u$ are the daily average of 20/07/2015).}

%% file: Compare_rd_ru_nash_mono.tex
\centering
\gdef\figheight{4cm}
\gdef\figwidth{6cm}
\begin{tabular}{ccc}
	$\bar\theta = 0.3, \gamma = 0.05$ & $\bar\theta = 0.1, \gamma = 0.05$ & $\bar\theta = 0.1, \gamma = 0.5$ \\
	
	{\footnotesize\begin{tikzpicture}
\begin{axis}[width=\figwidth,height=\figheight,xlabel=$r_u$,ylabel=$r_d$,
cycle list name=\mylist]
\addplot[fill=blue,fill opacity=0.2,draw opacity=0,line width=0pt] coordinates{
(1.55,0)(1.55,0.54)(1.56,0.55)(1.57,0.56)(1.58,0.57)(1.59,0.58)(1.6,0.59)(1.61,0.6)(1.62,0.61)(1.63,0.62)(1.64,0.63)(1.66,0.64)(1.67,0.65)(1.68,0.66)(1.69,0.67)(1.7,0.68)(1.71,0.69)(1.72,0.7)(1.73,0.71)(1.74,0.72)(1.75,0.73)(1.76,0.74)(1.78,0.75)(1.79,0.76)(1.8,0.77)(1.81,0.78)(1.82,0.79)(1.83,0.8)(1.84,0.81)(1.85,0.82)(1.86,0.83)(1.87,0.84)(1.89,0.85)(1.9,0.86)(1.91,0.87)(1.92,0.88)(1.93,0.89)(1.94,0.9)(1.95,0.91)(1.96,0.92)(1.97,0.93)(1.98,0.94)(2,0.95)(2.01,0.96)(2.02,0.97)(2.03,0.98)(2.04,0.99)(2.05,1)
(2.5,1)(2.5,0)};
\addplot[fill=red,fill opacity=0.2,draw opacity=0,line width=0pt] coordinates{
(1.55,0.55)(1.56,0.56)(1.57,0.57)(1.58,0.58)(1.59,0.59)(1.6,0.6)(1.61,0.61)(1.62,0.62)(1.63,0.63)(1.65,0.64)(1.66,0.65)(1.67,0.66)(1.68,0.67)(1.69,0.68)(1.7,0.69)(1.71,0.7)(1.72,0.71)(1.73,0.72)(1.74,0.73)(1.75,0.74)(1.77,0.75)(1.78,0.76)(1.79,0.77)(1.8,0.78)(1.81,0.79)(1.82,0.8)(1.83,0.81)(1.84,0.82)(1.85,0.83)(1.86,0.84)(1.88,0.85)(1.89,0.86)(1.9,0.87)(1.91,0.88)(1.92,0.89)(1.93,0.9)(1.94,0.91)(1.95,0.92)(1.96,0.93)(1.97,0.94)(1.99,0.95)(2,0.96)(2.01,0.97)(2.02,0.98)(2.03,0.99)(2.04,1)
(0,1)(0,0.55)};
\addplot[fill=black,fill opacity=0.4,draw opacity=0,line width=0pt] coordinates{(0,0)(0,0.54)(1.54,0.54)(1.54,0)};
\input{DataJuly.tex}
\draw (axis cs:1.65,0.5) node (Jul20) [anchor=south west]{\scriptsize 20/07};
\draw[->](Jul20)--(axis cs:1.6628,0.3759);
\draw (axis cs:0.6,0.18) node[rectangle,rounded corners=4pt,fill=white,fill opacity=.4,text opacity=1]{$P_n^{\text{opt}}=\mbox{NaN}$};
\draw (axis cs:0.45,0.85) node[rectangle,rounded corners=4pt,fill=white,fill opacity=.4,text opacity=1]{$P_n^{\text{opt}}=0$};
\draw (axis cs:2.3,0.6) node[rectangle,rounded corners=4pt,fill=white,fill opacity=.4,text opacity=1]{\begin{turn}{90}$P_n^{\text{opt}}=P_d$\end{turn}};

\end{axis}\end{tikzpicture}}

	&
	{\footnotesize\begin{tikzpicture}
\begin{axis}[width=\figwidth,height=\figheight,xlabel=$r_u$,ylabel=$r_d$,
legend style={at={(0.03,1)},anchor=north west},
legend entries = {},
cycle list name=\mylist]
\addplot[fill=blue,fill opacity=0.2,draw opacity=0,line width=0pt]  coordinates{
(1.55,0)(1.55,0.54)(1.56,0.55)(1.57,0.56)(1.58,0.57)(1.59,0.58)(1.6,0.59)(1.61,0.6)(1.62,0.61)(1.63,0.62)(1.64,0.63)(1.66,0.64)(1.67,0.65)(1.68,0.66)(1.69,0.67)(1.7,0.68)(1.71,0.69)(1.72,0.7)(1.73,0.71)(1.74,0.72)(1.75,0.73)(1.77,0.74)(1.78,0.75)(1.79,0.76)(1.8,0.77)(1.81,0.78)(1.82,0.79)(1.83,0.8)(1.84,0.81)(1.85,0.82)(1.87,0.83)(1.88,0.84)(1.89,0.85)(1.9,0.86)(1.91,0.87)(1.92,0.88)(1.93,0.89)(1.94,0.9)(1.95,0.91)(1.97,0.92)(1.98,0.93)(1.99,0.94)(2,0.95)(2.01,0.96)(2.02,0.97)(2.03,0.98)(2.04,0.99)(2.06,1)(2.5,1)(2.5,0)};
\addplot[fill=red,fill opacity=0.2,draw opacity=0,line width=0pt]  coordinates{
(1.55,0.55)(1.56,0.56)(1.57,0.57)(1.58,0.58)(1.59,0.59)(1.6,0.6)(1.61,0.61)(1.62,0.62)(1.63,0.63)(1.65,0.64)(1.66,0.65)(1.67,0.66)(1.68,0.67)(1.69,0.68)(1.7,0.69)(1.71,0.7)(1.72,0.71)(1.73,0.72)(1.74,0.73)(1.76,0.74)(1.77,0.75)(1.78,0.76)(1.79,0.77)(1.8,0.78)(1.81,0.79)(1.82,0.8)(1.83,0.81)(1.84,0.82)(1.86,0.83)(1.87,0.84)(1.88,0.85)(1.89,0.86)(1.9,0.87)(1.91,0.88)(1.92,0.89)(1.93,0.9)(1.94,0.91)(1.96,0.92)(1.97,0.93)(1.98,0.94)(1.99,0.95)(2,0.96)(2.01,0.97)(2.02,0.98)(2.03,0.99)(2.05,1)(0,1)(0,0.55)};
\addplot[fill=black,fill opacity=0.4,draw opacity=0,line width=0pt] coordinates{(0,0)(0,0.54)(1.54,0.54)(1.54,0)};
\input{DataJuly.tex}
\end{axis}\end{tikzpicture}}

	&
	{\footnotesize\begin{tikzpicture}
\begin{axis}[width=\figwidth,height=\figheight,xlabel=$r_u$,ylabel=$r_d$,
legend style={at={(0.03,1)},anchor=north west},
legend entries = {},
cycle list name=\mylist]
\addplot[fill=blue,fill opacity=0.2,draw opacity=0,line width=0pt] coordinates{
(1.8,0)(1.8,0.76)(1.82,0.77)(1.83,0.78)(1.84,0.79)(1.85,0.8)(1.86,0.81)(1.87,0.82)(1.88,0.83)(1.89,0.84)(1.9,0.85)(1.92,0.86)(1.93,0.87)(1.94,0.88)(1.95,0.89)(1.96,0.9)(1.97,0.91)(1.98,0.92)(1.99,0.93)(2.01,0.94)(2.02,0.95)(2.03,0.96)(2.04,0.97)(2.05,0.98)(2.06,0.99)(2.07,1)(2.5,1)(2.5,0)};
\addplot[fill=red,fill opacity=0.2,draw opacity=0,line width=0pt] coordinates{
(1.77,0.77)(1.78,0.78)(1.79,0.79)(1.81,0.8)(1.82,0.81)(1.83,0.82)(1.84,0.83)(1.85,0.84)(1.86,0.85)(1.88,0.86)(1.89,0.87)(1.9,0.88)(1.91,0.89)(1.92,0.9)(1.94,0.91)(1.95,0.92)(1.96,0.93)(1.97,0.94)(1.99,0.95)(2,0.96)(2.01,0.97)(2.02,0.98)(2.03,0.99)(2.05,1)(0,1)(0,0.77)};
\addplot[fill=black,fill opacity=0.4,draw opacity=0,line width=0pt] coordinates{(0,0)(0,0.76)(1.79,0.76)(1.79,0)};
\input{DataJuly.tex}
\end{axis}\end{tikzpicture}}

\\
	{\footnotesize\begin{tikzpicture}
\begin{axis}[width=\figwidth,height=\figheight,xlabel=$r_u$,ylabel=$r_d$,
legend style={at={(0,1)},anchor=north west},
cycle list name=\mylist]
\addplot[only marks, mark=o, blue] coordinates{	
                (1.74069767441860,0)  
                (1.20930232558140,0.335581395348837)  
                (1.20930232558140,0.340930232558140)  
                (1.20930232558140,	0.340930232558140)  
                (0.851627906976744,0.335581395348837)  
                (0.744186046511628,0.696976744186047)  
                (0.744186046511628,0.696976744186047)  
                (0.765116279069767,0.696976744186047)  
                (0.765116279069767,0.696976744186047)  
                (1.71534883720930,0.696976744186047)  
                (1.71534883720930,0.696976744186047)  
                (1.35581395348837,0.0465116279069767)  
                (1.35581395348837,0.441860465116279)  
                (1.51162790697674,0.441860465116279)  
                (1.51162790697674,0.133953488372093)  
                (1.58139534883721,0.133953488372093)  
                (1.58139534883721,0.0116279069767442)  
                (1.71860465116279,0)  
                (1.71860465116279,0)  
                (1.84093023255814,0)  
                (1.84093023255814,0)  
                (1.89534883720930,0)  
                (1.89534883720930,0)  
                (2,0)  
                (2,0)  
                (1.22395348837209,0)  
                (1.22395348837209,0)  
                (2.30232558139535,0)  
                (2.30232558139535,0)  
                (2.24581395348837,0)  
                (2.24581395348837,0)  
                (0,0)  
                (0,0)  
                (1.90697674418605,0)  
                (1.90697674418605,0)   
                (1.83720930232558,0)  
                (1.83720930232558,0)  
                (1.81395348837209,0)  
                (1.81395348837209,0)  
                (0,0)  
                (0,0)  
                (1.83720930232558,0)  
                (1.83720930232558,0)  
                (2.32558139534884,0.133953488372093)  
                (2.32558139534884,0)  
                (2.24581395348837,0.0627906976744187)  
                (2.24581395348837,0.133953488372093)  
		};
\addlegendentry{\tiny Price per half hour on 07/20/2015};	
\addplot[only marks,mark=square,red]
		coordinates {
		(1.6628,0.3759)
		(1.85406666666667,0.0656463414634150)  
		(1.82484146341464,0.0903937499999996)  
		(1.49025000000000,0.258639534883721)  
		(1.32151041666667,0.117562500000000)  
		(1.213708333333334,0.730402439024390)  
		(1.824835365853659,0.742287500000000)  
		};
\addlegendentry{\tiny Average daily price 07/20 - 07/26/2015}; 
\addplot[fill=blue,fill opacity=0.2,draw opacity=0,line width=0pt]  coordinates{
(0.95,0)(0.96,0.01)(0.97,0.02)(0.98,0.03)(0.99,0.04)(1,0.05)(1.02,0.06)(1.03,0.07)(1.04,0.08)(1.05,0.09)(1.06,0.1)(1.07,0.11)(1.08,0.12)(1.09,0.13)(1.1,0.14)(1.11,0.15)(1.12,0.16)(1.13,0.17)(1.14,0.18)(1.16,0.19)(1.17,0.2)(1.18,0.21)(1.19,0.22)(1.2,0.23)(1.21,0.24)(1.22,0.25)(1.23,0.26)(1.24,0.27)(1.25,0.28)(1.26,0.29)(1.27,0.3)(1.28,0.31)(1.3,0.32)(1.31,0.33)(1.32,0.34)(1.33,0.35)(1.34,0.36)(1.35,0.37)(1.36,0.38)(1.37,0.39)(1.38,0.4)(1.39,0.41)(1.4,0.42)(1.41,0.43)(1.43,0.44)(1.44,0.45)(1.45,0.46)(1.46,0.47)(1.47,0.48)(1.48,0.49)(1.49,0.5)(1.5,0.51)(1.51,0.52)(1.52,0.53)(1.53,0.54)(1.54,0.55)(1.56,0.56)(1.57,0.57)(1.58,0.58)(1.59,0.59)(1.6,0.6)(1.61,0.61)(1.62,0.62)(1.63,0.63)(1.64,0.64)(1.65,0.65)(1.66,0.66)(1.67,0.67)(1.69,0.68)(1.7,0.69)(1.71,0.7)(1.72,0.71)(1.73,0.72)(1.74,0.73)(1.75,0.74)(1.76,0.75)(1.77,0.76)(1.78,0.77)(1.79,0.78)(1.81,0.79)(1.82,0.8)(1.83,0.81)(1.84,0.82)(1.85,0.83)(1.86,0.84)(1.87,0.85)(1.88,0.86)(1.89,0.87)(1.9,0.88)(1.91,0.89)(1.93,0.9)(1.94,0.91)(1.95,0.92)(1.96,0.93)(1.97,0.94)(1.98,0.95)(1.99,0.96)(2,0.97)(2.01,0.98)(2.02,0.99)(2.04,1)(2.5,1)(2.5,0)};

\addplot[fill=red,fill opacity=0.2,draw opacity=0,line width=0pt] coordinates{
(0.93,0)(0.95,0.01)(0.96,0.02)(0.97,0.03)(0.98,0.04)(0.99,0.05)(1,0.06)(1.01,0.07)(1.02,0.08)(1.03,0.09)(1.04,0.1)(1.05,0.11)(1.06,0.12)(1.07,0.13)(1.09,0.14)(1.1,0.15)(1.11,0.16)(1.12,0.17)(1.13,0.18)(1.14,0.19)(1.15,0.2)(1.16,0.21)(1.17,0.22)(1.18,0.23)(1.19,0.24)(1.2,0.25)(1.22,0.26)(1.23,0.27)(1.24,0.28)(1.25,0.29)(1.26,0.3)(1.27,0.31)(1.28,0.32)(1.29,0.33)(1.3,0.34)(1.31,0.35)(1.32,0.36)(1.34,0.37)(1.35,0.38)(1.36,0.39)(1.37,0.4)(1.38,0.41)(1.39,0.42)(1.4,0.43)(1.41,0.44)(1.42,0.45)(1.43,0.46)(1.44,0.47)(1.46,0.48)(1.47,0.49)(1.48,0.5)(1.49,0.51)(1.5,0.52)(1.51,0.53)(1.52,0.54)(1.53,0.55)(1.54,0.56)(1.55,0.57)(1.56,0.58)(1.58,0.59)(1.59,0.6)(1.6,0.61)(1.61,0.62)(1.62,0.63)(1.63,0.64)(1.64,0.65)(1.65,0.66)(1.66,0.67)(1.67,0.68)(1.68,0.69)(1.7,0.7)(1.71,0.71)(1.72,0.72)(1.73,0.73)(1.74,0.74)(1.75,0.75)(1.76,0.76)(1.77,0.77)(1.78,0.78)(1.79,0.79)(1.81,0.8)(1.82,0.81)(1.83,0.82)(1.84,0.83)(1.85,0.84)(1.86,0.85)(1.87,0.86)(1.88,0.87)(1.89,0.88)(1.9,0.89)(1.92,0.9)(1.93,0.91)(1.94,0.92)(1.95,0.93)(1.96,0.94)(1.97,0.95)(1.98,0.96)(1.99,0.97)(2,0.98)(2.01,0.99)(2.03,1)(0,1)(0,0)};

\end{axis}\end{tikzpicture}}
	
	&
	{\footnotesize\begin{tikzpicture}
\begin{axis}[width=\figwidth,height=\figheight,xlabel=$r_u$,ylabel=$r_d$,
legend style={at={(0.03,1)},anchor=north west},
legend entries = {},
cycle list name=\mylist]
\addplot[fill=blue,fill opacity=0.2,draw opacity=0,line width=0pt] coordinates{
(1.19,0)(1.19,0.21)(1.2,0.22)(1.21,0.23)(1.22,0.24)(1.23,0.25)(1.24,0.26)(1.25,0.27)(1.26,0.28)(1.27,0.29)(1.28,0.3)(1.29,0.31)(1.3,0.32)(1.32,0.33)(1.33,0.34)(1.34,0.35)(1.35,0.36)(1.36,0.37)(1.37,0.38)(1.38,0.39)(1.39,0.4)(1.4,0.41)(1.41,0.42)(1.42,0.43)(1.43,0.44)(1.44,0.45)(1.45,0.46)(1.47,0.47)(1.48,0.48)(1.49,0.49)(1.5,0.5)(1.51,0.51)(1.52,0.52)(1.53,0.53)(1.54,0.54)(1.55,0.55)(1.56,0.56)(1.57,0.57)(1.59,0.58)(1.6,0.59)(1.61,0.6)(1.62,0.61)(1.63,0.62)(1.64,0.63)(1.65,0.64)(1.66,0.65)(1.67,0.66)(1.68,0.67)(1.69,0.68)(1.71,0.69)(1.72,0.7)(1.73,0.71)(1.74,0.72)(1.75,0.73)(1.76,0.74)(1.77,0.75)(1.78,0.76)(1.79,0.77)(1.81,0.78)(1.82,0.79)(1.83,0.8)(1.84,0.81)(1.85,0.82)(1.86,0.83)(1.87,0.84)(1.88,0.85)(1.89,0.86)(1.9,0.87)(1.92,0.88)(1.93,0.89)(1.94,0.9)(1.95,0.91)(1.96,0.92)(1.97,0.93)(1.98,0.94)(1.99,0.95)(2.01,0.96)(2.02,0.97)(2.03,0.98)(2.04,0.99)(2.05,1)(2.5,1)(2.5,0)};
\addplot[fill=red,fill opacity=0.2,draw opacity=0,line width=0pt] coordinates{
(1.19,0.22)(1.2,0.23)(1.21,0.24)(1.22,0.25)(1.23,0.26)(1.24,0.27)(1.25,0.28)(1.26,0.29)(1.27,0.3)(1.28,0.31)(1.29,0.32)(1.3,0.33)(1.32,0.34)(1.33,0.35)(1.34,0.36)(1.35,0.37)(1.36,0.38)(1.37,0.39)(1.38,0.4)(1.39,0.41)(1.4,0.42)(1.41,0.43)(1.42,0.44)(1.43,0.45)(1.44,0.46)(1.46,0.47)(1.47,0.48)(1.48,0.49)(1.49,0.5)(1.5,0.51)(1.51,0.52)(1.52,0.53)(1.53,0.54)(1.54,0.55)(1.55,0.56)(1.56,0.57)(1.58,0.58)(1.59,0.59)(1.6,0.6)(1.61,0.61)(1.62,0.62)(1.63,0.63)(1.64,0.64)(1.65,0.65)(1.66,0.66)(1.67,0.67)(1.68,0.68)(1.7,0.69)(1.71,0.7)(1.72,0.71)(1.73,0.72)(1.74,0.73)(1.75,0.74)(1.76,0.75)(1.77,0.76)(1.78,0.77)(1.8,0.78)(1.81,0.79)(1.82,0.8)(1.83,0.81)(1.84,0.82)(1.85,0.83)(1.86,0.84)(1.87,0.85)(1.88,0.86)(1.89,0.87)(1.91,0.88)(1.92,0.89)(1.93,0.9)(1.94,0.91)(1.95,0.92)(1.96,0.93)(1.97,0.94)(1.98,0.95)(2,0.96)(2.01,0.97)(2.02,0.98)(2.03,0.99)(2.04,1)(0,1)(0,0.22)};
\addplot[fill=black,fill opacity=0.4,draw opacity=0,line width=0pt] coordinates{(0,0)(0,0.21)(1.18,0.21)(1.18,0)};
\input{DataJuly.tex}
\end{axis}\end{tikzpicture}}

	&
	{\footnotesize\begin{tikzpicture}
\begin{axis}[width=\figwidth,height=\figheight,xlabel=$r_u$,ylabel=$r_d$,
legend style={at={(0.03,1)},anchor=north west},
legend entries = {},
cycle list name=\mylist]
\addplot[fill=blue,fill opacity=0.2,draw opacity=0,line width=0pt] coordinates{
(1.51,0)(1.51,0.5)(1.52,0.51)(1.53,0.52)(1.54,0.53)(1.55,0.54)(1.56,0.55)(1.57,0.56)(1.58,0.57)(1.6,0.58)(1.61,0.59)(1.62,0.6)(1.63,0.61)(1.64,0.62)(1.65,0.63)(1.66,0.64)(1.67,0.65)(1.68,0.66)(1.69,0.67)(1.71,0.68)(1.72,0.69)(1.73,0.7)(1.74,0.71)(1.75,0.72)(1.76,0.73)(1.77,0.74)(1.78,0.75)(1.8,0.76)(1.81,0.77)(1.82,0.78)(1.83,0.79)(1.84,0.8)(1.85,0.81)(1.86,0.82)(1.87,0.83)(1.89,0.84)(1.9,0.85)(1.91,0.86)(1.92,0.87)(1.93,0.88)(1.94,0.89)(1.95,0.9)(1.96,0.91)(1.98,0.92)(1.99,0.93)(2,0.94)(2.01,0.95)(2.02,0.96)(2.03,0.97)(2.04,0.98)(2.05,0.99)(2.07,1)(2.5,1)(2.5,0)};
\addplot[fill=red,fill opacity=0.2,draw opacity=0,line width=0pt]  coordinates{
(1.48,0.54)(1.49,0.55)(1.5,0.56)(1.51,0.57)(1.52,0.58)(1.54,0.59)(1.55,0.6)(1.56,0.61)(1.57,0.62)(1.58,0.63)(1.59,0.64)(1.61,0.65)(1.62,0.66)(1.63,0.67)(1.64,0.68)(1.65,0.69)(1.66,0.7)(1.68,0.71)(1.69,0.72)(1.7,0.73)(1.71,0.74)(1.72,0.75)(1.74,0.76)(1.75,0.77)(1.76,0.78)(1.77,0.79)(1.78,0.8)(1.8,0.81)(1.81,0.82)(1.82,0.83)(1.83,0.84)(1.84,0.85)(1.86,0.86)(1.87,0.87)(1.88,0.88)(1.89,0.89)(1.9,0.9)(1.92,0.91)(1.93,0.92)(1.94,0.93)(1.95,0.94)(1.97,0.95)(1.98,0.96)(1.99,0.97)(2,0.98)(2.02,0.99)(2.03,1)(0,1)(0,0.54)};
\addplot[fill=black,fill opacity=0.4,draw opacity=0,line width=0pt] coordinates{(0,0)(0,0.53)(1.5,0.53)(1.50,0)};
\input{DataJuly.tex}
\end{axis}\end{tikzpicture}}

\end{tabular}
\caption{Comparison of variable regions on $r_d\times r_u$ plane and the $x$ chosen, $t = 0.03$, $C_B = 50$, $\rho_d = \rho_u = 0.48$}	

%% file: Appendix_Proof.tex
%
%
\appendices

\section{}\label{appx:proof1}

\short{\begin{theorem}\label{theo:T_s}
The \simplecharging{} station has a unique best-response price as follows:
{\tiny
\begin{subequations}\label{eq:prop1}
  \begin{align}[left ={T_s^{br}(T_r)  =\empheqlbrace}] 
 	 & t+(P_d-P_A)\frac{\bar\theta}{C_B}  & \mbox{ if }  T_r < (t+(P_d-P_A)\frac{\bar\theta}{C_B})\frac{P_A}{P_d} \label{eq:Ts_a}\\
	 & t+P_d\frac{\bar\theta}{C_B} & \mbox{ if } T_r > (t+P_d\frac{\bar\theta}{C_B})\frac{P_A}{P_d} \label{eq:Ts_b}\\
	 & T_r\frac{P_d}{P_A}  & \mbox{ otherwise} \label{eq:Ts_c}
   \end{align}
\end{subequations}}  
\end{theorem}}

\begin{IEEEproof}

The function~\eqref{eq:R_s} is continuous in $T_s$, and is differentiable over the intervals $(-\infty,\frac{P_d}{P_A}T_r)$ and $(\frac{P_d}{P_A}T_r,+\infty)$, with partial derivatives
{\tiny\begin{equation*}
\frac{\partial R_s}{\partial T_s} = \begin{cases}
(1+(T_s-t)(-\frac{C_B}{\bar\theta P_d})) C_B \exp(-\frac{C_BT_s}{\bar\theta P_d}) & T_s < \frac{P_d}{P_A}T_r\\
(1+(T_s-t)(-\frac{C_B}{\bar\theta(P_d-P_A)})) C_B \exp(-\frac{C_B(T_s-T_r)}{\bar\theta(P_d-P_A)}) & T_s >\frac{P_d}{P_A}T_r
\end{cases}
\end{equation*}}

We observe from~\eqref{eq:R_s} that the revenue is negative if $T_s<t$ and positive for $T_s\geq t$, hence we can restrict our attention to $T_s\geq t$.
In that region, each derivative is first strictly positive, then null at one point, then strictly negative, hence:
\begin{itemize}
\item if $t\geq \frac{P_d}{P_A}T_r$ only the interval $[t,+\infty)$ needs to be considered, and there is a unique revenue-maximizing price $t+(P_d-P_A)\frac{\bar\theta}{C_B}$ given by the first-order condition (note that it is an interior solution in the interval).

\item if $t<\frac{P_d}{P_A}T_r$, the revenue has a unique maximum on each of the intervals $[t,\frac{P_d}{P_A}T_r]$ and $[\frac{P_d}{P_A}T_r,+\infty)$:
\begin{itemize}
\item on the interval $[t,\frac{P_d}{P_A}T_r]$, the optimal price is 
$
t+P_d\frac{\bar\theta}{C_B}
$
if $t+P_d\frac{\bar\theta}{C_B} \leq \frac{P_d}{P_A}T_r$ (interior solution), and $\frac{P_d}{P_A}T_r$ otherwise (corner solution);
\item on the interval $[\frac{P_d}{P_A}T_r,+\infty]$, the optimal price is $t+(P_d-P_A)\frac{\bar\theta}{C_B}$
if $t+(P_d-P_A)\frac{\bar\theta}{C_B}  \geq \frac{P_d}{P_A}T_r$ (interior solution), and $\frac{P_d}{P_A}T_r$ otherwise (corner solution).
\end{itemize}
\end{itemize}

\end{IEEEproof}


\section{}\label{appx:proof2}

\begin{IEEEproof}
\begin{table*}[htbp]
   \centering
   \caption{Solution of $\frac{\partial R_r}{\partial T_r} = 0$ in different circumstances }
   \label{tab:Conditions}
   \begin{tabular}{|c|c|c||c|c|c|c|} 
   \hline
   \multicolumn{3}{|c||}{\multirow{2}{*}{Conditions on $E_r$}} & \multicolumn{2}{|c|}{$\frac{\partial R_r}{\partial T_r}$} & \multirow{2}{*}{Solution of $\frac{\partial R_r}{\partial T_r} = 0$} & \multirow{2}{*}{$T_r^{br}$}\\
   \cline{4-5}
   \multicolumn{3}{|c||}{}& $T_r = 0^-$ 
   & $T_r = 0^+$
   & & \\
   \hline
   		
   \multirow{3}{*}{$E_r < \frac{\bar\theta P_A}{C_B}$}
			& \multirow{2}{*}{$E_r<E_{r,2}$} & $E_{r,1} < E_r$ & $>0$ & $<0$ & None & $0$ \\
  \cline{3-7}			
			&& $E_r\le E_{r,1}$ & $>0$ & $\ge0$ & $\in\big[0,\min\{\frac{\bar\theta P_A}{C_B}-E_r,\frac{P_A}{P_d}T_s\}\big)$ & The solution of $\frac{\partial R_r}{\partial T_r} = 0$\\
  \cline{2-7}			
			& \multicolumn{2}{c||}{$E_{r,2}\le E_r$} &  $\le 0$ & $<0$ & $\in(-E_r,0]$ & The solution of $\frac{\partial R_r}{\partial T_r} = 0$\\
  \hline			
   \multirow{2}{*}{$\frac{\bar\theta P_A}{C_B} \le E_r$} & \multicolumn{2}{c||}{$E_r<E_{r,2}$} & $>0$ & $<0$ & None & $0$\\
   \cline{2-7}
   			& \multicolumn{2}{c||}{$E_{r,2}\le E_r$} & $\le 0$ & $<0$ & $\in(-E_r,0]$ & The solution of $\frac{\partial R_r}{\partial T_r} = 0$\\
    \hline			
   \end{tabular}
\end{table*}


In the first place, it is non-trivial to verify that~\eqref{eq:prop2} it is a function, i.e. 
$$\nexists T_s\in\mathbb{R}_{\ge0}: T_s\le-E_r\frac{P_d}{P_A} \mbox{ and } E_{r,1}(T_s) \le E_r \le E_{r,2}(T_s).$$ This is true because $\forall T_s\in\mathbb{R}_{\ge0}$, we have $0 < E_{r,1}(T_s) < E_{r,2}(T_s).$

The \regcharging{} station has nonnegative revenue when $-E_r \le T_r \le \frac{P_A}{P_d}T_s$ from~\eqref{eq:R_r}, so the following is a prerequisite:
{\tiny
\begin{equation*}\label{eq:I+}
E_r \ge -\frac{P_A}{P_d}T_s
\end{equation*}}
When this condition is not met, the \regcharging{} station would rather leave the market by setting a price sufficiently high i.e. $T_r^{br}\ge\frac{P_A}{P_d}T_s$ such that no client would come.

If $E_r \ge -\frac{P_A}{P_d}T_s$, and $-E_r \le T_r \le \frac{P_A}{P_d}T_s$, we further examine the partial derivative of the revenue function~\eqref{eq:R_r}:
{\tiny
\begin{subequations}\label{eq:Rr}
   \begin{align}[left ={\frac{\partial R_r}{\partial T_r} = \empheqlbrace}]
  &  C_B\big(1-\exp[-\frac{C_B(T_s-T_r)}{\bar\theta(P_d-P_A)}][1+\frac{C_B(T_r+E_r)}{\bar\theta(P_d-P_A)}]\big) &\mbox{ if }T_r<0\label{eq:d_Rr_R-}\\
  &  C_B\{\exp(-\frac{C_BT_r}{\bar\theta P_A})[1-\frac{C_B(T_r+E_r)}{\bar\theta P_A}] & \label{eq:d_Rr_R+1}\\
  & ~~-\exp[-\frac{C_B(T_s-T_r)}{\bar\theta(P_d-P_A)}][1+\frac{C_B(T_r+E_r)}{\bar\theta(P_d-P_A)}]\}& \mbox{ if } T_r>0
   \label{eq:d_Rr_R+2}
   \end{align}
\end{subequations}}

We begin with putting the boundaries of $T_r$ into~\eqref{eq:Rr}:
{\tiny\begin{align}
& \frac{\partial R_r}{\partial T_r}|_{T_r = -E_r} >0 & \mbox{ if } E_r\ne0 \label{eq:partialTrmin}\\
& \frac{\partial R_r}{\partial T_r}|_{T_r = 0^-} > 0~;~\frac{\partial R_r}{\partial T_r}|_{T_r = 0^+} >0 & \mbox{ if } E_r = 0 \label{eq:partialTr0}\\ 
& \frac{\partial R_r}{\partial T_r}|_{T_r = \frac{P_A}{P_d}T_s} < 0.& \label{eq:partialTrmax}
\end{align}}
Then within those bounds, we say~\eqref{eq:Rr} is strictly decreasing on $\big(\min\{0,-E_r\},\max\{0,\min\{\frac{\bar\theta P_A}{C_B}-E_r,\frac{P_A}{P_d}T_s\}\}\big)$ because:
\begin{itemize}
\item when $T_r < 0$, \eqref{eq:d_Rr_R-} is strictly decreasing;
\item when $T_r > 0$, noticing that~\eqref{eq:d_Rr_R+1} is positive iff $T_r < \frac{\bar\theta P_A}{C_B}-E_r$ and is strictly decreasing when positive, meanwhile~\eqref{eq:d_Rr_R+2} is strictly decreasing and negative, $\frac{\partial R_r}{\partial T_r}$ is strictly decreasing for $T_r\in(0,\max\{0,\min\{\frac{\bar\theta P_A}{C_B}-E_r,\frac{P_A}{P_d}T_s\}\})$;
\item when $T_r = 0$, $\frac{\partial R_r}{\partial T_r}|_{T_r = 0^-}>\frac{\partial R_r}{\partial T_r}|_{T_r = 0^+}$.
\end{itemize}

From the monotony of $\frac{\partial R_r}{\partial T_r}$, we know that there is either a unique solution of $\frac{\partial R_r}{\partial T_r} = 0$ or none. Table~\ref{tab:Conditions} summarizes the conditions for each of them to occur, together with the intervals wherein lies those possible solutions. 

Jointly consider the first order optimality condition of $\frac{\partial R_r}{\partial T_r}=0$ and the boundary values in~\eqref{eq:partialTrmin}~\eqref{eq:partialTr0} and~\eqref{eq:partialTrmax}, we conclude that $R_r$ achieves the maximum either at the unique solution of $\frac{\partial R_r}{\partial T_r} = 0$ (if it exist) or at $T_r = 0$, as stated in the last column of table~\ref{tab:Conditions}, thus the optimality of the best-respond price in~\eqref{eq:prop2} is proved. 

\end{IEEEproof}


\section{}\label{appx:proof3}

\short{\begin{theorem}\label{theo:Nash}

The pricing game defined in~\ref{game:Def} has either a unique Nash equilibrium or a unique Pareto-dominant one when there exist infinite number of Nash equilibria. The equilibria prices in different circumstances are:
{\tiny
\begin{subequations}
\begin{align}[left ={N^E: \empheqlbrace}]
& T_r = -E_r;~T_s = -\frac{P_d}{P_A}E_r \nonumber \\
&~~~ \mbox{ if } E_r \le -\frac{P_A}{P_d}[t+(P_d-P_A)\frac{\bar\theta}{C_B}] \label{eq:N1}\\
& T_r \in(0,\min\{\frac{\theta P_A}{C_B}-E_r,\frac{P_A}{P_d}T_s)\};~T_s = t+(P_d-P_A)\frac{\bar\theta}{C_B} \nonumber\\
&~~~ \mbox{ if }-\frac{P_A}{P_d}[t+(P_d-P_A)\frac{\bar\theta}{C_B}] < E_r < E_{r,1}\big(t+(P_d-P_A)\frac{\bar\theta}{C_B}\big)\label{eq:N2}\\
& T_r = 0;~T_s = t+(P_d-P_A)\frac{\bar\theta}{C_B} \nonumber\\
&~~~ \mbox{ if }E_{r,1}\big(t+(P_d-P_A)\frac{\bar\theta}{C_B}\big) \le E_r \le E_{r,2}\big(t+(P_d-P_A)\frac{\bar\theta}{C_B}\big) \label{eq:N3}\\
& T_r \in(-E_r,0);~T_s = t+(P_d-P_A)\frac{\bar\theta}{C_B} \nonumber \\
&~~~ \mbox{ if }E_{r,2}\big(t+(P_d-P_A)\frac{\bar\theta}{C_B}\big) < E_r \label{eq:N4}
\end{align}
\end{subequations}}

\end{theorem}}

\begin{IEEEproof}

We prove the existence and uniqueness of the Nash equilibrium through exhaustively combine~\eqref{eq:Ts_a}~\eqref{eq:Ts_b}~\eqref{eq:Ts_c} and \eqref{eq:Tr_a}~\eqref{eq:Tr_b}~\eqref{eq:Tr_c}, in order to find possible intersections between the two best-response prices.

First of all, \eqref{eq:Ts_b} can be easily excluded since according to Proposition~\ref{theo:T_r}, $T_r^{br}(T_s)\le T_s\frac{P_A}{P_d}$, conflict with the condition in~\eqref{eq:Ts_b}. 

Then we check whether~\eqref{eq:Ts_a} intersect with~\eqref{eq:prop2}. Putting $T_s = t+(P_d-P_A)\frac{\bar\theta}{C_B}$ into~\eqref{eq:prop2} gives: 
{\tiny
\begin{subequations}\label{eq:Ts_across}
  \begin{align}[left ={T_r^{br}\empheqlbrace}] 
	& =\big(t+(P_d-P_A)\frac{\bar\theta}{C_B}\big)\frac{P_A}{P_d} & \mbox{ if } t+(P_d-P_A)		\frac{\bar\theta}{C_B} \le -E_r\frac{P_d}{P_A} \label{eq:Ts_across1}\\
	& < \big(t+(P_d-P_A)\frac{\bar\theta}{C_B}\big)\frac{P_A}{P_d} & \mbox{ otherwise.}\label{eq:Ts_across2}
  \end{align}
\end{subequations}}
Note that~\eqref{eq:Ts_across1} corresponds to the first case in~\eqref{eq:prop2} while~\eqref{eq:Ts_across2} the rest. Comparing the values in~\eqref{eq:Ts_across1} with the condition in~\eqref{eq:Ts_a}, we can further rule out the~\eqref{eq:Ts_a}\eqref{eq:Tr_a} pair. 

As a result, if $-\frac{P_A}{P_d}[t+(P_d-P_A)\frac{\bar\theta}{C_B}] < E_r$, \eqref{eq:Ts_a} has an intersection with~\eqref{eq:prop2}, which provides a Nash equilibrium. Depending on the value of $E_r$, the price profile of this equilibrium falls in different segments, as expressed in~\eqref{eq:N2}, \eqref{eq:N3} and~\eqref{eq:N4}. Otherwise, when $E_r \le -\frac{P_A}{P_d}[t+(P_d-P_A)\frac{\bar\theta}{C_B}]$, we end up with infinite intersections between~\eqref{eq:Ts_c} and~\eqref{eq:Tr_a}. Among all the possible equilibria ($T_s \le -E_r\frac{P_d}{P_A}\mbox{, }T_r = T_s\frac{P_A}{P_d}$), $T_s = -E_r\frac{P_d}{P_A}\mbox{, }T_r = T_s\frac{P_A}{P_d}$ Pareto dominates the rest because the \regcharging{} station is indifferent towards the choices of $T_r$ given $T_r = T_s\frac{P_A}{P_d}$, whereas the \simplecharging{} station prefers $T_s = -E_r\frac{P_d}{P_A}$ rather than $T_s < -E_r\frac{P_d}{P_A}$. This preference is from:
{\tiny$$
\frac{\text{d} R_s(T_r = T_s\frac{P_A}{P_d})}{\text{d}T_s}|_{T_s \le t+(P_d-P_A)\frac{\bar\theta}{C_B}} >0
$$}
The conditions for each equilibrium to occur are exclusive and cover all possible circumstances.
\end{IEEEproof}
